\documentclass[preprint,12pt]{elsarticle}
\usepackage[margin=2.5cm]{geometry} 
\usepackage{setspace}
\usepackage{amsthm}
\theoremstyle{definition}

\usepackage{graphicx}
\usepackage{dcolumn}
\usepackage{multirow}
\usepackage{subfigure}
\usepackage{times,mathptm}
\usepackage{float}
\usepackage{color}
\usepackage{amsmath,amsfonts}
\usepackage{mathptmx}
\usepackage{mathrsfs}
\usepackage{bbm}
\usepackage{bm}
\usepackage{xfrac}

\newcommand{\beq}{\begin{equation}}
\newcommand{\eeq}{\end{equation}} 
\newcommand{\bea}{\begin{eqnarray}}
\newcommand{\eea}{\end{eqnarray}} 

\newcommand{\Om}{\Omega}
\newcommand{\up}{\uparrow}
\newcommand{\dn}{\downarrow}

\newcommand{\x}{\mathbf{x}}

\newcommand{\ua}{\uparrow}
\newcommand{\da}{\downarrow}

\renewcommand{\d}{\delta}

\newcommand{\p}{\psi}
\renewcommand{\b}{\beta}
\renewcommand{\a}{\alpha}

\renewcommand{\ni}{\noindent}
\newcommand{\bx}{\mathbf{x}}

\newcommand{\bk}{\mathbf{k}}

\newcommand{\vx}{{\vec{x}}}

\newcommand{\vk}{{\vec{k}}}

\newcommand{\m}{\mu}

\renewcommand{\r}{\rho}
\newcommand{\e}{\epsilon}

\renewcommand{\k}{\kappa}

\newcommand{\bp}{\mathbf p}

\newcommand{\oh}{\frac{1}{2}}

\newcommand{\dg}{\dagger}
\newcommand{\non}{\nonumber}

\newcommand{\rf}[1]{(\ref{#1})}
\newcommand{\ra}{\rightarrow}

\renewcommand{\vec}[1]{\bm #1}

\usepackage{ulem}

\journal{Annals of Physics}
\begin{document}
\begin{frontmatter}

\title{Hartree-Fock with Nambu spinors, and d-wave condensation in the 2D Hubbard model} 

\bigskip
\bigskip

\author[1]{Kazue Matsuyama} 
\ead{kazuem@sfsu.edu}
\author[1]{Jeff Greensite}
\ead{greensit@sfsu.edu}
\address[1]{Physics and Astronomy Department, San Francisco State University, \\ 1600 Holloway Ave,
San Francisco, CA 94132 USA}
\bigskip
%
\begin{abstract}

 
     The usual Hartree-Fock approximation to the Hubbard model is based on eigenstates of
the electron number operator.  But this formulation is not unique.  A different (and inequivalent) version,
formulated in terms of Nambu two-component spinors, is based on eigenstates of the difference between the numbers of spin up and spin down electrons, with electron density dependent on parameters $U,t,t'$ and chemical potential $\mu$.  The advantage
of this formulation is that electron pairing condensates can be directly computed. We show that in
the ground states away from half-filling, obtained in this ``Nambu'' Hartree-Fock approximation, a discrete rotation
symmetry is spontaneously broken, and the condensates exhibit the expected d-wave form in momentum
space.   We also show that the Mott insulator electron configuration is obtained in this formulation at large $U/t$ and half-filling,
and roughly locate the boundary, in the hole doping$-$$U/t$ plane, between a region of local antiferromagnetism and
stripe/domain formation, and the region of d-wave condensation.

 \end{abstract}

\end{frontmatter}

\bibliographystyle{elsarticle-num} 

\section{\label{Intro} Introduction}

    The application of the Hartree-Fock method to the Hubbard model in two and three space dimensions has a long history, dating back to 1985 \cite{Hirsch} (or even earlier \cite{Penn}), and a sample of later work is found in 
\cite{Poilblanc,Zaanen,Machida,Schulz1,Schulz2,Ichimura,Verges,Inui,Dasgupta,Xu}. The strengths and limitations of this method are well known, and are described in a number of reviews \cite{Scalettar,Powell,Lechermann,Imada,Fazekas}.  Briefly, the main successes are, first, an understanding of the emergence of ferromagnetism in the context of the Hubbard model, which goes back to the very early work of \cite{Hirsch},
and the relation to the Stoner criterion.  Second is the prediction of stripes in certain parameter ranges \cite{Poilblanc,Zaanen}, a feature later observed experimentally.  There are, however, many shortcomings related to the neglect, in the mean-field approach, of electron correlations.  For one thing, predictions of transition temperatures and critical exponents are often wildly inaccurate \cite{Powell}, and we will have nothing to say about that here.  But there is another
serious deficiency of the Hartree-Fock approach that we will address in this article.  This is the difficulty in computing a pairing condensate, particularly the d-wave condensate associated with cuprate superconductivity.  Since the Hubbard model is considered to be a candidate for understanding cuprate superconductivity, that is a very serious shortcoming of the Hartree-Fock approach.  
  
      In this article we will reformulate the Hubbard
model in terms of Nambu two-component spinors, and apply the Hartree-Fock technique to the two-dimensional Hubbard Hamiltonian
in this formulation. The advantage, as we show here, is that the emergence of 
$\mbox{d-wave}$ condensation is natural and even, at $U=0$, rather trivial. Obviously, in any formulation, the
Hartree-Fock method is still only an approximation.  But within that approximation, and with appropriate caveats, 
we are able to study the emergence of d-wave condensation in terms of the parameters of the Hamiltonian and the associated electron density, up to large values of $U/t$.  We also find a Mott insulator configuration at large $U/t$
and half-filling, i.e.\ localized electrons in which all nearest neighbors of an electron of spin up (down) are electrons of spin down (up), along with a band gap.  We will see that this configuration is obtained, in the Nambu Hartree-Fock formulation, in an unexpected way.

    In section \ref{NambuForm} we reformulate the Hubbard model in Nambu spinor variables and from there derive the corresponding Hartree-Fock Hamiltonian.  This section also contains details about our iterative procedure for finding the ground state
of this Hamiltonian.  In section \ref{U0soln} the model is solved analytically at $U=0$ and half-filling, where we see a pairing condensate emerge rather trivially, and also solved numerically at $U=0$ away from half-filling. Section \ref{SSB} presents some results at small non-zero $U/t$,
including the emergence of a d-wave condensate associated with spontaneous symmetry breaking.  We also display stripe (or domain) patterns at moderate $U/t$.  In section \ref{Mott} we present our results for the electron configuration at large $U/t$ and half-filling, which turns out to be the Mott insulator configuration,  and discuss in some detail the anti-correlation of local antiferromagnetism and d-wave condensation, along with a rough plot of the boundary between these two regions.  In section \ref{Compare} we compare the energy expectation value of the full Hubbard Hamiltonian in ground states obtained by the standard and Nambu Hartree-Fock approaches, at various $U,t'$ values.  All computations in this article are carried out at zero temperature. Conclusions are in section \ref{Conclude}.

\bigskip

\section{\label{NambuForm} Hartree-Fock in Nambu spinor variables}

    We begin with the Hubbard model Hamiltonian 
 \bea
 H &=& -t \sum_{<xy>} (c^\dg(x,\ua) c(y,\ua) + c^\dg(x,\da) c(y,\da)) 
       -t' \sum_{[xy]} (c^\dg(x,\ua) c(y,\ua) + c^\dg(x,\da) c(y,\da)) \non \\
 & &  + U\sum_x c^\dg(x,\ua) c(x,\ua) c^\dg(x,\da) c(x,\da)
      - \mu \sum_x  (c^\dg(x,\ua) c(x,\ua) + c^\dg(x,\da) c(x,\da)) \ ,
\label{Hub}
 \eea
 where $<xy>$ and $[xy]$ denote nearest and next-nearest neighbors respectively.
The essence of the Hartree-Fock approximation is to replace this operator with an operator $H_{eff}$ which is only
quadratic in electron operators $c,c^\dg$.  Let $|\Om\rangle$ be a state (usually the ground state) of interest.  If we strip away two fermionic operators from $H_{eff}$ and $H$ (by the double
commutators below), then $H_{eff}$ goes to a constant while the reduced $H$ is still quadratic in fermion operators.  The best one can do is to require that the constant is equal to the expectation value of the
reduced $H$ in the state of interest.  More precisely, we define $H_{eff}$ from the requirement that, for all $x,s$ and $x',s'$,
\bea
\{c^\dg(x,s), [H_{eff}, c(x',s')] \} &=& \langle \Om |\{ c^\dg(x,s), [ H, c(x',s')] \} | \Om \rangle \non \\
\{ c^\dg(x,s), [H_{eff}, c^\dg(x',s')] \} &=& \langle \Om |\{ c^\dg(x,s), [ H, c^\dg(x',s')]\} | \Om \rangle \non \\
\{ c(x,s), [H_{eff}, c(x',s')] \} &=& \langle \Om |\{ c(x,s), [ H, c(x',s')] \} | \Om \rangle \ , \non \\
\label{dcom}
\eea
where $[...], \{...\}$ represent commutators and anticommutators respectively.
Defined in this way, $H_{eff}$ of course depends on $|\Om\rangle$ (and vice-versa).  The result is
 \bea
  \lefteqn{H_{eff}(\Om)} \non \\
   &=& -t \sum_{<xy>} (c^\dg(x,\ua) c(y,\ua) + c^\dg(x,\da) c(y,\da))
   -t' \sum_{[xy]} (c^\dg(x,\ua) c(y,\ua) + c^\dg(x,\da) c(y,\da)) \non \\
   & &     - \mu \sum_x  (c^\dg(x,\ua) c(x,\ua) + c^\dg(x,\da) c(x,\da))\non \\
  & &+ U \sum_x \bigg\{ \langle \Om |c^\dg(x,\ua) c(x,\ua)|\Om \rangle c^\dg(x,\da) c(x,\da) 
 +  \langle \Om |c^\dg(x,\da) c(x,\da)|\Om \rangle c^\dg(x\ua) c(x\ua) \non \\
  & &  -  \langle \Om |c^\dg(x,\da) c(x,\ua)|\Om \rangle c^\dg(x,\ua) c(x,\da)
 - \langle \Om |c^\dg(x,\ua) c(x,\da)|\Om \rangle c^\dg(x,\da) c(x,\ua) \non \\
 & &  -  \langle \Om | c^\dg(x,\ua) c^\dg(x,\da)|\Om\rangle c(x,\ua) c(x,\da)
   -   \langle \Om | c(x,\ua) c(x,\da)|\Om\rangle c^\dg(x,\ua) c^\dg(x,\da) \bigg\} \non \\
  & &  = K - \m \tilde{N} + H_{direct} + H_{ex} + \mbox{electron number changing terms} \ .
\label{Heff1}
 \eea 
 One could also add a constant so that $\langle \Om| H_{eff} |\Om\rangle =  \langle \Om| H |\Om\rangle$,
 but this will not be important for what follows.
 If we restrict our attention to states $|\Om\rangle$ which are eigenstates of total electron number
 \beq
          \tilde{N} = \tilde{N}_\up + \tilde{N}_\dn = \sum_x (c^\dg(x,\ua)c(x,\ua) + c^\dg(x,\da)c(x,\da) ) \ ,
 \eeq
 then it is obvious that such states cannot be eigenstates of $H_{eff}$.  In the usual approach one simply drops (or ignores) the electron number-changing terms, and then 
 ${H_{eff}(\Om) |\Om\rangle = E| \Om\rangle}$ becomes a self-consistency condition, where
 \beq
         |\Om \rangle = \prod_{i=1}^M \bigg( \sum_{x_i,s_i} \phi_i(x_i,s_i) c^\dg(x_i,s_i) \bigg) |0\rangle
 \label{Om1}
 \eeq
is also an eigenstate of particle number.  The procedure for finding numerically the one-particle wave functions 
$\phi_i(x,s)$ has been described by many authors, e.g.\ \cite{Poilblanc,Zaanen,Machida,Schulz1,Schulz2,Ichimura,Verges,Inui,Dasgupta,Imada,Fazekas,Xu}, and most recently by us in \cite{Matsuyama:2022kam}.  This is the standard Hartree-Fock treatment of the Hubbard model.

On the other hand there is nothing sacred about eigenstates of particle number, and in this article we
consider re-expressing the Hubbard Hamiltonian in terms of Nambu spinors
\bea
 \psi(x) &=& \left[ \begin{array}{c} c(x,\ua) \cr c^\dg(x,\da) \end{array} \right] 
    ~~~,~~~    \psi^\dg(x) = [c^\dg(x,\ua) , c(x,\da) ] \ ,\non \\
 \eea
 and the Hubbard Hamiltonian is
 \bea
 H   &=& -t \sum_{<xy>} \{\p^\dg_1(x) \p_1(y) - \p^\dg_2(x) \p_2(y)\} 
              -t' \sum_{[x]} \{\p^\dg_1(x) \p_1(y) - \p^\dg_2(x) \p_2(y)\}  \non \\
    & &+ U \sum_x \p^\dg_1(x) \p_1(x) 
       -U \sum_x \p^\dg_1(x) \p_1(x) \p^\dg_2(x) \p_2(x) \non \\
     & &   - \m  \sum_{x} \{\p^\dg_1(x) \p_1(x) + 1 - \p^\dg_2(x) \p_2(x)\} \ ,
 \eea
 where $\psi_1,\psi_2$ are the upper and lower components, respectively, of the Nambu spinor $\psi$.
 The effective action, quadratic in the fermion operators, is obtained by either the double commutator
 approach of eq.\ \rf{dcom} (replacing $c,c^\dg$ by $\p,\p^\dg$), or by simply rewriting \rf{Heff1} in terms
 of Nambu spinors.  Either way, dropping irrelevant constants, we find
 \bea
  \lefteqn{H_{eff}(\Om)} \non \\
  &=&  
  -t \sum_{<xy>} ( \psi_1^\dg(x) \psi_1(y) - \psi_2^\dg(x) \psi_2(y))  
  -t' \sum_{[xy]} ( \psi_1^\dg(x) \psi_1(y) - \psi_2^\dg(x) \psi_2(y))  \non \\
  & &
  - \mu \sum_x  ( \psi_1^\dg(x) \psi_1(x) - \psi_2^\dg(x) \psi_2(x))  \non \\
 & &  + U \sum_x  \bigg\{ \psi_1^\dg(x) \psi_1(x) 
 - \langle \Om|\psi^\dg_2 \psi_2|\Om\rangle \psi^\dg_1 \psi_1  - \langle \Om|\psi^\dg_1 \psi_1|\Om\rangle \psi^\dg_2 \psi_2 \non \\
& &  - \langle \Om|\psi_2 \psi_1|\Om\rangle \psi_1^\dg \psi^\dg_2
 -\langle \Om|\psi^\dg_1 \psi_2^\dg|\Om\rangle \psi_2 \psi_1  
  + \langle \Om|\psi_1^\dg \psi_2|\Om\rangle \psi^\dg_2 \psi_1 + 
  \langle \Om|\psi^\dg_2 \psi_1|\Om\rangle \psi_1^\dg \psi_2 \bigg\}\ .
 \label{Heff2}
 \eea
 This time we will restrict our attention to eigenstates of the $\psi$ number operator
 on a periodic $L\times L$ lattice,
\bea
        N_\psi &=& \sum_{x} (\psi_1^\dg(x) \psi_1(x) + \psi_2^\dg(x) \psi_2(x)) \non \\
            &=& \sum_{x} (c^\dg(x,\ua) c(x,\ua) - c^\dg(x,\da) c(x,\da) + 1) \non \\
            &=& N_{\ua} - N_{\da} + L^2 \ ,
\eea
 and this means that we drop terms proportional to $\psi \psi$ and $\psi^\dg \psi^\dg$,
and compute eigenstates of the resulting $H_{eff}$ of the form
 \beq
         |\Om \rangle = \prod_{i=1}^M \bigg( \sum_{x_i,\a_i} \phi_i(x_i,\a_i) \psi^\dg_{\a_i}(x_i) \bigg) |0\rangle \ ,
 \label{Om2}
 \eeq
 where now the Fock vacuum is defined by $\psi_\a(x) |0\rangle = 0$, and $M$ is an eigenvalue of $N_\psi$.  In the multi-particle state 
 $|\Om\rangle$ the one-particle states 1 to $M$ are occupied, and states $M+1$ to $2L^2$ are
 unoccupied. Note that the Fock vacuum is not equivalent
 to the $|0\rangle$ state in \rf{Om1}.  
 
 The state $|\Om\rangle$ defined in \rf{Om2} is obviously an eigenstate of $N_\psi$ with eigenvalue $M$, which means it is an eigenstate of the {\it difference} $N_\ua - N_\da$.  In this article we will restrict our computations to the states with $N_\ua=N_\dn$
 which means that we consider only states with $M=L^2$, each of which will have a different electron density $\langle \Om| (N_\ua + N_\da) |\Om\rangle$.  This density can be adjusted by varying the chemical potential.  Even restricting to equal numbers of up and down spin
 electrons, the local magnetization may be antiferromagnetic, ferromagnetic, or negligible, see section \ref{ferro} below.
 
 In the numerical calculations, $t'=0$ unless otherwise indicated.

    The operators $H_{eff}$ in eqs.\ \rf{Heff1} and \rf{Heff2} are identical up to constants, and in the standard approach one drops the electron number-changing terms proportional to $cc$ and $c^\dg c^\dg$.  In the Nambu spinor approach we propose to drop instead the terms proportional to $\psi \psi$ and $\psi^\dg \psi^\dg$. After dropping those terms, the truncated $H_{eff}$ operators in the standard and Nambu spinor formulations are simply different. In effect we are trading the exchange terms, which are kept in the standard approach, for the electron number-changing terms, which are kept in this Nambu spinor approach. Either truncation amounts to an approximation on top of an approximation. Let $|\Om^S\rangle, |\Om^N\rangle$ denote the Hartree-Fock ground states in the standard and Nambu formulations respectively. Then we judge which is closer to the true ground state of the Hubbard model by a comparison of the energy expectation values $\langle \Om^S|H|\Om^S\rangle$ and $\langle \Om^N|H|\Om^N\rangle$, where $H$ is the full Hubbard Hamiltonian \rf{Hub}.  
 
 The great advantage of the Nambu approach is that once the state $|\Om\rangle$ in \rf{Om2} is determined, one can easily investigate the momentum space distribution of the condensate via
 \bea
 P({\bf k}) &=& {1\over L^2} \langle \Om | c^\dg({\bf k},\ua) c^\dg({-\bf k},\da) | \Om \rangle \non \\
 &=& {1\over L^2} \langle \Om | \psi^\dg_1({\bf k}) \psi_2({\bf k}) | \Om \rangle \ ,
 \eea
 whereas these quantities simply vanish in the electron number eigenstates.  In the standard approach one could try to investigate the expectation values of pairs of well-separated pairing operators $\langle (cc)_x (c^\dg c^\dg)_y \rangle$ with $|x-y| \gg 1$, which is non-zero in electron number eigenstates, but it could be that by dropping the number-changing terms in the Hamiltonian one is losing something essential regarding pairing.  
 
The downside is that by dropping the $\psi \psi$ and $\psi^\dg \psi^\dg$ terms, we are discarding the electron spin exchange terms, and the question is whether that is justified.  Here it should be noted that some of the early work on the Hartree-Fock method applied to the
Hubbard model, e.g.\  \cite{Hirsch,Poilblanc} as well as some of the later work, e.g.\ \cite{Xu}, also drop the exchange term, by making the (incomplete) mean-field approximation
\beq
           c^\dg_\ua c_\ua c^\dg_\dn c_\dn \longrightarrow  \langle c^\dg_\ua c_\ua \rangle c^\dg_\dn c_\dn
           +  c^\dg_\ua c_\ua \langle c^\dg_\dn c_\dn \rangle -  \langle c^\dg_\ua c_\ua \rangle \langle c^\dg_\dn c_\dn \rangle
\eeq
and the results of this incomplete approximation do not seem to be altered drastically by inclusion of the exchange terms.  And, of course,
even the Hartree-Fock computations which retain the exchange term are dropping, without justification, 
the number-changing terms.   We can do a little better by asking whether the Nambu approach recovers one of the main achievements of the standard Hartree-Fock treatment of the 2D Hubbard model, which is
the discovery of stripes and domains.  This question is investigated below in section \ref{SSB}, where show that such domains survive
after discarding the exchange terms in the Nambu Hartree-Fock approach.
 
  In any case, since the standard approach has been thoroughly investigated while the Nambu spinor approach has not (at least not to our knowledge), the properties of the latter will be the focus of our work in this article.

Introducing the notation
\bea
 \r(x,\a\b) &=& \langle \Om| \psi^\dg_\a(x) \psi_\b(x) |\Om\rangle  
\eea
we have, after dropping the $\psi \psi$ and $\psi^\dg \psi^\dg$ terms,
\bea
 \lefteqn{H_{eff}(\Om)} \non \\
  &=& -t \sum_{<xy>} ( \psi_1^\dg(x) \psi_1(y) - \psi_2^\dg(x) \psi_2(y)) -t' \sum_{[xy]} ( \psi_1^\dg(x) \psi_1(y) - \psi_2^\dg(x) \psi_2(y)) \non \\
   & & - \m \sum_{x} ( \psi_1^\dg(x) \psi_1(x) - \psi_2^\dg(x) \psi_2(x)) \non \\
 & & + U \sum_x  \bigg\{ [1-\r(x,22)] \psi^\dg_1(x) \psi_1(x)  - \r(x,11) \psi^\dg_2(x) \psi_2(x) 
   + \r(x,12) \psi^\dg_2 \psi_1 + \r(x,21)\psi_1^\dg \psi_2 \bigg\} \non \\
  &=& \sum_x \sum_y \psi^\dg_\a(x) [H]_{x\a,y\b} \psi_\b(y) \ .
\label{Heff3}
 \eea
 From this we can read off the non-zero elements of the ${2L^2 \times 2L^2}$ matrix $[H]_{x\a,y\b}$
 \begin{itemize}
\item $x=y$
\bea
       H_{x1,x1} &=& U(1 - \r(x,22)) - \m \non \\
       H_{x2,x2} &=& -U\r(x,11) + \m \non \\
       H_{x1,x2} &=& U\r(x,21) \non \\
       H_{x2,x1} &=& U\r(x,12)
\eea
\item  $x,y$ nearest neighbors
\bea
       H_{x1,y1} &=& -t   ~~~,~~~       H_{x2,y2} =  t
\eea
\item  $x,y$ next-nearest neighbors
\bea
       H_{x1,y1} &=& -t'   ~~~,~~~       H_{x2,y2} =  t'
\eea
\end{itemize}

The iterative procedure begins with some random initialization of $\r(x,\a\b)$.  The details of the initialization shouldn't
matter much, and they certainly don't at small $U$.  Still, as we emphasized in  \cite{Matsuyama:2022kam} there are a
vast number of self-consistent ``ground states'' of the effective Hamiltonian, with very nearly the same energy, which is 
a situation very reminiscent of a spin glass.
At large $U$, we find that the ``zero occupancy'' initialization usually leads to slightly lower ground state energies as
compared to other initializations.  The zero occupancy initialization begins with setting all $\rho(x,11)=0, \rho(x,22)=1$, which corresponds to having no electrons of either spin on any site.  A stochastic element is introduced by initializing 
$\rho(x,12)=\rho(x,21)$ to a random number, different at each point, in the range $[-1,1]$.  The next step is to solve numerically the sparse
matrix eigenvalue equation
\beq
            [H]_{x\a,y\b} \phi_i(y,\b) = \e_i \phi_i(x,\a) \ ,
\label{eval}
\eeq
and insert the appropriate $\phi_i$ into \rf{Om2}, which gives a new estimate for the $\rho(x,\a\b)$
\beq
 \rho^{new}(x,\a\b)= \sum_{i=1}^{L^2} \phi_{i}^*(x,\a) \phi_{i}(x,\b) \ .
\eeq
Rather than simply adopting $\rho^{new}$ as the new estimate for $\rho$, we have found that convergence is improved
at large $U$ by choosing
\beq
      \rho(x,\a\b) = \rho^{old}(x,\a\b) + \kappa(\rho^{new}(x,\a\b) - \rho^{old}(x,\a\b))
\eeq
where $\rho^{old}(x,\a\b)$ was the $\rho$ obtained at the previous iteration.  We found, by trial and error, that good convergence at large $U$ is obtained with the choice $\k=0.2$, and that variations of $\k$ around this value do not affect the final results. 
The new $\rho(x,\a\b)$ give new matrix elements  $[H]_{x\a,y\b}$, and again we solve the eigenvalue equation
\rf{eval} and compute another set of  $\rho(x,\a\b)$, repeating these steps until the average value of 
$(\rho(x,\a\b)-\rho^{old}(x,\a\b))^2$  is less than $10^{-10}$. At that point
we compute observables.

   Although we are interested here in the Nambu Hartree-Fock ground state \rf{Om2} for given $U,t,t',\m$ (and we set $t=1$ throughout in our numerical work), we note in passing that
there are many eigenstates of $N_\psi$ of the same form, obtained by
introducing the set of indices $\{n(i), i=1,2,...,M\}$, and states
\beq
     |\Om \rangle = \prod_{i=1}^M \left\{\sum_{x_i,s_i} \phi_{n(i)}(x_i,\a_i) \psi_{\a_i}^\dg(x_i) \right\}|0\rangle \ ,
\label{ni}
\eeq
with
\beq
 \rho(x,\a\b)= \sum_{i=1}^{M} \phi_{n(i)}^*(x,\a) \phi_{n(i)}(x,\b) \ .
\eeq
Simple combinatorics tells us that there are $\sim 2^{2L^2}$ states with $N_\psi=L^2$.  Here we only investigate
the lowest energy state \rf{Om2} in this set, which corresponds to ${\{n(i)=i, i=1,2,...,L^2\}}$.  

Let us first consider the analytically soluble case, which is ${U=0}$.  There we already see, in this Nambu spinor formulation, a distribution which is strongly reminiscent of d-wave pairing.

\section{\label{U0soln} Pairing at $U=t'=0$}

  At $U=0$ there is no mixing of the $\psi_1$ and $\psi_2$ operators, and at $t'=0$ the eigenstates/eigenvalues of the
$[H]$ matrix are the usual plane waves
 \bea
      \phi^{1}_\bk(x) &=& {1\over L} \left[ \begin{array}{c} 
                  e^{i(k_x x + k_y y)} \cr 0 \end{array} \right] ~~,~~ \e^1_\bk = -2t (\cos(k_x)+\cos(k_y)) -\m \non \\
      \phi^{2}_\bk(x) &=& {1\over L} \left[ \begin{array}{c} 
                  0 \cr e^{i(k_x x + k_y y)}  \end{array} \right]     ~~,~~ \e^2_\bk = 2t (\cos(k_x)+\cos(k_y) ) + \m \ . \non \\           
 \eea
The ground state will consist of occupying all the negative energy states.
which means that the following states have occupation number one (i.e.\ belong in the product \rf{Om2})
\bea
\begin{array}{cl}
  \phi^{1}_\bk & \text{for all} ~~ 2t (\cos(k_x)+\cos(k_y)) + \m > 0 ~~~ \mbox{(region I)}\cr
  \phi^{2}_\bk & \text{for all} ~~ 2t (\cos(k_x)+\cos(k_y)) + \m < 0 ~~~ \mbox{(region II)}\end{array} \ . \non \\
\label{occ}
\eea
All other energy eigenstates have occupation number zero except for the zero energy states with momenta
satisfying
\beq
      2t (\cos(k_x)+\cos(k_y) ) + \m = 0 \ ,
\label{fermi}
\eeq
assuming, on a finite lattice, that there are momenta $(k_x,k_y)$ which satisfy this condition.
We will, with perhaps a slight abuse of language, refer to these momenta as the ``Fermi surface,'' assuming they
exist.  If not, then we use the term ``Fermi boundary'' to refer to a contour separating regions I and II. 

   Let us begin with the simplest case of $\m=0$.  In that case, momenta satisfying the condition 
$\cos(k_x)+\cos(k_y) = 0$ occupy, in the square    
domain $-\pi  \le k_x,k_y < \pi$, a diamond-shaped contour consisting of four lines
\bea
       k_x + k_y = \pm \pi ~~,~~ k_x - k_y = \pm \pi \ .
\eea
Because the total number of states is $L^2$ (recall that this corresponds to equal numbers of up and down electrons, not the total number of electrons), each momentum along the Fermi surface must have occupation number one,
but the state at each momentum can be any superposition of $\phi^1_k$ and $\phi^2_k$.  In other words, at
$U=t'=\m=0$ the ground state is degenerate.

   Now let us consider the pairing correlator
\bea
        P(\bp) &=& {1\over L^2} \sum_x \sum_{x'} \langle \Om | c^\dg(x,\ua) c^\dg(x',\da) |\Om \rangle e^{i \bp \cdot(\bx-\bx')} \non \\
        &=&  {1\over L^2} \sum_x \sum_{x'} \langle \Om | \psi^\dg_1(x) \psi_2(x') |\Om \rangle e^{i \bp \cdot(\bx-\bx')} \non \\ 
        &=& {1\over L^2} \sum_x \sum_{x'}  \sum_{i=1}^{L^2} \phi_i^*(x,1) \phi_i(x',2) e^{i \bp \cdot(\bx-\bx')} 
         \non \\
         &=&  {1\over L^2}  \sum_{i=1}^{L^2} \phi_i^*(p,1) \phi_i(p,2) \ .
\label{pc}
\eea
From the previous discussion it is obvious that, in the ground state, $P(\bp)=0$ everywhere except, perhaps, along the Fermi surface, where $\phi(x,\a)$ might have nonzero components at both $\a=1$ and $\a=2$.  But because the Hamiltonian is invariant under an interchange of the $x$ and $y$ directions, it must be true that on a finite volume $P(\bp)$ is also invariant under the interchange of momentum components $p_x \leftrightarrow p_y$,  and this invariance is obviously incompatible with anything other than s-wave pairing.  Any other pairing requires a spontaneous breaking of interchange symmetry (or symmetry under 90$^\circ$ rotations).  The standard numerical method to detect the spontaneous breaking of some symmetry from computations on a finite volume is to insert a small explicit symmetry breaking term, and extrapolate to infinite volume.  So we add to $[H]_{x\a,x'\b}$ a symmetry breaking term with non-zero elements
\bea
[V]_{x1,x'2} &=&  [V]_{x'2,x1} \non \\
     &=& h (\d_{x,x'+\hat{e}_x} + \d_{x,x'-\hat{e}_x} - \d_{x,x'+\hat{e}_y} - \d_{x,x'-\hat{e}_y}) \ .
\eea
This is sufficient to change the condensate pattern dramatically for any finite $h$, no matter how small, and
in fact we find that
\bea
P(\bk) =  \left\{ \begin{array}{cl} 
       \oh \text{sign}(\cos k_x - \cos k_y) & \bk \in \text{Fermi surface} \cr
       0 & \text{otherwise} \end{array} \right. \ ,
\label{dwave}
\eea
 a result which is strongly reminiscent of d-wave pairing.
 
 \begin{figure}[bth]
 \center
 \includegraphics[scale=0.45]{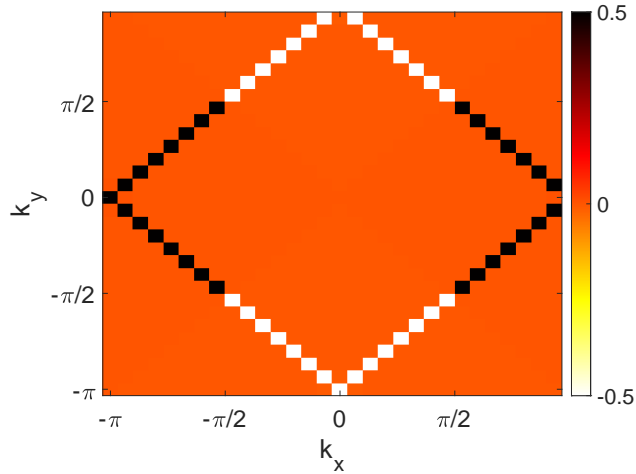}
 \caption{$P(\bk)$ vs $(k_x,k_y)$ at $U=0$ and $\mu=0$, with a tiny explicit symmetry breaking parameter
 $h=10^{-4}$,  on a $30\times 30$ lattice.}
 \label{pair0}
 \end{figure}

   This result follows from simple degenerate perturbation theory.  We compute, for all $\bk$ on the Fermi surface,
the non-zero matrix elements
 \bea
 (\phi^1_{\bk}|V|\phi^2_{\bk'}) &=& (\phi^2_{\bk'}|V|\phi_{\bk}^1) \non \\
   &=& 2h(\cos k_x - \cos k_y) \d_{\bk \bk'} \ .
 \label{V1}
 \eea
 Because the matrix is only non-zero at $\bk=\bk'$, it splits into $2\times 2$ submatrices which are readily
 diagonalized and we find, to lowest order in $h$,
 the new eigenstates of $[H+V]$, with $\bk$ on the Fermi surface
\bea
        \phi_\bk^+(x)  &=& {1\over \sqrt{2}} (\phi_\bk^1(x) + \phi_\bk^2(x)) \non \\
        \phi_\bk^-(x)   &=& {1\over \sqrt{2}} (\phi_\bk^1(x) - \phi_\bk^2(x)) \ ,
 \label{phipm}
 \eea
 with corresponding energies
 \bea
 \e_\bk^+ &=& h(\cos(k_x) - \cos(k_y)) \non \\
  \e_\bk^- &=& -h(\cos(k_x) - \cos(k_y)) \ .
 \eea
 For a given $\bk$ on the Fermi surface only one of the two states $\phi_\bk^{\pm}$ can contribute to the ground state shown in \rf{Om2}, as
 explained previously, and this is the state of lowest energy, which in turn depends on $\bk$.  Therefore, 
 from \rf{pc}, the pairing correlator is  
 \bea
   \hspace{-10pt} P(\bk) = {1\over L^2} \sum_\vx \sum_{\vx'} \sum_{\bk}^{\text{Fermi}} \bigg( \phi_{\bk}^{*+}(\vx,1)\phi_{\bk}^{+}(\vx',2) \theta(-\e_\bk^+)
            + \phi_{\bk}^{*-}(\vx,1)\phi_{\bk}^{-}(\vx',2) \theta(-\e_\bk^-) \bigg) e^{i\bk\cdot(\vx_1-\vx_2)} \ ,
\label{Pk}
\eea
where $\theta(x)$ is the Heaviside step function, and we have defined
\beq
\sum_{\bk} \equiv \sum_{n_x=-L/2}^{L/2-1} \sum_{n_y=-L/2}^{L/2-1} ~~~\mbox{with}
~~~ k_x \equiv {2\pi n_x \over L}, ~ k_y \equiv {2\pi n_x \over L}  \ ,
\eeq
and
$\sum_{\bk}^{\text{Fermi}}$  
represents the sum over momenta along the Fermi surface.
Substitution of \rf{phipm} into \rf{Pk} gives
the result for the pairing correlator stated in \rf{dwave}.

If we wish to construct states with density different from one, then in this formalism it is necessary to choose
a nonzero chemical potential $\m$ (see below), or else to consider a state $|\Om\rangle$ different from the ground state (i.e.\ a set $n(i)$ in \rf{ni} with $n(i) \ne i$).

     The result \rf{Pk} is illustrated in Fig.\ \ref{pair0} on a $30\times 30$ lattice, at $h=0.0001$.  The colors shown
indicate the value of $P(\bk)$, arrived at by a numerical solution of the problem.\footnote{Since the answer for
$P(\bk)$ is known analytically, this is also a modest test of our Matlab code.}  The Fermi surface
is the diamond-shaped contour shown, whose interior is region I, and exterior is region II.  The condensate is
only non-zero along this contour, and changes sign as predicted in \rf{dwave}.  Changing the sign of $h$ interchanges the black and white colors.  This pattern depends only on the sign and not the magnitude of $h$.  Only at $h=0$
do we see that $P(k)=0$ everywhere. 

   Now this pairing is not exactly d-wave pairing; it is proportional to the sign of $\cos k_x - \cos k_y$, rather than the quantity itself.  Moreover, there is no suggestion that the states
along the Fermi surface are localized; they are simply plane waves.  Nevertheless, it is certainly worth noting how natural d-wave condensation appears in the Nambu formalism, even for the free $U=\m=0$ case.  Whether the 
Nambu Hartree-Fock ground state, where the d-wave form of the condensate is readily understood, is energetically
preferred over the standard Hartree-Fock ground state is a matter of comparing energy expectation values $\langle H \rangle$.

This comparison is easy to make at $U=\m=t'=0$.   In the standard case, the lowest energy state is obtained at half-filling, i.e.\ ${N=L^2}$ particles, with equal numbers of
spin up and spin down states.  Each momentum is doubly occupied, by a spin up and spin down state,
for momenta satisfying
\beq
     \cos k_x + \cos k_y > 0 \ .
\eeq
Then we have
\bea
       E^S &=& -2t \sum_{n_x=-L/2}^{L/2-1} ~ \sum_{n_y=-L/2}^{L/2-1} 2 \bigg(\cos{2\pi n_x\over L} + 
       \cos{2\pi n_y\over L}  \bigg) \non \\
       & & \qquad \times \theta(\cos{2\pi n_x\over L} + 
       \cos{2\pi n_y\over L} ) \non \\
               &\ra& -4t L^2 \int_{-\pi}^\pi {dk_x\over 2\pi}  \int_{-\pi}^\pi {dk_y\over 2\pi} 
               (\cos k_x + \cos k_y ) \theta(\cos k_x + \cos k_y) \non \\
               &=& -1.6212 \times t L^2 \ ,
\eea

For the Nambu Hartree-Fock state, states are single occupied as described in \rf{occ}, which leads, at $h=0$, to
\bea
      E^N &=& -t  \sum_{n_x=-L/2}^{L/2-1} ~ \sum_{n_y=-L/2}^{L/2-1}  2\bigg(\cos{2\pi n_x\over L} + 
       \cos{2\pi n_y\over L}  \bigg) \non \\
       & & \qquad \times \theta(\cos{2\pi n_x\over L} + 
       \cos{2\pi n_y\over L} ) \non \\
       & & -t  \sum_{n_x=-L/2}^{L/2-1} ~ \sum_{n_y=-L/2}^{L/2-1} (-1) 2 \bigg(\cos{2\pi n_x\over L} + 
       \cos{2\pi n_y\over L}  \bigg) \non \\
       & & \qquad \times \theta(-\cos{2\pi n_x\over L} -
       \cos{2\pi n_y\over L} ) \non \\ 
      &=& -2t \sum_{n_x=-L/2}^{L/2-1} ~ \sum_{n_y=-L/2}^{L/2-1}  \bigg|\cos{2\pi n_x\over L} + 
       \cos{2\pi n_y\over L}  \bigg| \non \\
              &\ra&  - 2t L^2 \int_{-\pi}^\pi {dk_x\over 2\pi}  \int_{-\pi}^\pi {dk_y\over 2\pi} 
               |\cos k_x + \cos k_y | \non \\
              &=& -1.6212 \times t L^2 \ .
\eea
So the energies of the standard and Nambu Hartree-Fock ground states are identical at $U=\m=t'=h=0$.  This is not a surprise, since at $U=0$ the Hubbard model is soluble analytically, and the ground state energies must agree in either
formulation.   To compare ground state energies beyond the $U=\m=0$ limit, we must compute 
$\langle \Om |H| \Om \rangle$ numerically, at the same density 
\beq
         d = {1\over L^2} \langle \Om| (N_\up + N_\dn) |\Om\rangle \ ,
\eeq
at any given parameters $U,\m, t' \ne 0$.

\begin{figure}[h]
\center
\hspace{-10pt}
\subfigure[~h=0.03]  
{
\includegraphics[scale=0.35]{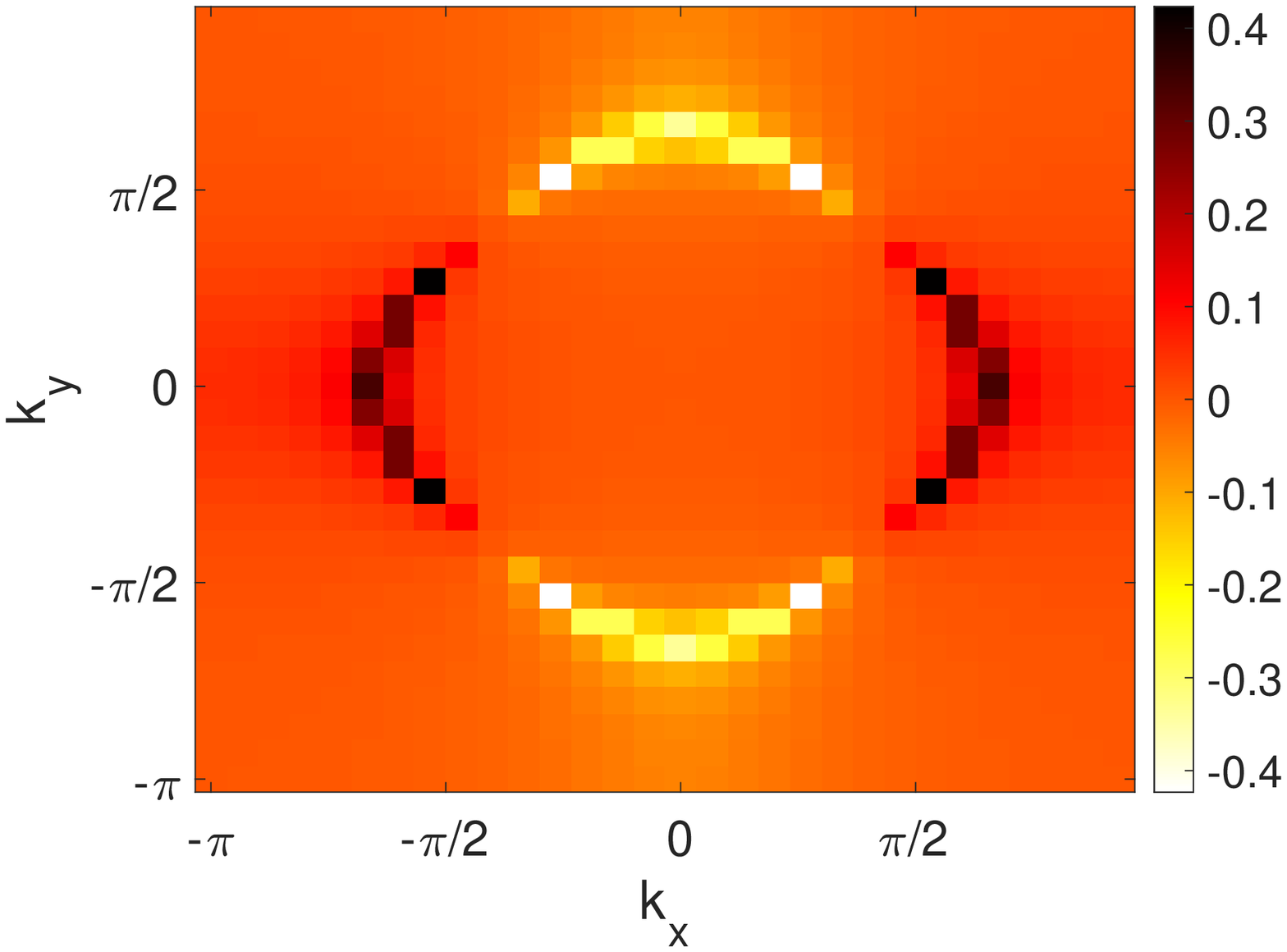} 
\label{x1}
}
\bigskip
\bigskip
\subfigure[~h=0.01] 
{
\hspace{10pt}
 \includegraphics[scale=0.35]{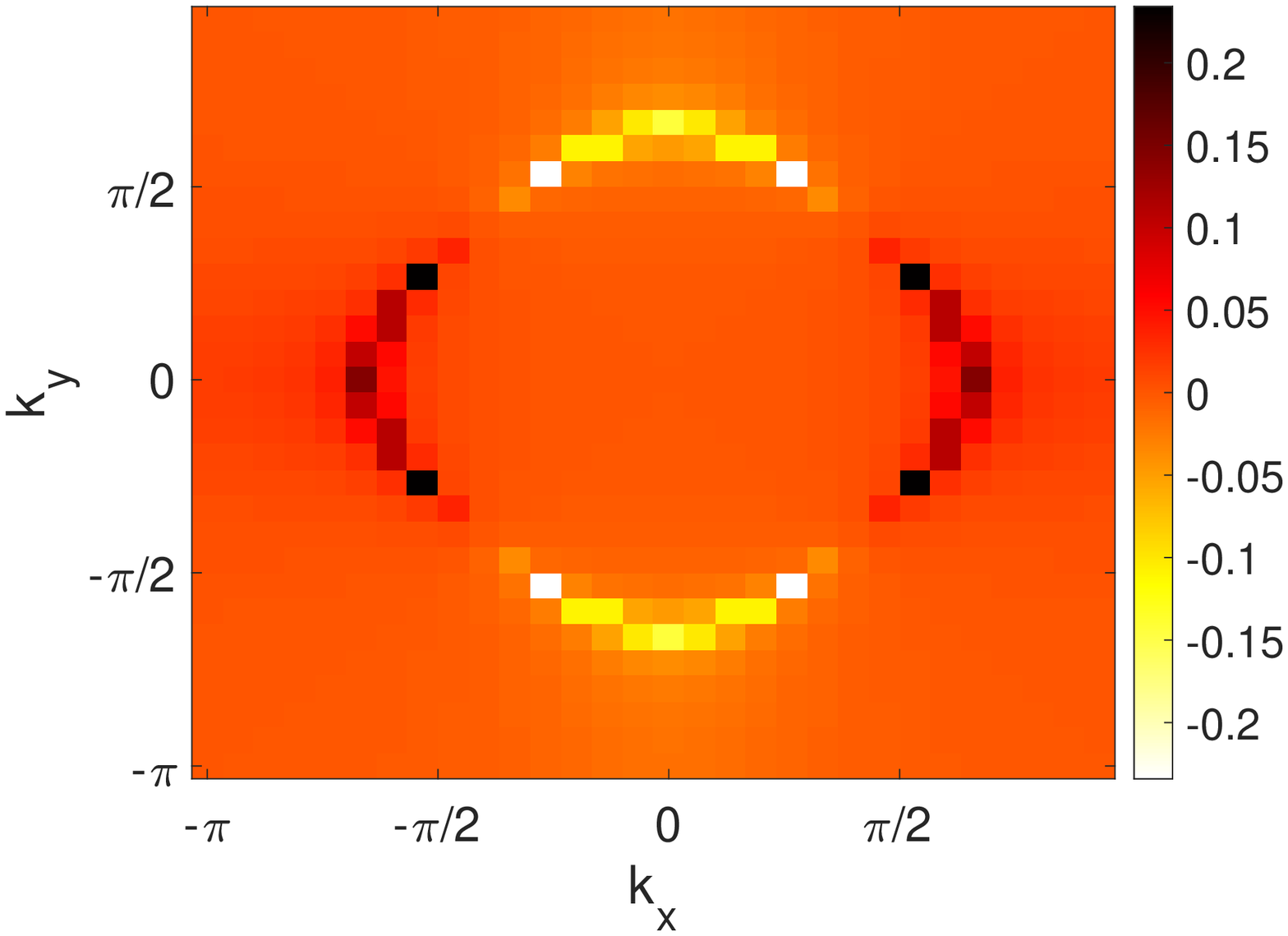}
\label{x2}
 }
\subfigure[~h=0.001] 
{
\includegraphics[scale=0.35]{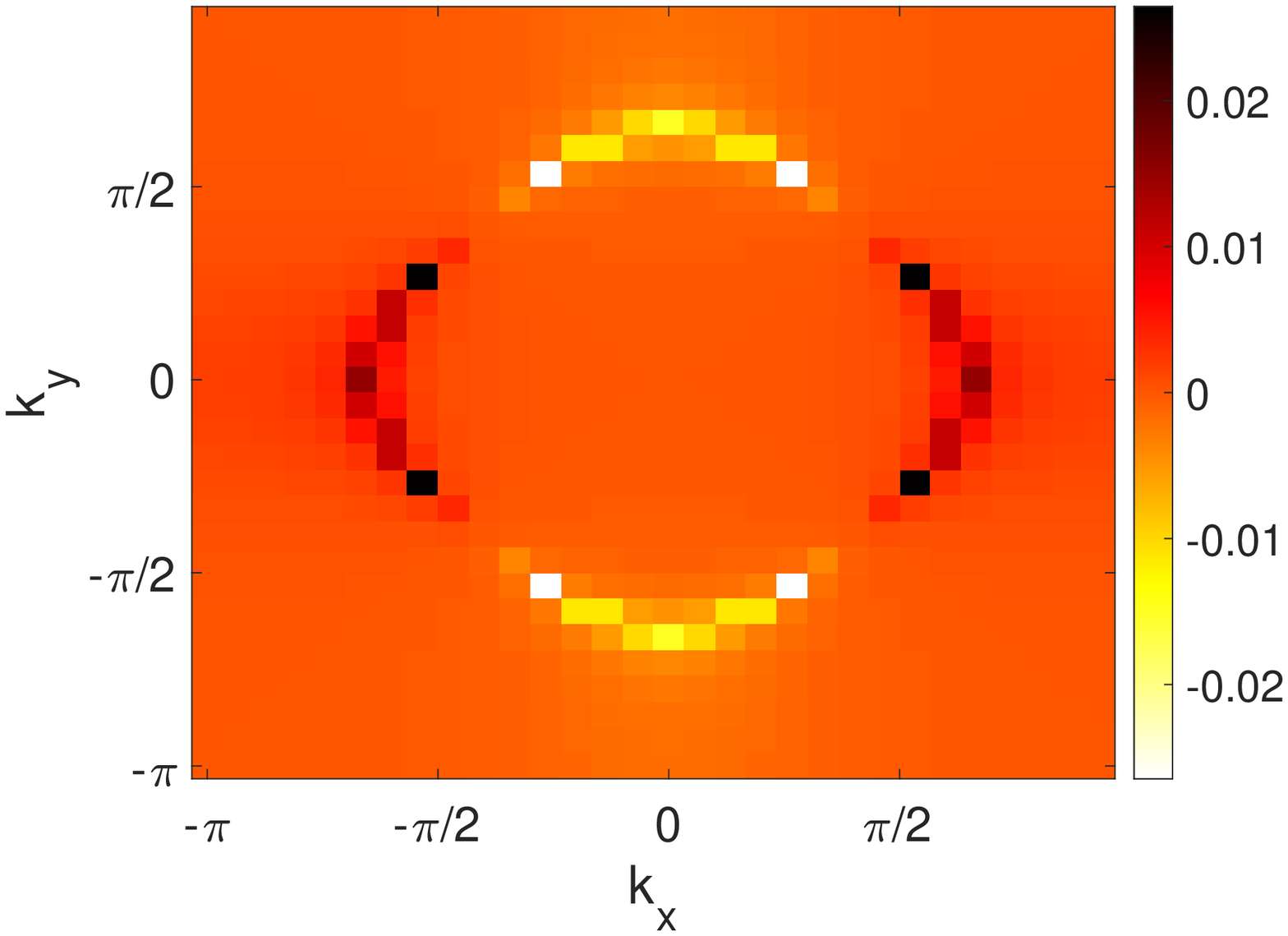}
\label{x3}
}
\caption{$P(\bk)$ vs $(k_x,k_y)$ at $U=0$ and density $d=0.6$.  As $h\ra0$ at fixed lattice volume (here $30\times 30$),  $P(k)\ra0$ (note the color bar scale to the right of each plot). (a) $h=0.03$; (b) $h=0.01$;
 (c) $h=0.001$.}
\label{vary_h}
\end{figure}

\begin{figure*}[htb]
\center
 \hspace{-10pt}
\subfigure[~$20\times 20$]  
{
 \includegraphics[scale=0.35]{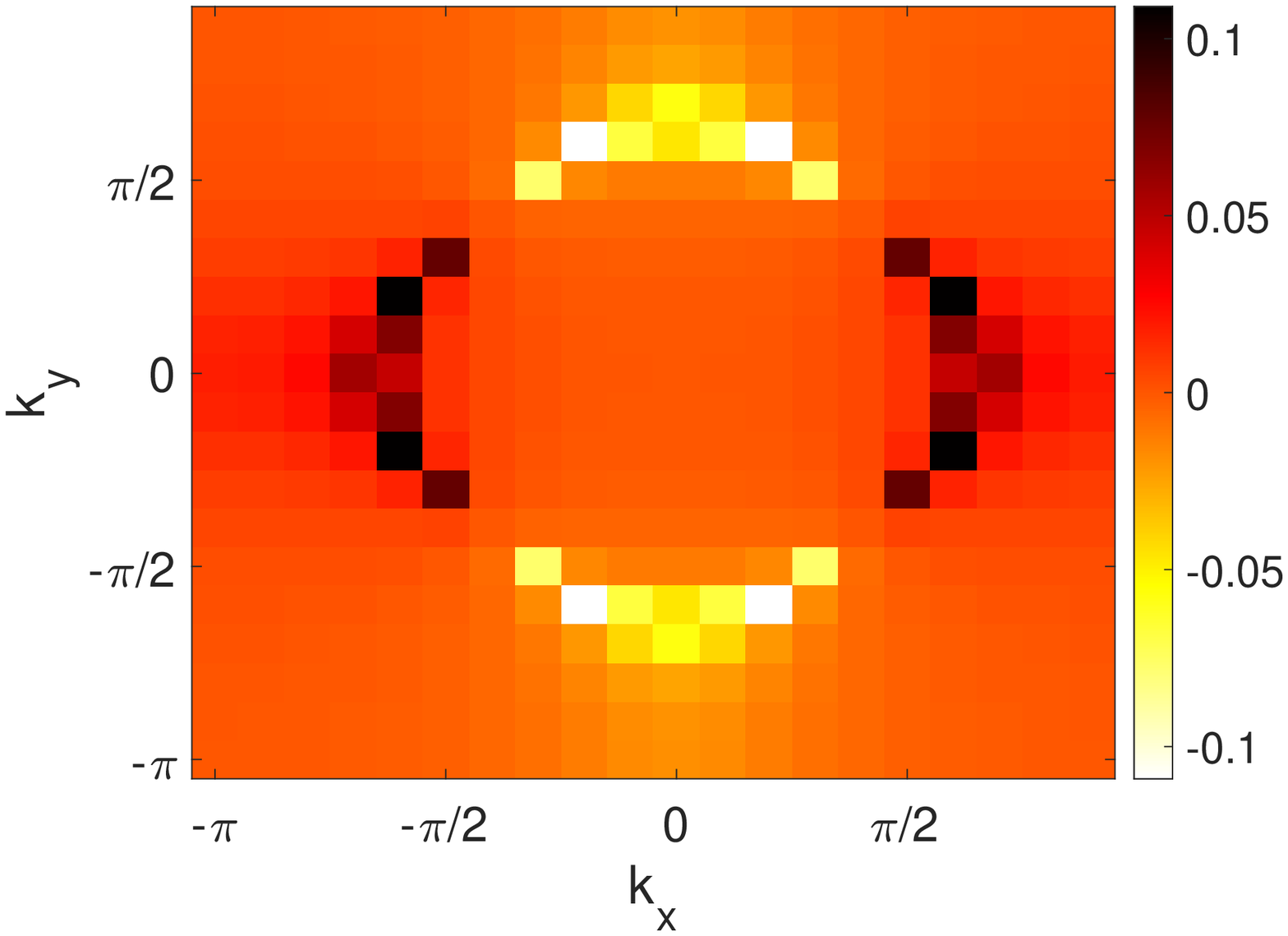} 
\label{xx1}
}
\bigskip
\bigskip
\subfigure[~$30\times 30$] 
{
  \hspace{10pt}
 \includegraphics[scale=0.35]{pair30U0h0p01mu-1p1.eps}
\label{xx2}
 }
\subfigure[~$40\times 40$] 
{
\includegraphics[scale=0.35]{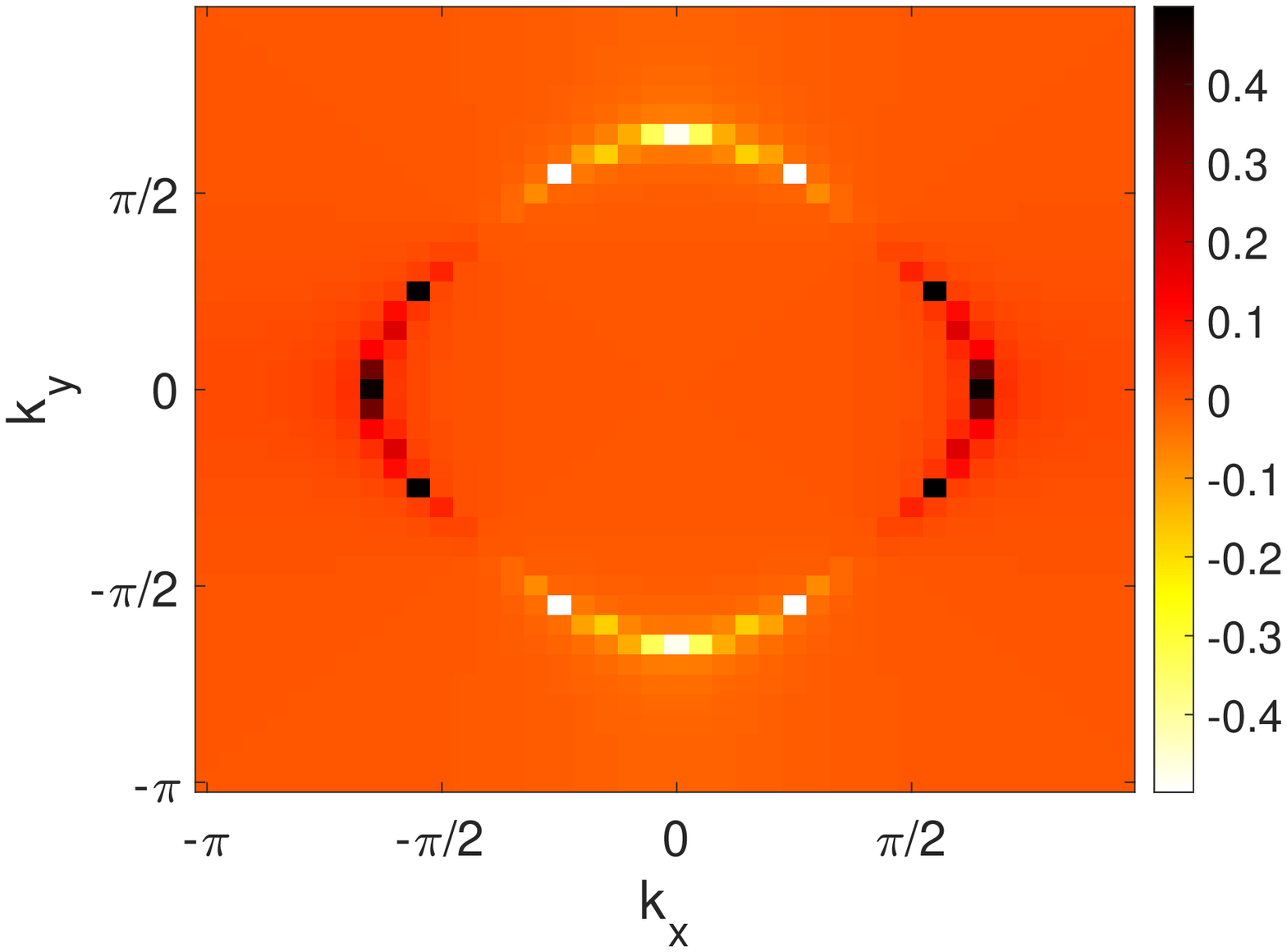}
\label{xx3}
}
\caption{$P(\bk)$ vs $(k_x,k_y)$ at $U=0$ and density $d=0.6$.  As lattice volume increases at fixed $h$ (here $h=0.01$), $P(k)$ converges, near the Fermi surface, to a range $[-0.4:0.4]$. (Again note the scale at the right.)  Lattice volumes (a) $20\times 20$; (b) $30\times 30$; (c) $40\times 40$.}
\label{vary_L}
\end{figure*}
  
   Before going on to $U \ne 0$, let us first consider $\m \ne 0$ at $U=0$.  This case is significantly more complicated than the $\m=0$ case, so much so that we resort to a numerical calculation.  The origin of the complication is that for general $\m$ we do not have an exact ground state degeneracy due to degeneracy of states along the Fermi
surface, because in general the condition \rf{fermi} does not have a solution at $\m\ne 0$ on a finite lattice with discrete
momenta.  But that does not mean that the effect of the perturbing term $V$ is negligible at small $h$.  
The point is that as $L \ra \infty$, the energy difference between $\phi^I$ and $\phi^{II}$ states on either side of the Fermi boundary goes to zero, so even if $h$ is very small, this is countered (in an ordinary perturbative approach) by very small energy denominators, providing the $L^2$ volume is large enough.
    
   The evidence, a sample of which is shown in Figures \ref{vary_h} and \ref{vary_L}, indicates that the condensate,
always concentrated near the Fermi surface, disappears as $h\ra 0$ at fixed lattice volume, but increases to some limit
at fixed $h$ and increasing lattice volume.  It seems likely that the condensate tends to finite non-zero 
values near the Fermi boundary in the thermodynamic $L \ra \infty$ limit, followed by the $h\ra 0$ limit.  This is strongly reminiscent
of the spontaneous breaking of a symmetry in a system at finite temperature.  We emphasize that this behavior, with
a roughly d-wave distribution, is seen already in the free theory, where the only feature affecting particle occupation number comes from Fermi statistics.

\section{\label{SSB} Pairing and spontaneous symmetry breaking at small $\mathbf U$}

\begin{figure*}[h!]
\center
\subfigure[~$P(\bk)$] 
{
 \includegraphics[scale=0.35]{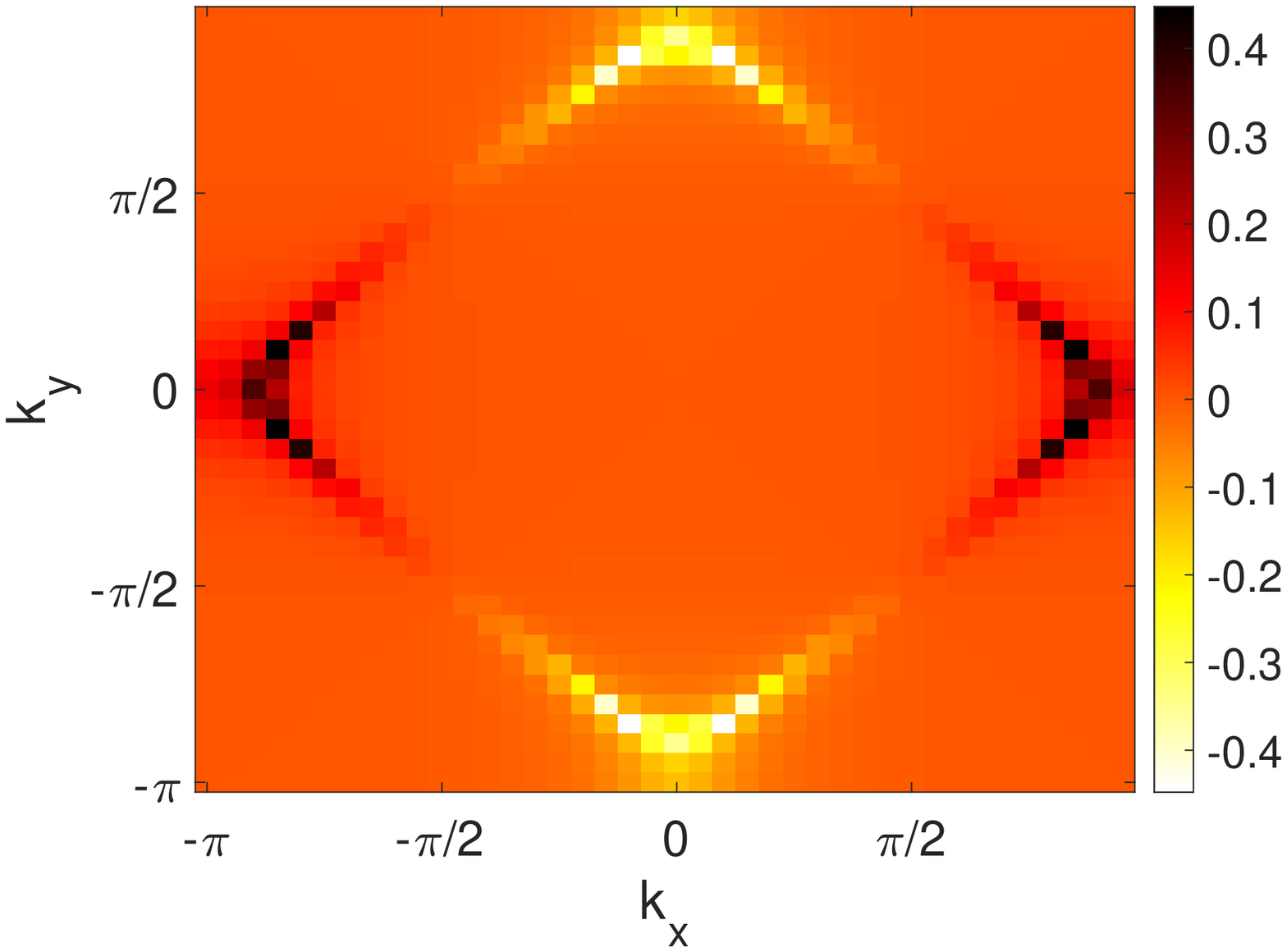} 
\label{pair40}
}
\subfigure[~$\widetilde{P}(\bk)$] 
{
 \includegraphics[scale=0.35]{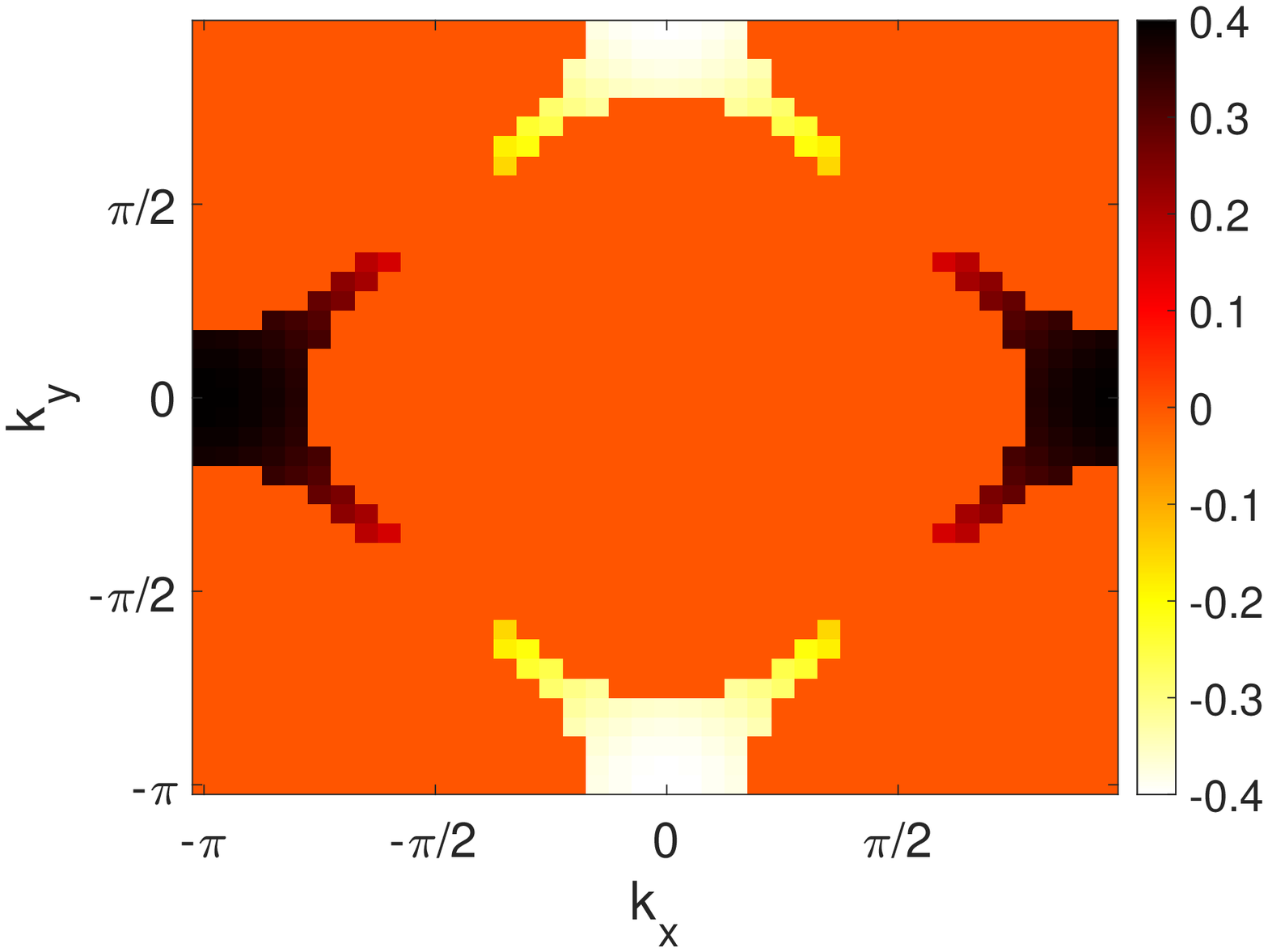} 
\label{dwave40}
}
\caption{A comparison of the pairing condensate $P(\vk)$, computed at $U=0.3,t'=\m=0,h=0.01,d=0.923$, with
the d-wave form \rf{dform}, set to zero where the magnitude of the condensate is less than $0.05$.
(a) pairing condensate; (b) d-wave form.}
\label{Pkcomp}
\end{figure*}

    Since we have observed a condensate reminiscent of a d-wave pattern already at $U=0$, it is not surprising to see
condensations of this kind also at small $U$.  A typical result for $P(\bk)$ is shown in Fig.\ \ref{pair40}.  This result is for parameters $U=0.3, t'=0, \m=0$, which results in a density $d=0.923$.  Next to that, in Fig.\ \ref{dwave40}, we plot a function $ \widetilde{P}(\bk)$ proportional to the d-wave
form, but set to zero where the condensate $|P(\bk)|<0.05$:
\beq
       \widetilde{P}(\bk) = 0.2(\cos(k_x)-\cos(k_y))\theta(|P(\bk)|-0.05) \ .
\label{dform}
\eeq
The scaling constant $0.2$ is introduced so that $P(\bk)$ and $\widetilde{P}(\bk)$ have the same range, and are more easily compared.  It may be worth noting here that while $P(\bk)$ vanishes outside the Fermi surface at $U=\mu=t'=0$, at other parameters $P(\bk)$ is strongly peaked at the Fermi surface, but still slightly 
non-zero in the vicinity of the Fermi surface, as seen by a close inspection of the various figures shown
below.

\begin{figure*}[htb]
\center
 \hspace{-10pt}
\subfigure[~$20\times 20$]   
{
 \includegraphics[scale=0.35]{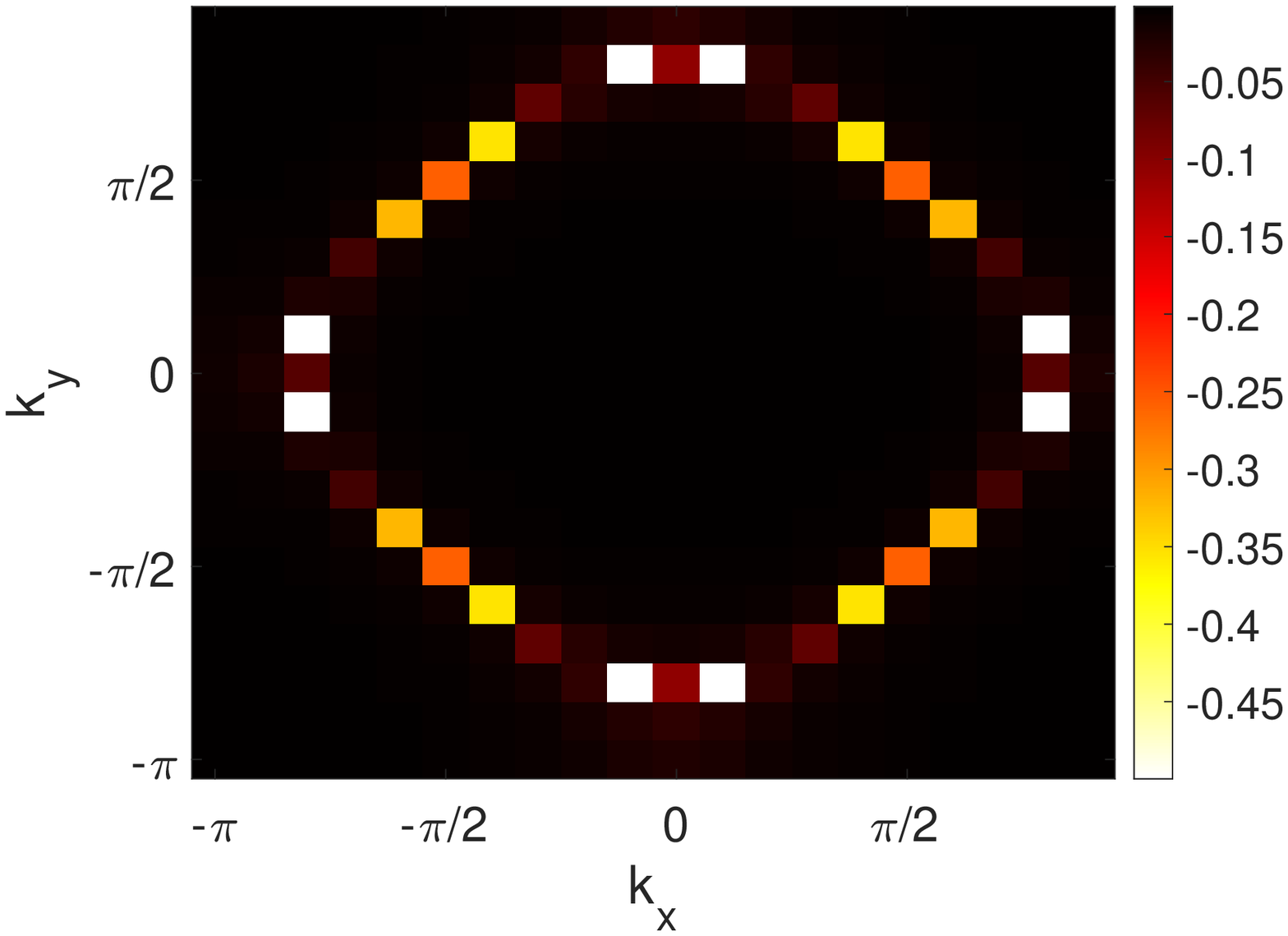} 
\label{h001L20}
}
\bigskip
\bigskip
\subfigure[~$30\times 30$]   
{
 \hspace{10pt}
 \includegraphics[scale=0.35]{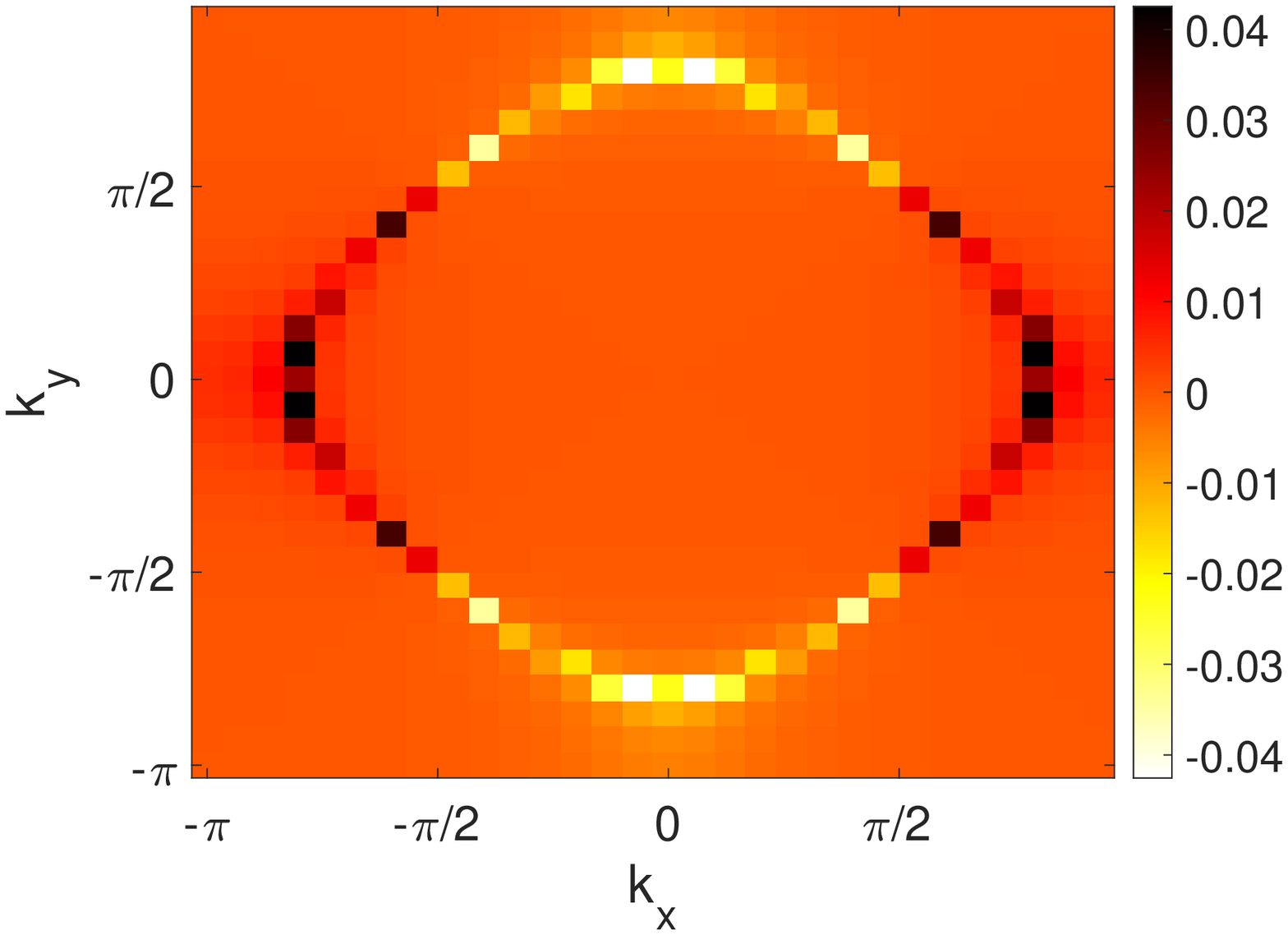} 
\label{h001L20}
}
\subfigure[~$40\times 40$]   
{
 \includegraphics[scale=0.35]{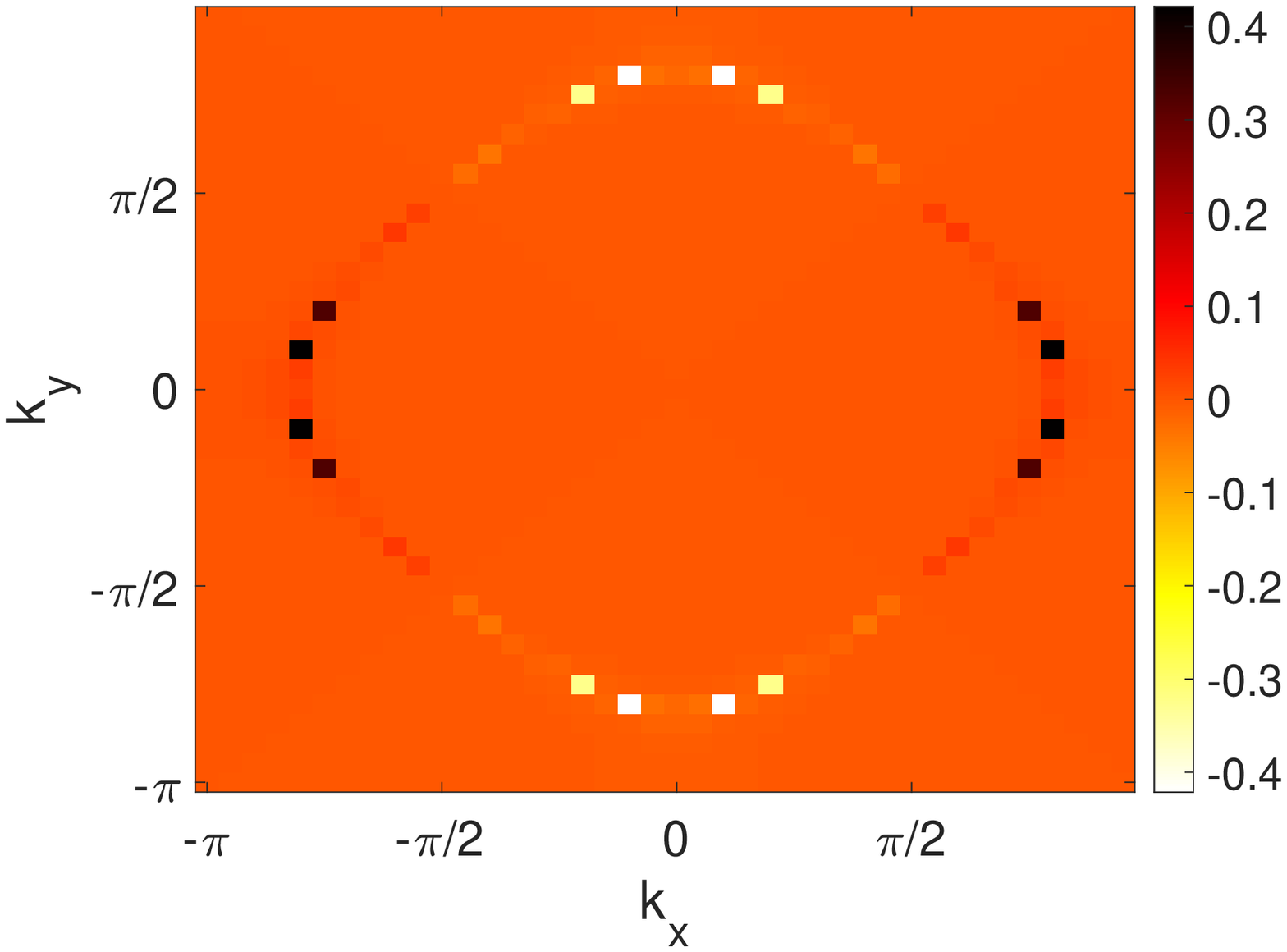} 
\label{h001L20}
}
\caption{Convergence to a d-wave pattern at density 0.9, $U=0.5,~t'=0.1,~h=0.001$ with increasing
volume.  Subfigures (a) $20^2$, (b) $30^2$, (c) $40^2$ lattice volumes.  Note that the amplitude of
the d-wave pattern on the $40^2$ volume is greater, by an order of magnitude, than the amplitude on
the $30^2$ volume.}  
\label{symbreak}
\end{figure*}

  We have already presented evidence, in Figs.\ \ref{vary_h} and \ref{vary_L}, that the pairing condensate
is due to spontaneous symmetry breaking (SSB), rather that a simple linear response to a perturbation, which in our case is proportional to $h$.  Those figures were obtained at $U=0$,  but the same pattern is seen at finite $U$ and fixed
density.  As one illustration, we display in Fig.\ \ref{symbreak} the condensate in the Nambu Hartree-Fock ground state at density $d=0.9$ and $U=0.5,t'=0.1$, at $h=0.001$ and volumes $20^2, 30^2, 40^2$.
As seen in the figure, there is no d-wave pattern at all on the $20\times 20$ volume.  A d-wave pattern with a rather small amplitude emerges at the $30 \times 30$ volume, and the amplitude grows (by a factor of 10) on going to the  $40\times 40$ volume.  This is consistent with a picture of d-wave condensation as spontaneous symmetry breaking of a discrete rotation symmetry, where the proper definition is that an order parameter (in this case the condensate) has a finite value in the broken phase in the limit of volume $V \ra \infty$, followed by
$h\ra 0$.

  This point is worth illustrating again, a little more systematically. SSB requires that the order parameter is finite in the thermodynamic limit,
followed by $h\ra 0$.  Of course we can only work on finite lattices, and there is no SSB at $h=0$ on a
finite volume $V$.  SSB is seen nevertheless if the following conditions are satisfied: (i) for any fixed $h$,
no matter how small, the order parameter rises to its asymptotic value as volume increases; and (ii) for small
$h$, those asymptotic values are almost independent of $h$.  If behavior of that kind is seen, then the 
finite value of the order parameter cannot be attributed to a linear response to the explicit symmetry breaking perturbation.

The order parameter in question is the magnitude of $P(k)$ in the region of the Fermi line.  One could
choose $\langle c_\uparrow^\dg(x) c_\downarrow^\dg(y) \rangle$ to be an order parameter, with either 
$x=y$, or $x,y$ nearest neighbors, or next-nearest neighbors.  But because $\langle c_\uparrow^\dg(x) c_\downarrow^\dg(y) \rangle$ for any $x,y$ can be obtained
from an appropriate Fourier transform of $P(k)$, and because the characteristic d-wave symmetry in $P(k)$ (if it exists) is clear at a glance, we prefer to work with this observable, and will refer to it here as an order parameter. 
 
\begin{figure*}[htb]
\center
\subfigure[~vary $V$]  
{
\includegraphics[scale=0.55]{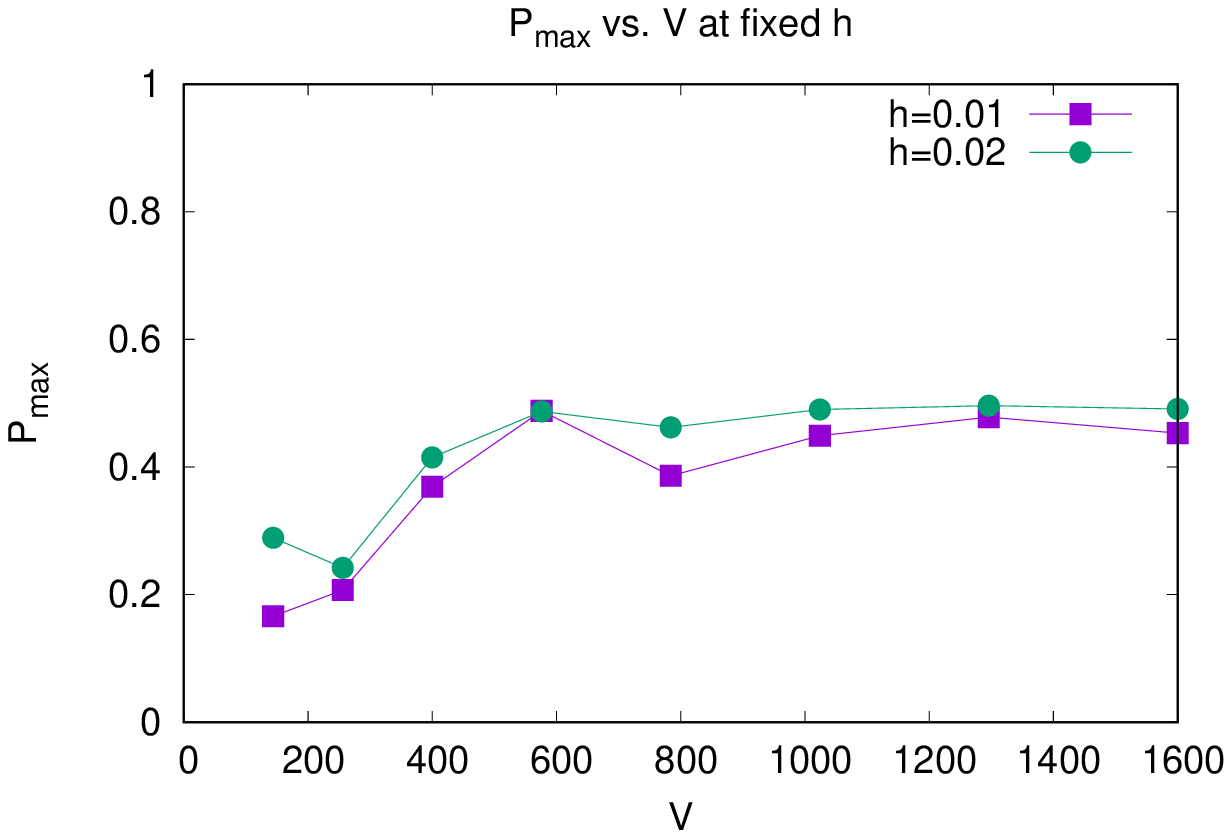} 
\label{x1}
}
\subfigure[~vary $h$] 
{
\includegraphics[scale=0.55]{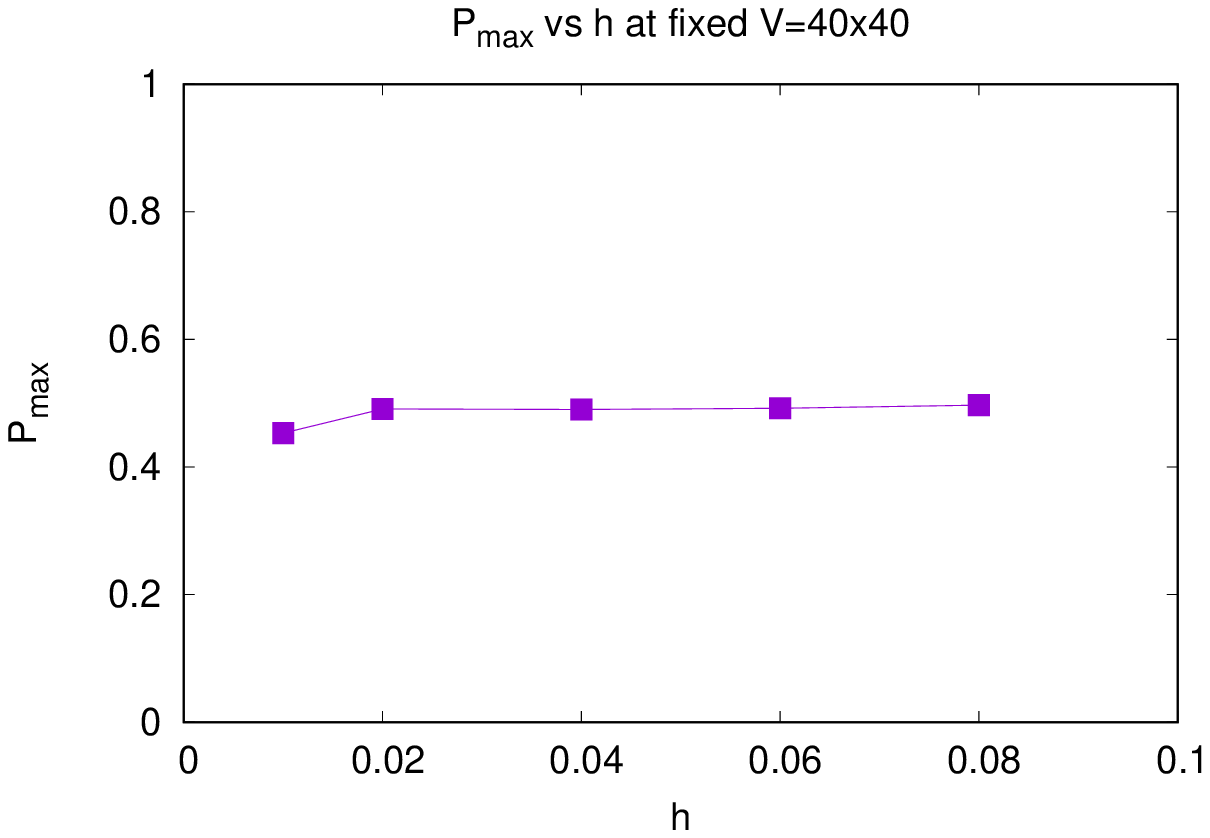}
\label{x2}
}
\caption{The maximum amplitude $P_{max}$ of $P(\vk)$ in a d-wave pattern, at $U=1, t'=0$ and
density $d=0.85$. (a) $P_{max}$ vs.\ volume
$V$, at $h=0.01$ and $0.02$; (b) $P_{max}$ vs $h$ at fixed volume $40\times 40$.}
\label{vary_Vh}
\end{figure*}

    The evidence that pairing is truly due to SSB is shown in Fig.\ \ref{vary_Vh}.  In this figure we display
the maximum value $P_{max}$ of $P(k)$, noting that for a d-wave the minimum of $P(k)$  is 
$P_{min} = -P_{max}$ (as we have checked). The calculation is at $U=1, t'=0$ and density $d=0.85$. What we find, at $h=0.01$ and $h=0.02$, is that at small volumes there is no d-wave pattern at all. On a $12 \times 12$ volume there emerges a d-wave pattern of modest amplitude $P_{max}$, as seen in Fig.\ \ref{x1}, with the amplitude
growing rapidly in the range of $12\times 12$ to $24 \times 24$ to an asymptotic value, roughly the same for both $h$ values, around $P_ {max} \approx 0.48$ or so.  In Fig.\ \ref{x2} we plot the values of 
$P_{max}$ vs.\ $h$ on a $40\times 40$ lattice volume, and here we see that there is very little variation in 
$P_{max}$ with the strength of the explicit breaking term, over a fairly wide range of $h$.   
The growth of $P_{max}$ with volume at fixed $h$ for small volumes, and the negligible dependence of $P_{max}$ on $h$ at sufficiently large volumes, is characteristic of spontaneous symmetry breaking, and inconsistent with the interpretation of the finite order parameter as merely a linear response to a perturbation. 

\subsection{\label{ferro} Stripes and domains at moderate $U$}

One of the great successes of the standard Hartree-Fock treatment was the discovery \cite{Zaanen} of stripe patterns in the
2D Hubbard model. But in our Nambu formulation we have dropped the exchange term, containing operators $c^\dg_\up c_\dn +$ h.c.,
in favor of retaining electron number-changing terms such as $c^\dg_\up c^\dg_\dn$.  The advantage of the Nambu
formulation is, as we have seen, the ease of observing d-wave condensation as a spontaneous symmetry breaking effect.
But dropping the exchange term must surely introduce some quantitative error. The question is whether the neglect
of this term is so serious that it introduces a qualitative error.  Specifically, does the absence of the exchange term
eliminate the stripe and rectangular domain order seen at moderate $U$ in the standard formulation?

\begin{figure*}[h!]
\center
\subfigure[~density 0.77]  
{
\includegraphics[scale=0.45]{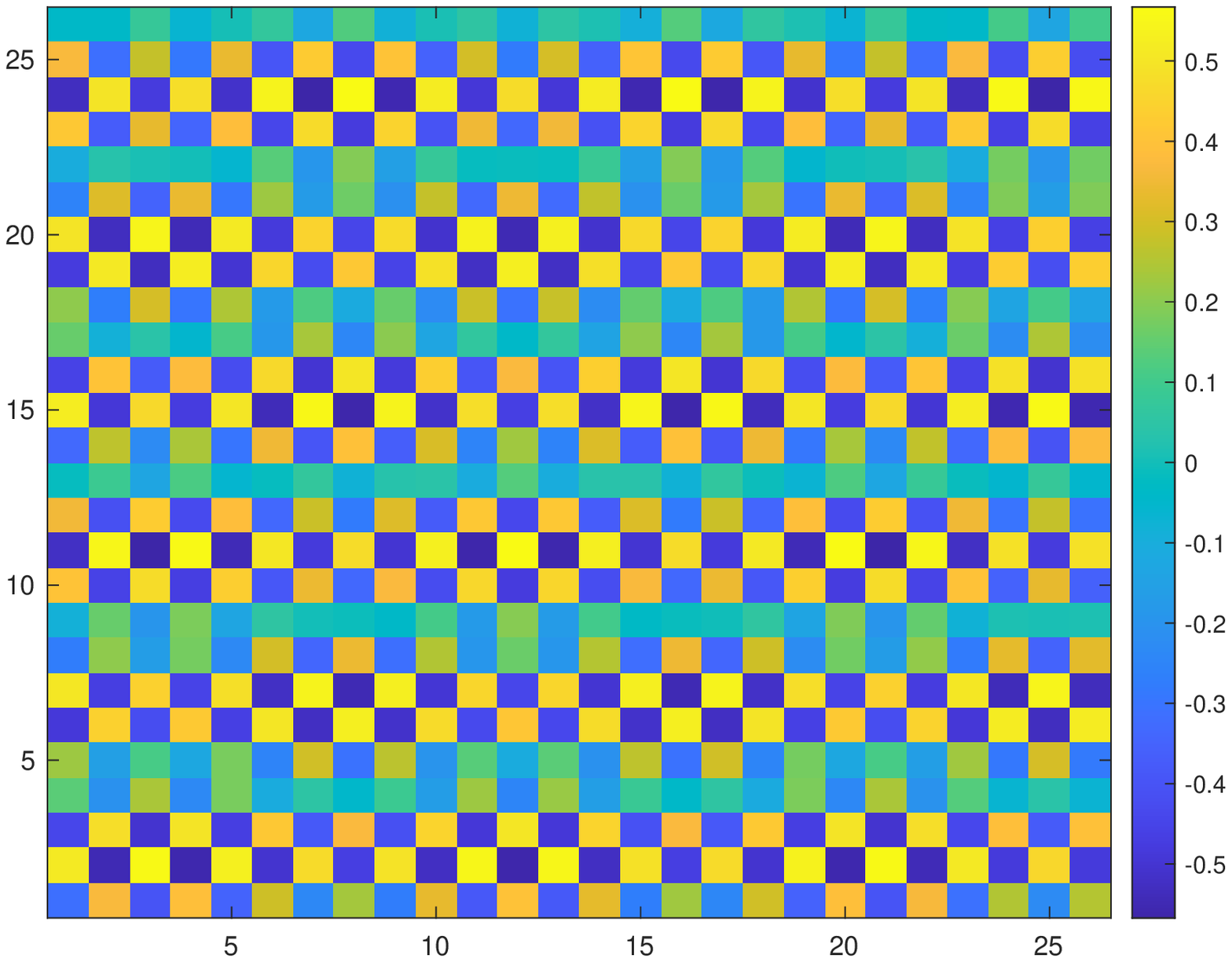} 
\label{stripes}
}
\bigskip
\bigskip
\subfigure[~density 0.81]  
{
\includegraphics[scale=0.45]{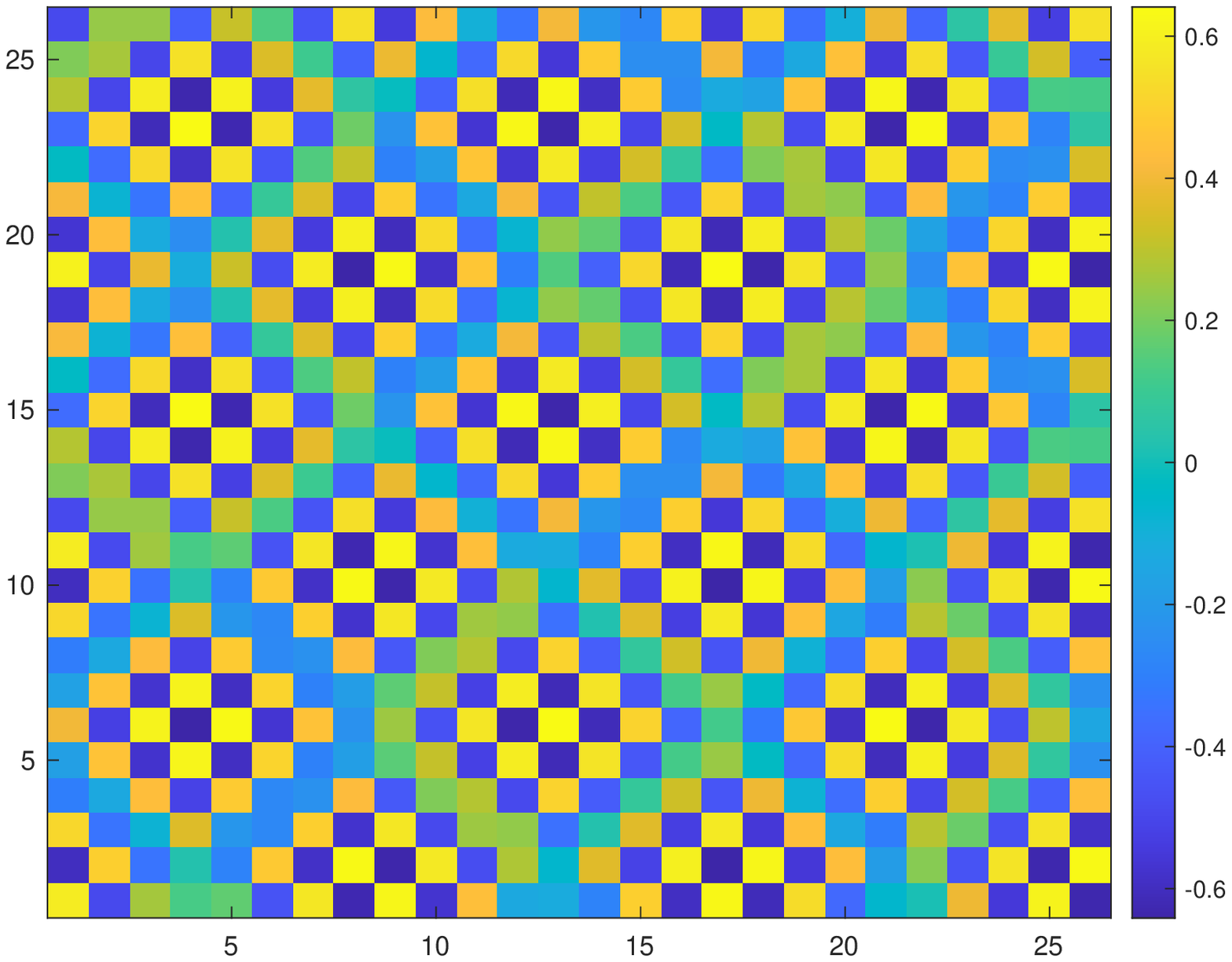} 
\label{stripes}
}
\subfigure[~half-filling] 
{
\includegraphics[scale=0.45]{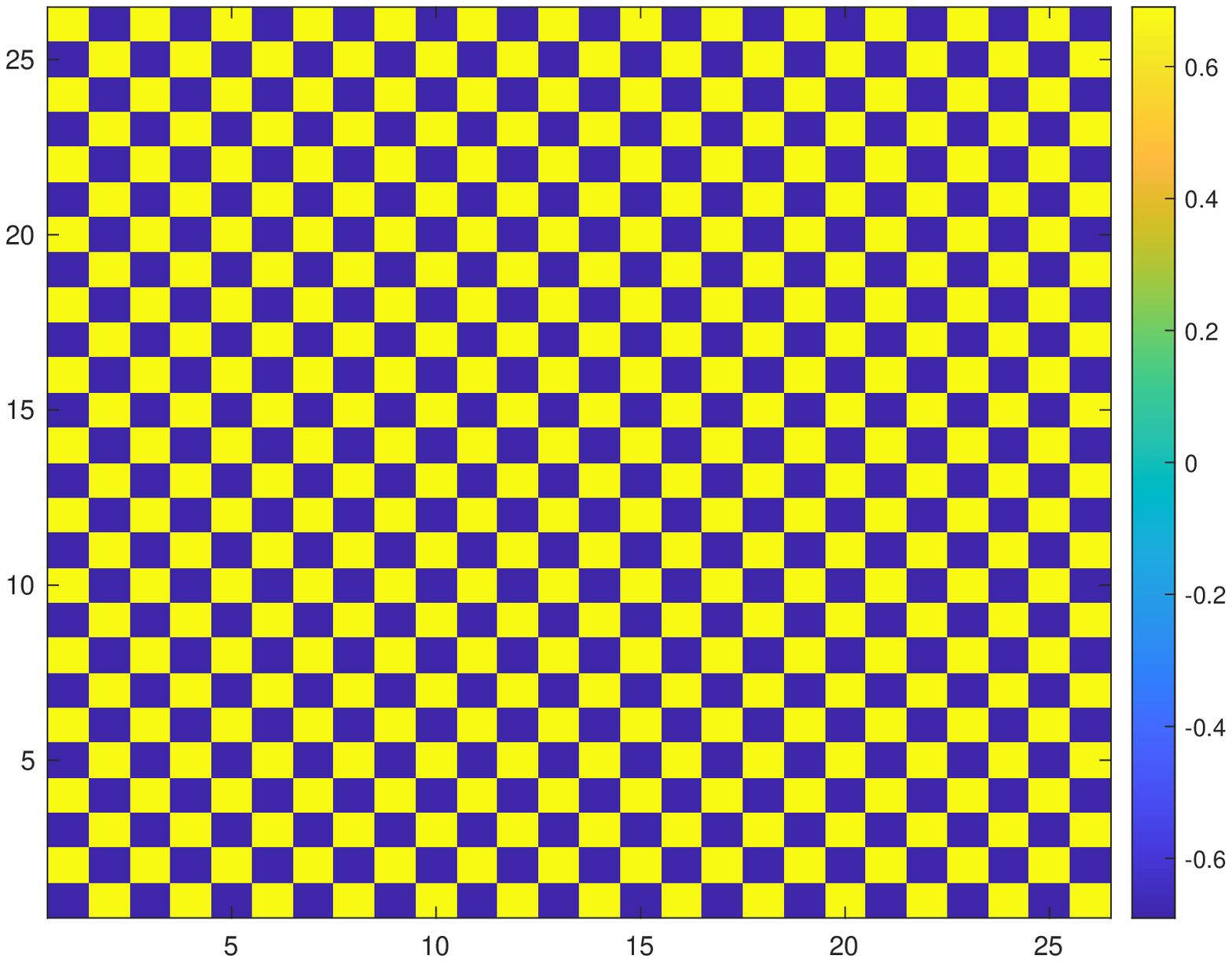}
\label{check}
}
\caption{$D(x)$ at $U=4$ and densities (a) 0.77, (b) 0.81, (c) half-filling.}
\label{U4}
\end{figure*}

    As it turns out, the stripe and domain patterns seen in standard Hartree-Fock at moderate $U$ values are also seen in Nambu Hartree-Fock.  This is illustrated at $U=4$ in Fig.\ \ref{U4}, where we plot the spin density at every site
 of the $26\times 26$ lattice
 \bea
         D(x) &=& \langle c^\dg_\up(x) c_\up(x)\rangle  - \langle  c^\dg_\dn(x) c_\dn(x)\rangle \non \\
                 &=& \langle \psi_1^\dg(x) \psi_1(x) \rangle  + \langle \psi_2^\dg(x) \psi_2(x) \rangle - 1 
 \eea   
 at electron densities 0.77, 0.81, and half-filling.
 These figures can be compared to the standard Hartree-Fock results at $U=3$ seen  in Figs.\ 10(b), 10(c), and 10(d) in 
our ref.\ \cite{Matsuyama:2022kam}.  There are some quantitative differences, but the picture is qualitatively very similar. This is at least one indication that dropping the exchange term may not be so serious, at least at the qualitative level.

   We note that one can use $D(x)$ to define a local magnetization
\beq
          m = {1\over 2L^2} \sum_x \sum_{\m=1}^2 D(x) D(x+\hat{\m})
\label{mag}
\eeq
which may be antiferromagnetic ($m<0$), ferromagnetic ($m>0$), or neither if $m$ is negligible.  The patterns in Fig.\ \ref{U4},
as well as Figs.\ \ref{D}, \ref{g99}, \ref{g81} below, correspond to states with locally antiferromagnetic order, while we find that in d-wave states with $P_{max} \ge 0.4$ the magnetization $m$ is negligible.  

\section{\label{Mott} Pairing and the Mott state at large $\bf U$}

\subsection{The Mott configuration}
    The Mott insulator is obtained in the 2D Hubbard model at half filling, fixed non-zero $t$, at $U \ra \infty$.  Large $U$ 
suggests a strongly correlated system, and of course it is correlations which are neglected in any mean field approach.  So
we might expect that nothing resembling the Mott insulator state could be obtained from a Hartree-Fock approach.  This, we
will see, is not entirely accurate.

     First consider $U/t \gg 1$ at half-filling, and disregard $t$.  It is not hard to see that in the ground state there can be no double occupancy, which means that each site is occupied by a single electron of either up or down spin.  This is equally true in the Hartree-Fock approach, where there exist self-consistent solutions with number densities, at each site, satisfying
 \bea
      \mbox{either}~~ & &   \langle c^\dg_\up(x) c_\up(x)\rangle = 1 ~,~  \langle c^\dg_\dn(x) c_\dn(x) \rangle= 0 
  \non \\
      \mbox{or} ~~ & &  \langle c^\dg_\up(x) c_\up(x)\rangle = 0 ~,~ \langle  c^\dg_\dn(x) c_\dn(x)\rangle = 1 \ ,
 \eea
which implies a spin density
\beq
         D(x) = \pm 1 \ .
\eeq
However, there is an enormous degeneracy in the ground state, since the pattern of up or down spins at each site is irrelevant, subject only to the constraint of half-filling.
 On the other hand,  for finite $t \ll U$, one can show from a perturbative calculation in $t/U$ that the ground state is obtained for equal numbers of up and down spins arranged in a checkerboard pattern, with all adjacent electrons having opposite spins
 \cite{Scalettar}. We will refer to this as the ``Mott configuration,'' because of its association with the Mott insulator.  Checkerboard 
 patterns at half-filling have been seen in the standard Hartree-Fock treatment at intermediate $U/t$, e.g.\ Fig.\ 10(c) in   \cite{Matsuyama:2022kam} and Fig.\ 1 in \cite{Xu}, and we have seen that this is also obtained in the Nambu formulation,
 see Fig.\  \ref{check}.  At $U/t \gg 1$, we require not only a checkerboard, but also that $D(x)=\pm 1$.
\begin{figure*}[h!]
\center
\subfigure[~D(x)]  
{
\includegraphics[scale=0.45]{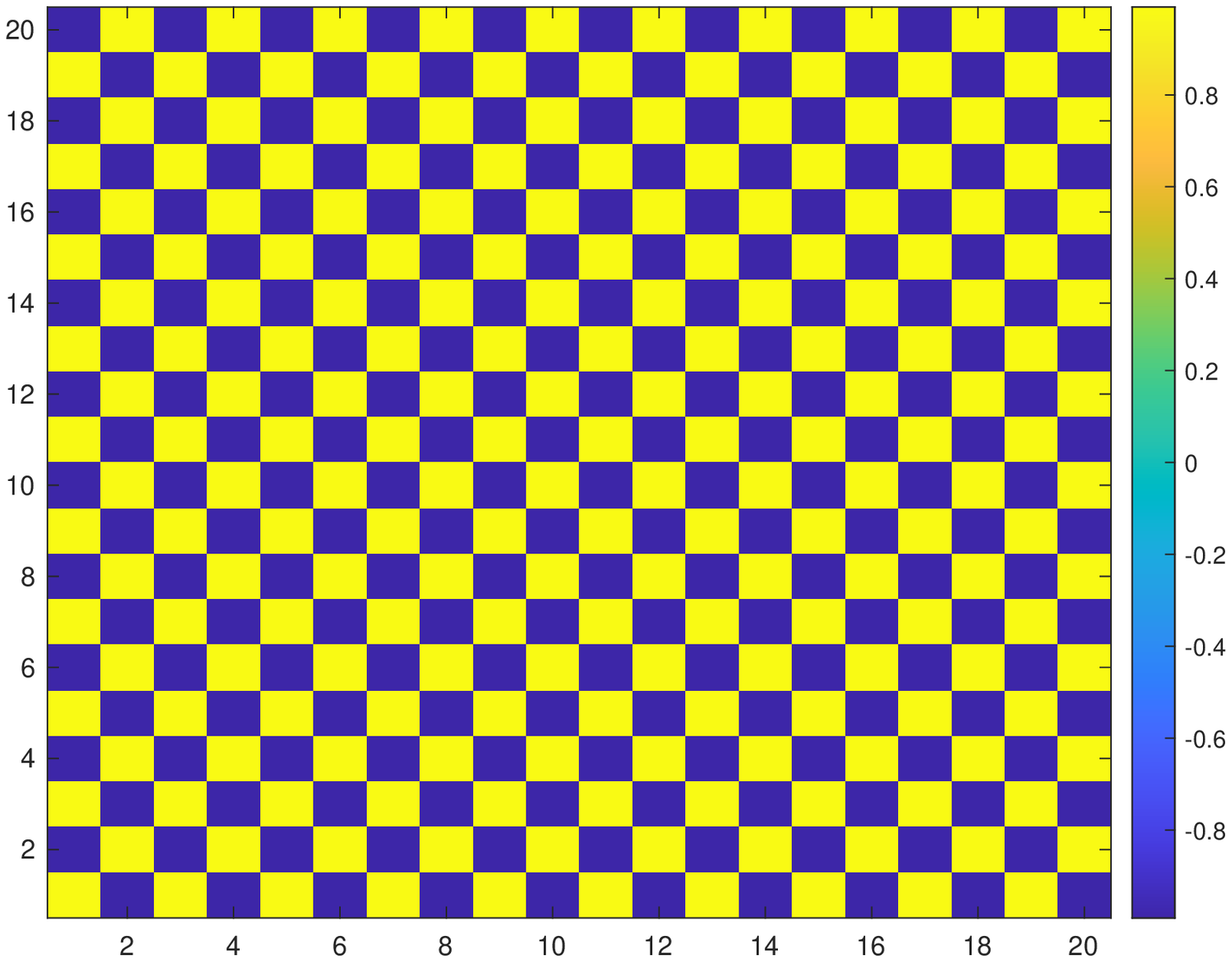} 
\label{D}
}
\subfigure[~IPR] 
{
\includegraphics[scale=0.4]{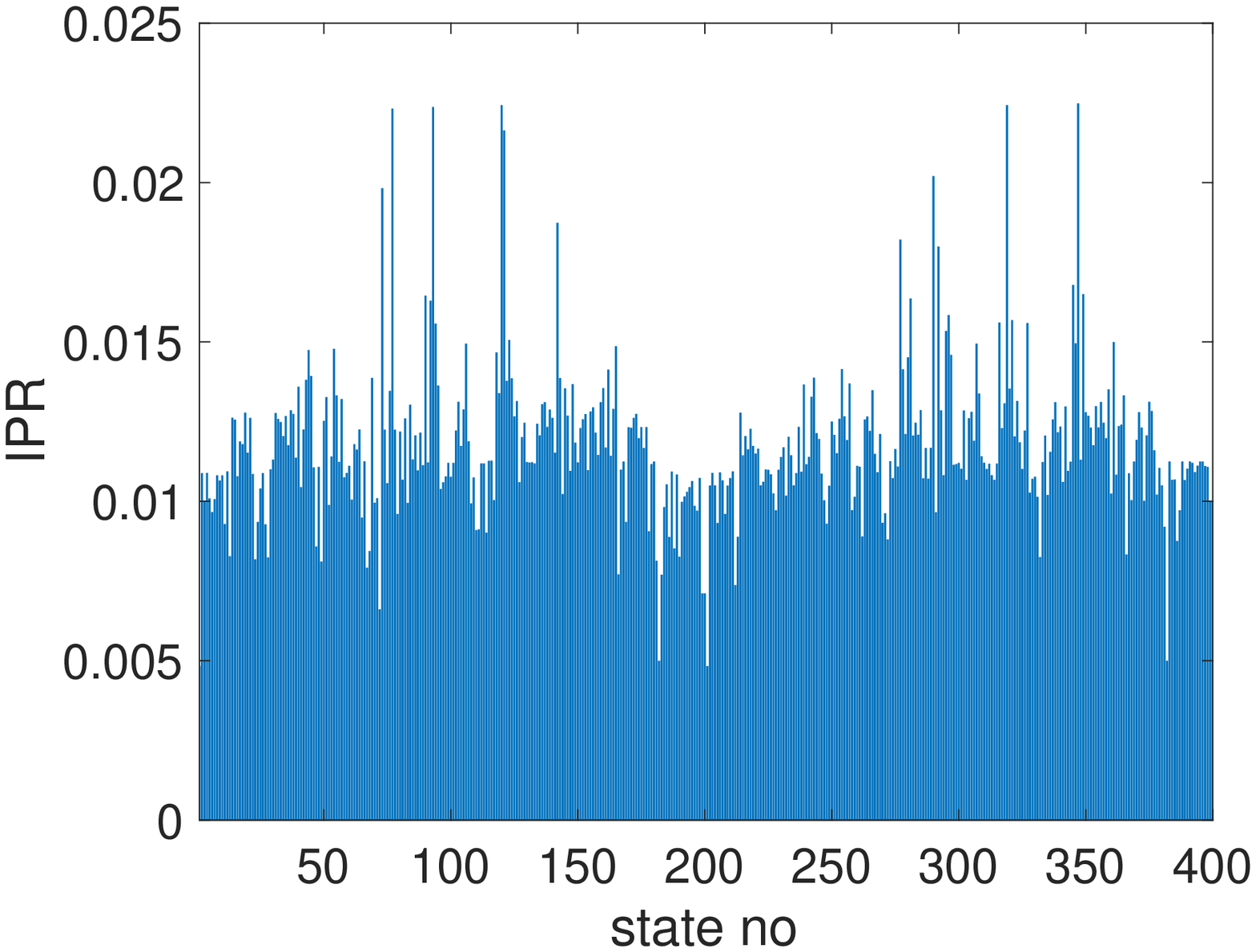}
\label{ipr}
}
\caption{Exact half-filling at $U=30$ on $20\times 20$ lattice; $\mu=15.3$.  (a) spin density D(x); (b) IPR values of the filled 
one particle wavefunctions $\phi_i(x)$.}
\label{U30}
\end{figure*}

     It turns out that the Nambu Hartree-Fock Hamiltonian, solved by the iterative approach described previously, does obtain the Mott pattern at large $U$ and exactly half-filling, as seen in Fig.\ \ref{D}, which was obtained at $U=30$ and $t=1$.
Electrons are  indeed localized in a checkerboard pattern with $D(x)=\pm 1$, but not quite in the way one might have expected.  First of all, $D(x)=-1$, which corresponds to a spin down electron at site $x$, can
only be obtained if the Nambu variable wave functions $\phi_i(x,s)$ are all negligible at that site.  On the other hand, $D(x)=+1$,
which corresponds to a spin up electron, is not obtained from a single localized $\phi_i(x,s)$ which gives the whole contribution,
but rather from a sum of contributions from many delocalized wave functions, each of which gives a small contribution
$\phi^*_i(x,s)\phi_i(x,s)$ at site $x$ but which sum to give a spin-up electron occupancy at site $x$.  In other words, while
electrons are completely localized at large $U$, the one-particle wave functions are not.  This is clearly seen in a plot of the Inverse
Participation Ratio (IPR) for each wave function, shown in Fig.\ \ref{ipr}.  An IPR=1 implies that the wave function is completely localized, while IPR = 1/(lattice volume) implies a wave function whose magnitude is spread evenly over the lattice.  From Fig.\ \ref{ipr} it is clear that the Nambu one-particle wave functions are very unlocalized.

     As $U$ is reduced, we still find a checkerboard pattern at half-filling, an example is Fig.\ \ref{check} at $U=4$.  However, in a
comparison of half-filling at $U=30$ (Fig.\ \ref{D}) with the corresponding figure at $U=4$, we find that $D(x)=\pm 1$ at
the higher value of $U=30$, while $D(x)=\pm 0.6$ at $U=4$.  When $D(x)=\pm 1$, we can say with certainty that there is no
double occupancy, and specify the spin of each electron found at each site in that ground state.  This is not the case at $U=4$, where at each site there is a higher probability ($\approx 80\%$) of finding the more likely spin at that site, but still a significant probability 
($\approx 20\%$) of finding the opposite spin.

\begin{figure}[h!]
\center
\includegraphics[scale=0.4]{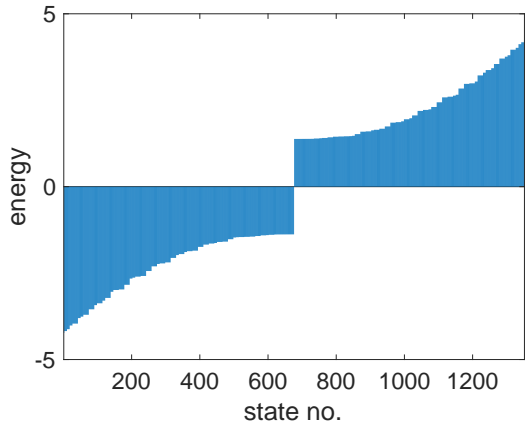}
\caption{Energy gap in the Mott  state at half-filling; $U=4$ on a $26\times 26$ lattice.  The histogram shows the eigenvalues $\e_i$ of
the one-particle eigenstates $\phi_i(x)$ defined in eq.\ \rf{eval}.  States with negative eigenvalues are occupied, and states with positive
eigenvalues are unoccupied. }
\label{gap}
\end{figure}

     In the checkerboard pattern at exactly half-filling, for all cases we have studied, there is an energy gap between the occupied
band of one-particle states, and the band of unoccupied states, as is typical for an insulator.  In the Nambu Hartree-Fock treatment, this
gap grows with $U$.  An example, at $U=4$ on a $26\times 26$ lattice, is shown in Fig.\ \ref{gap}, which displays the energies, i.e.\ the eigenvalues $\e_i$ of the effective Hamiltonian in eq.\ \rf{eval}, of the one-particle states. The states of negative energy are occupied, and states of positive energy are unoccupied.

 \subsection{Pairing}
 
\begin{figure}[h!]
\center
\subfigure[~$D(x),~d=0.999$]  
{
\includegraphics[scale=0.45]{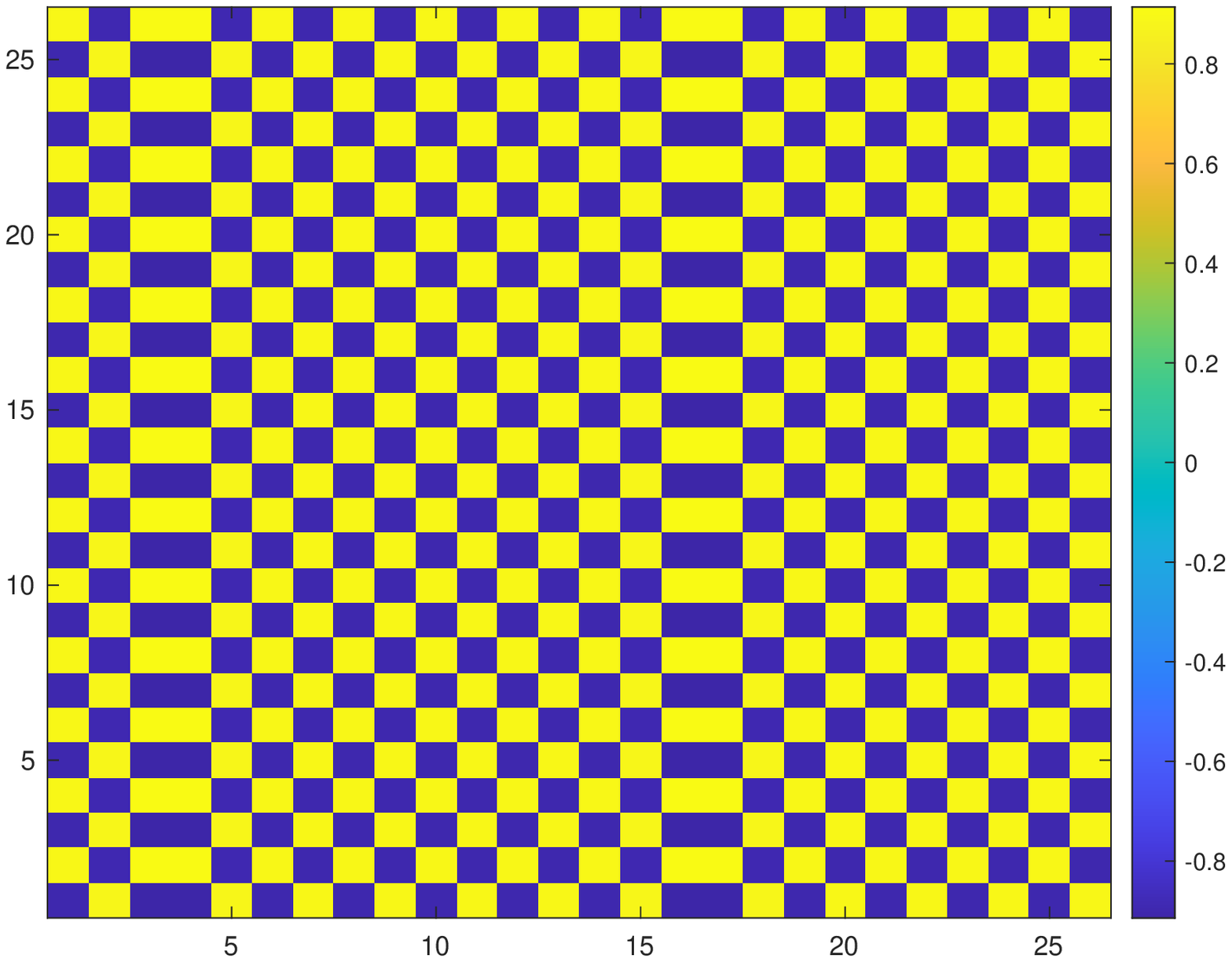} 
\label{g99}
}
\bigskip
\subfigure[~$P(k),~d=0.999$] 
{
\includegraphics[scale=0.4]{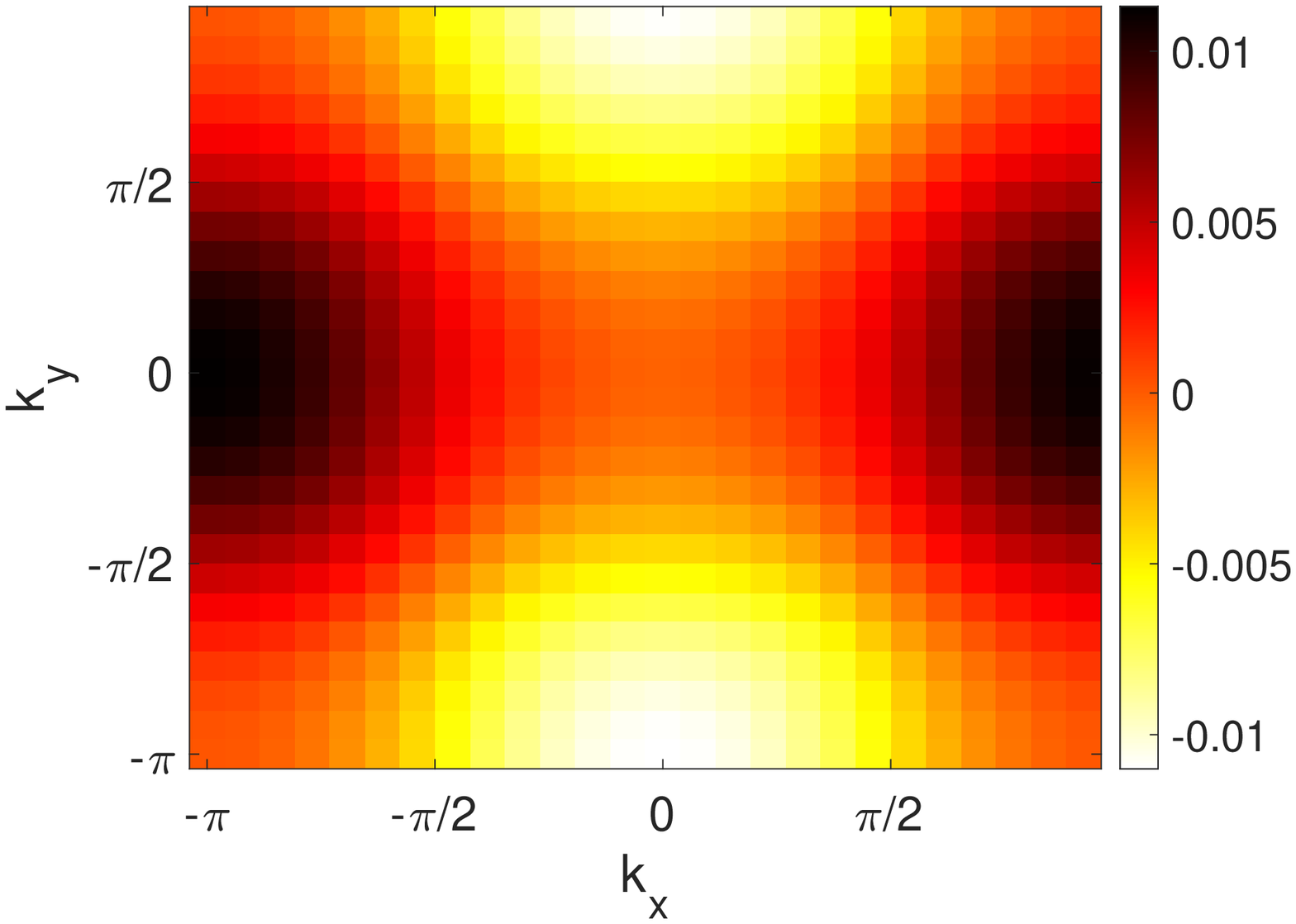}
\label{p99}
}
\bigskip
\subfigure[~$D(x), ~d= 0.81$]  
{
\includegraphics[scale=0.45]{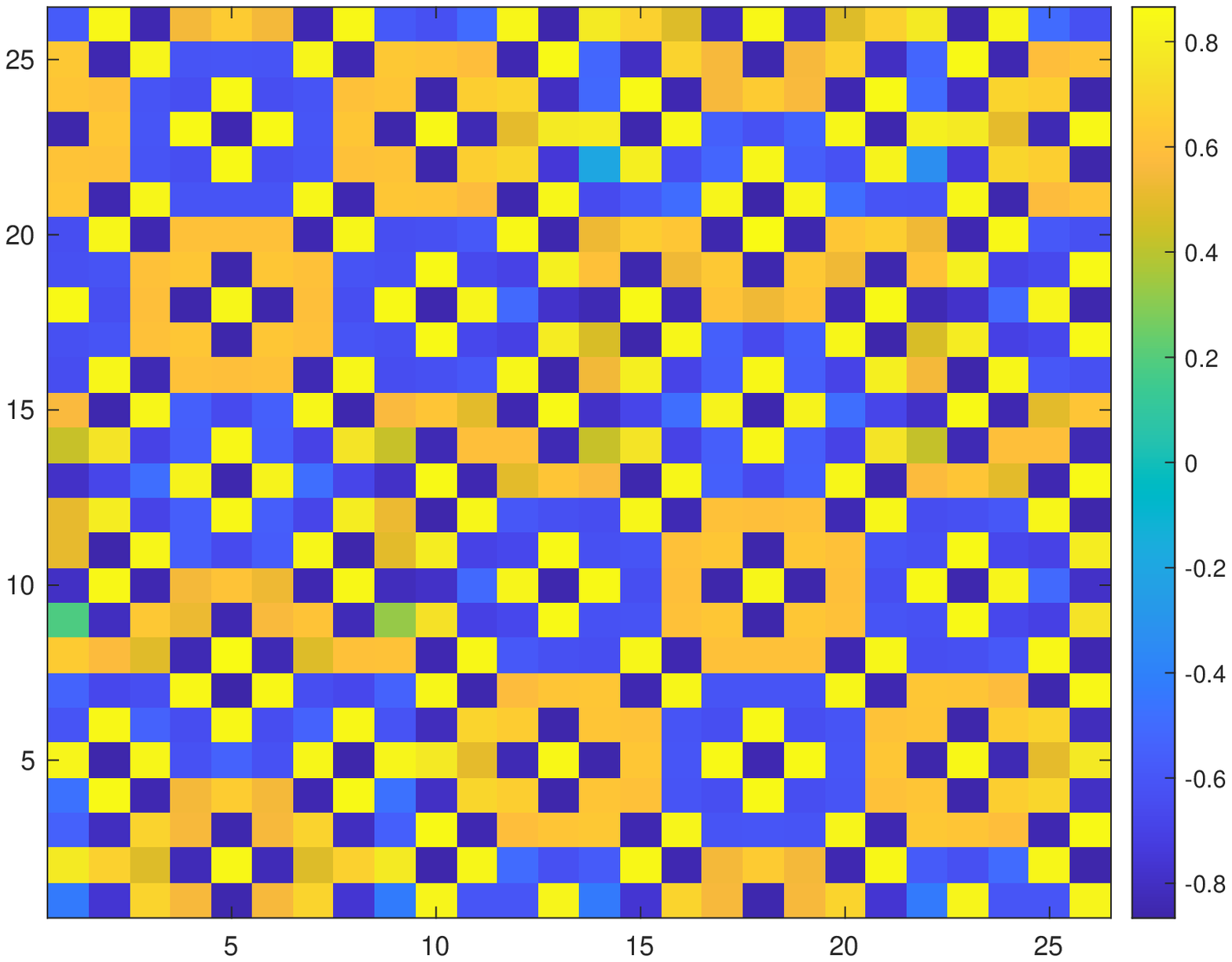} 
\label{g81}
}
\subfigure[~$P(k),~d=0.81$] 
{
\includegraphics[scale=0.4]{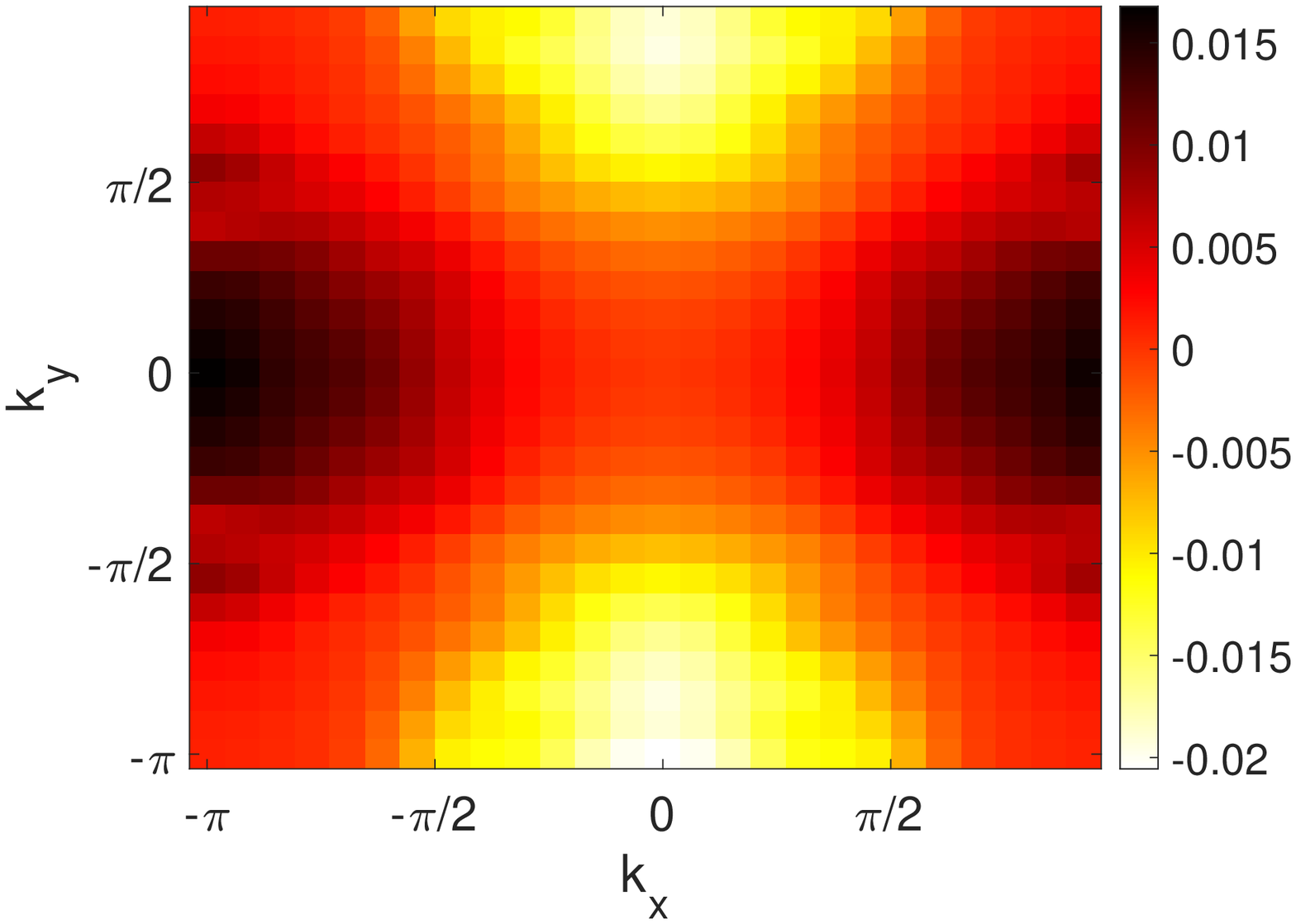}
\label{p81}
}
\subfigure[~$D(x),~d=0.31$]  
{
\includegraphics[scale=0.45]{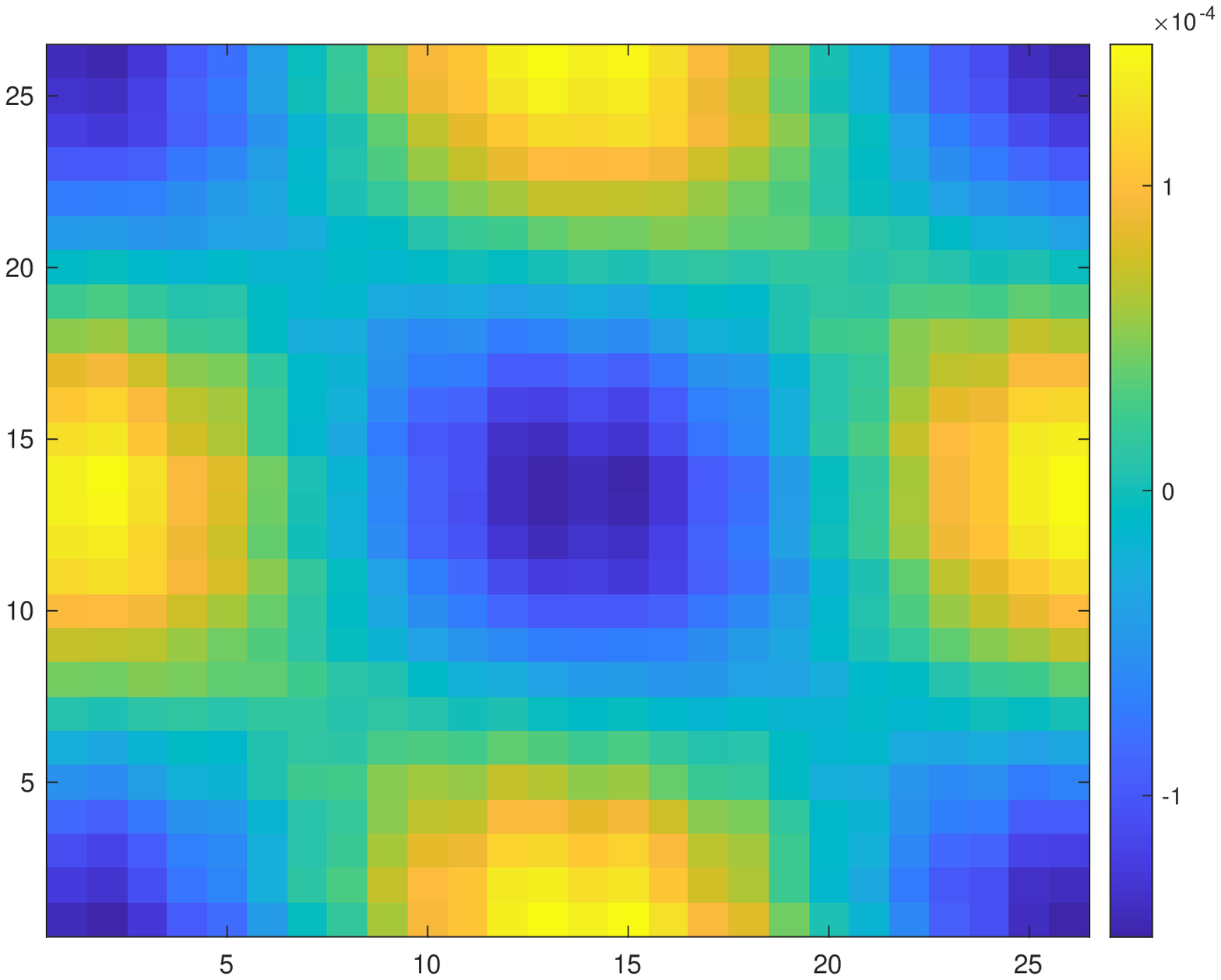} 
\label{g31}
}
\subfigure[~$P(k),~d=0.31$] 
{
\includegraphics[scale=0.4]{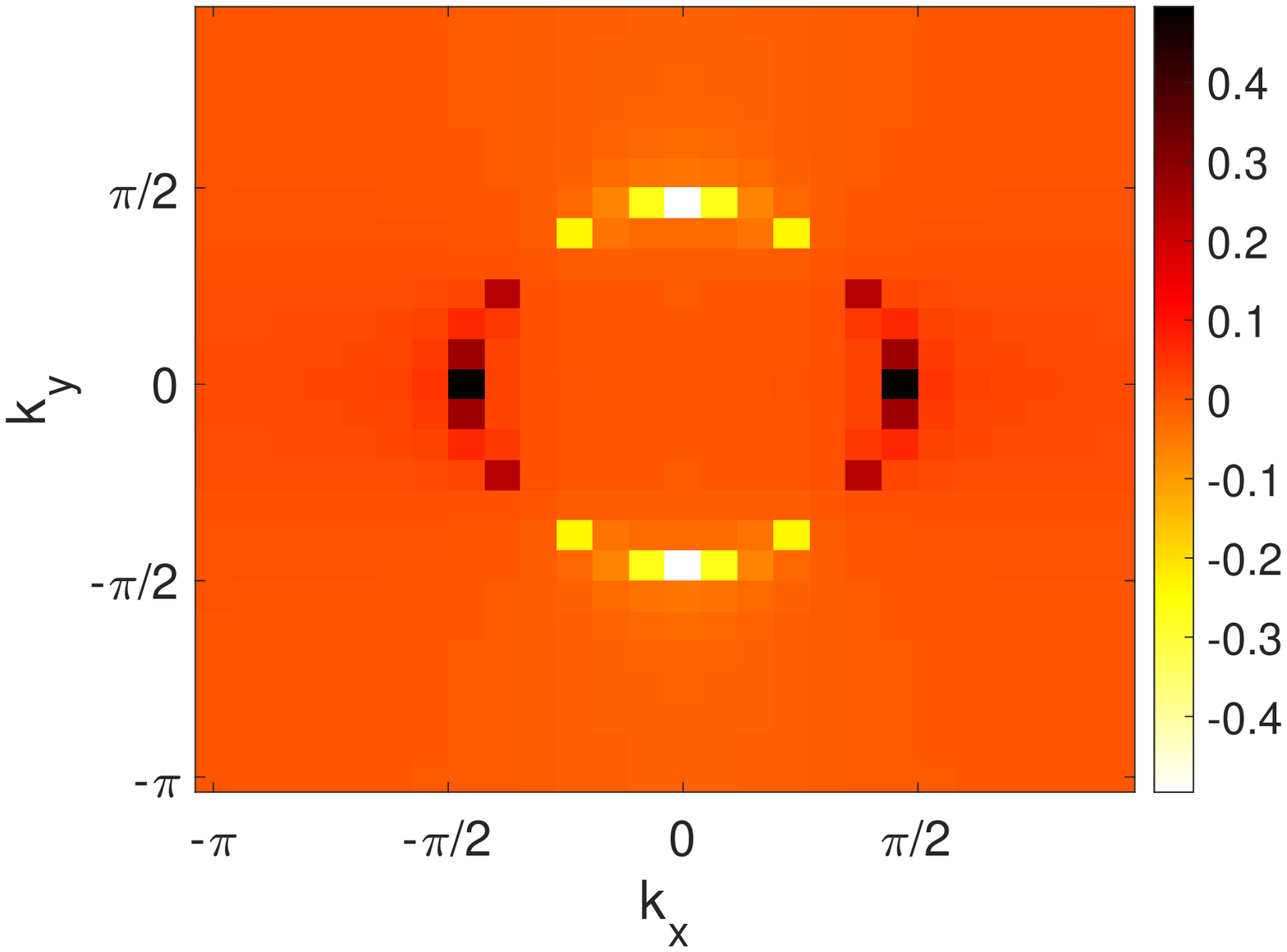}
\label{p31}
}
\caption{Spin densities $D(x)$ and pairing condensate $P(k)$ at coupling $U=8$ and densities
$d=0.999,0.81,0.31$ on a $26\times 26$ lattice.  Note that the spin densities are negligible, of order $10^{-4}$, at high hole doping
($d=0.31$).  The point is that spin density and d-wave condensation appear to be anti-correlated.}
\label{U8}
\end{figure}

    Let us next consider a coupling $U=8$, which may be more relevant to cuprates.  At very near half-filling (density $d=0.999$), we find 
find an almost-checkerboard pattern, Fig.\ \ref{g99}.  
There is no indication whatever, at half-filling, of a d-wave condensate in a plot of $P(k)$ at this same density; the small amplitude of $P(k)$ in Fig.\ \ref{p99} can be attributed to the explicit symmetry-breaking term with parameter $h=0.02$.  As the density is reduced we see some other geometric pattern emerge at $d=0.81$ (Fig.\ \ref{g81}),
but again the small amplitude of $P(k)$ at this density can be attributed to the symmetry breaking term.  What is interesting is
that at a sufficiently low density (high level of hole doping), we find that a d-wave condensate emerges, whose amplitude is
far greater than the magnitude of the symmetry breaking term (Fig.\ \ref{p31}). The spin density, however is negligible ($O(10^{-4})$)
at $d=0.31$ (Fig.\ \ref{g31}), which suggests that the condensate and the spin density may be anti-correlated.

We find that the emergence of a d-wave condensate at high doping is a general feature at moderate to high $U$ values.
The procedure for identifying a d-wave condensate is the same as at small $U$.
We have a small symmetry breaking term in the Hamiltonian proportional to $h$.  At half-filling this term has no discernible
effect.  At smaller densities, and small volumes, we may find $P_{max}=-P_{min}$ significantly greater than $h$.  The test for
spontaneous symmetry breaking is, as before, the behavior $P_{max}=-P_{min}$ with increasing volume.  Convergence to
a limit in which $P_{max}=-P_{min}$ is much greater than $h$, as volume increases, is the signal of spontaneous symmetry breaking, and the existence of a d-wave condensate.  It should be noted
that while the checkerboard and domain patterns are associated with an antiferromagnetic local magnetization, $m<0$, the local
magnetization is negligible in states with a d-wave condensate.
 
\clearpage

It may also be of interest that when there is a d-wave condensate we find that $\rho(x,12)=\rho(x,21)$ converges to zero, which is natural for a d-wave symmetry.  Yet we initialize with non-zero random values of this quantity; the point being that a random
start allows the iterative procedure to explore more of the landscape of self-consistent solutions, as we discussed previously
in \cite{Matsuyama:2022kam}.    

\subsection{Local antiferromagnetism and d-wave condensates}

   Generally speaking, local antiferromagnetism (AF),  i.e.\ $m<0$, and d-wave condensation are anti-correlated.  In the d-wave
region, $m$ is many orders of magnitude less than one and the amplitude $P_{max}=-P_{min}$ of the condensate is $> 0.4$. 
In the region of local antiferromagnetism,  ${-m \sim O(0.1)}$ while $P_{max}$ is on the order of the explicit symmetry breaking parameter $h$, or else there is no d-wave pattern at all.  However, this anticorrelation is not perfect, particularly at small $U$.  On a finite volume and fixed $h$, we find that $m$ decreases to a negligible value, and $P_{max}$ increases to $\approx 0.47$ as doping increases, but there is some range where $m$ is non-negligible, and $P_{max}$ is significantly greater than $h$. So the boundary between the AF and d-wave regions is not sharp, at least at finite volume.  It is more like a crossover.  At larger $U$
the region of AF and d-wave coexistence narrows, and the boundary becomes sharper.

\begin{figure}[h!]
\center
\subfigure[~half-filling]  
{
\includegraphics[scale=0.35]{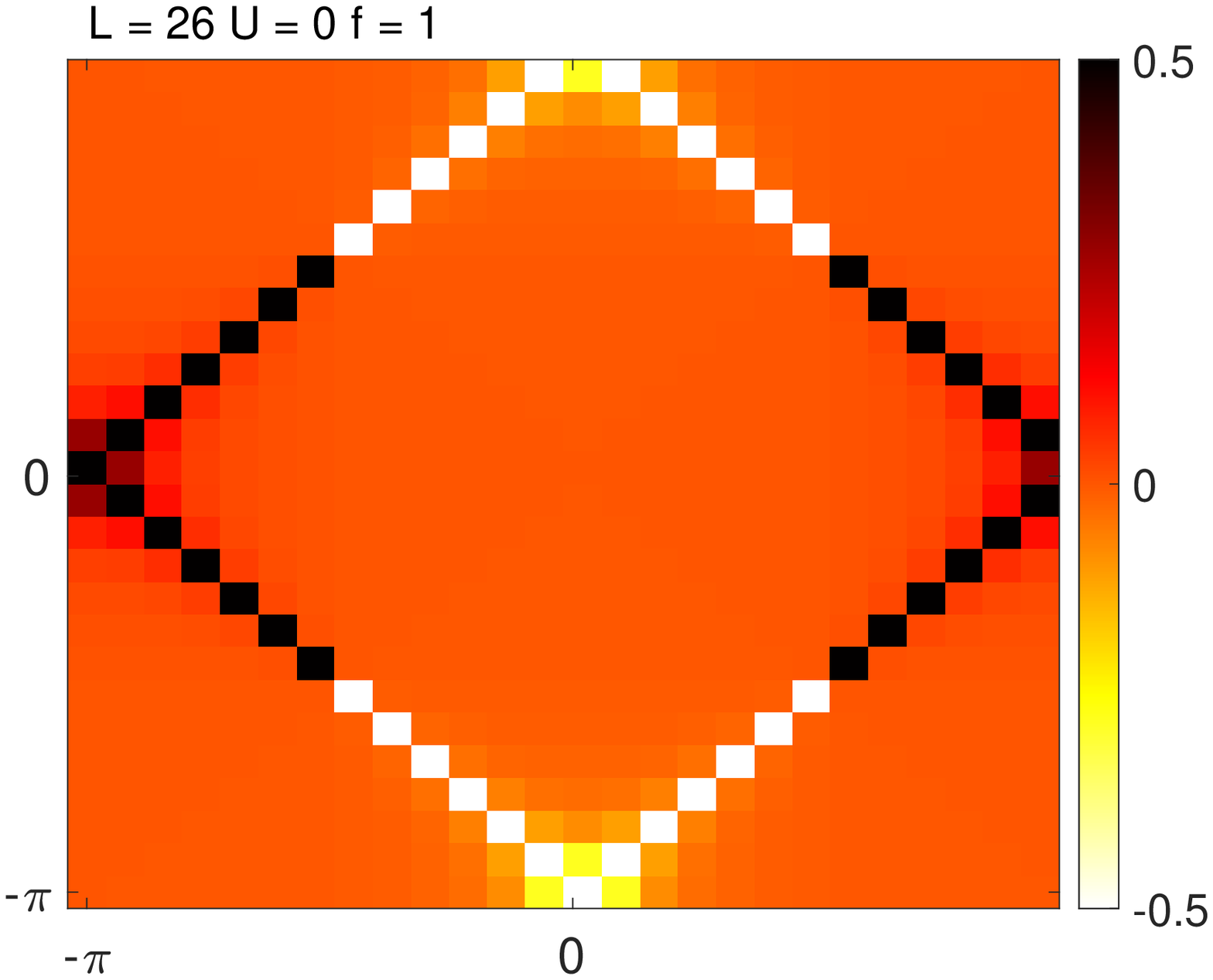} 
}
\subfigure[~density 0.9] 
{
\includegraphics[scale=0.35]{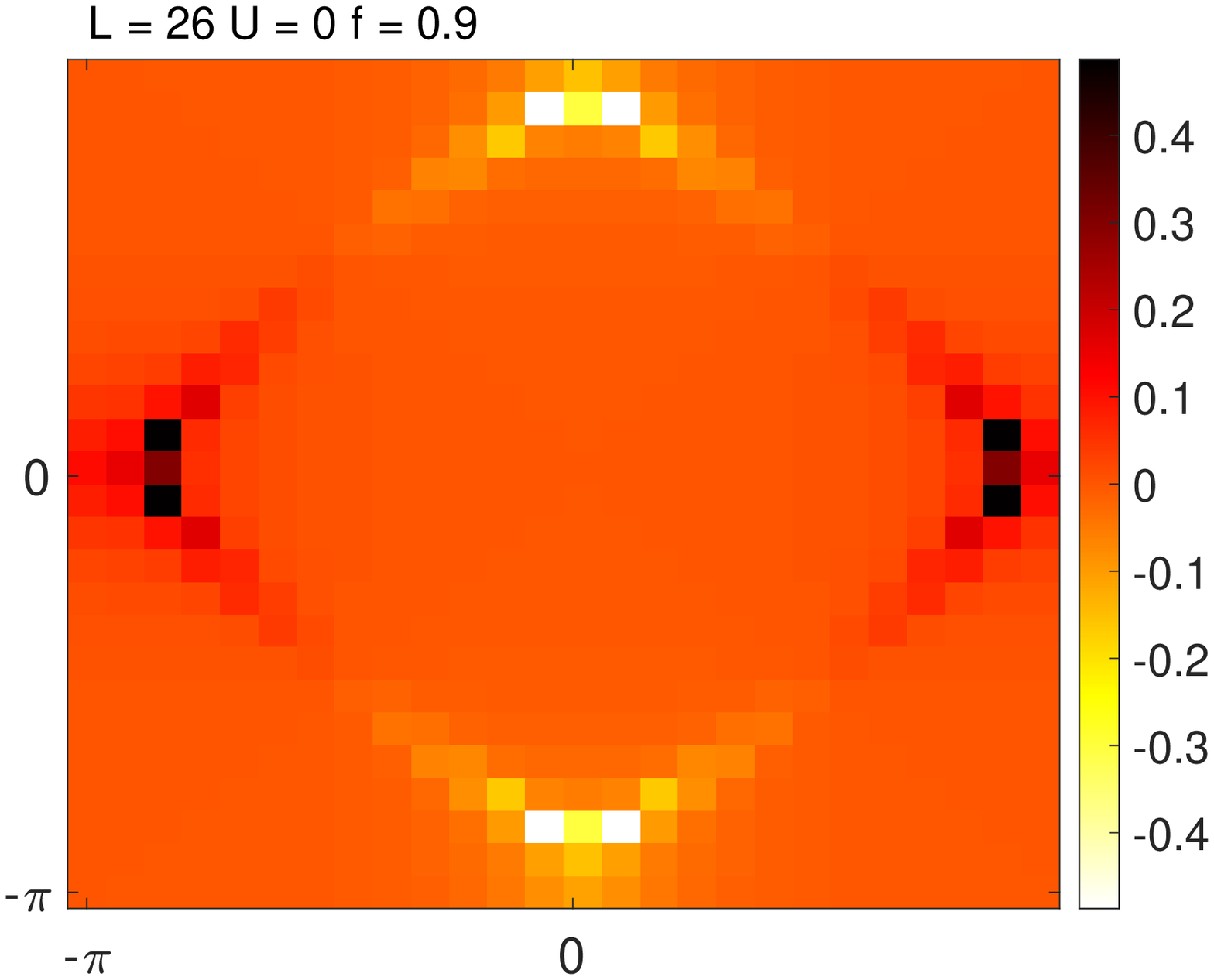} 
}
\subfigure[~half-filling] 
{
\includegraphics[scale=0.35]{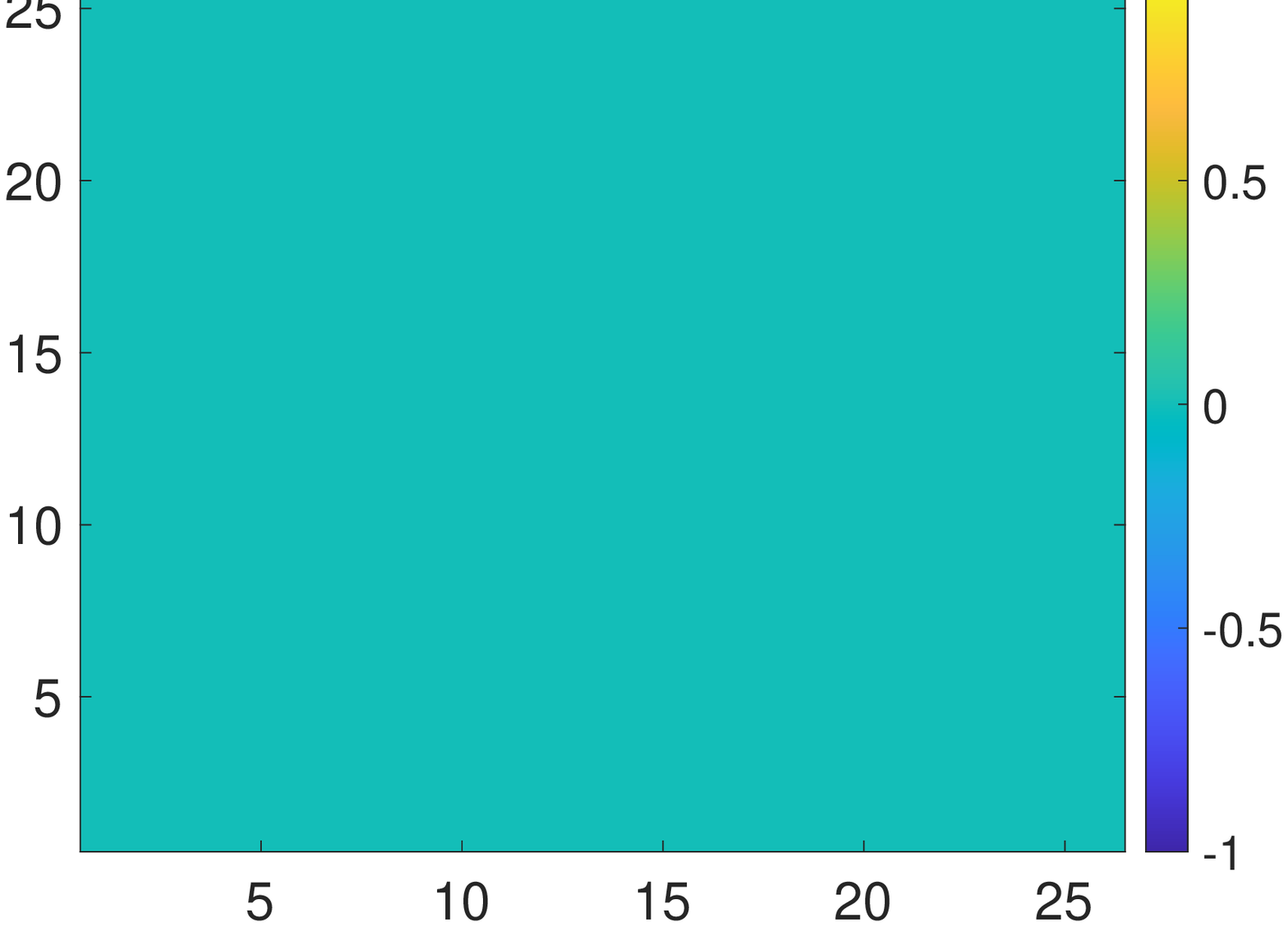} 
}
\subfigure[~density 0.9] 
{
\includegraphics[scale=0.35]{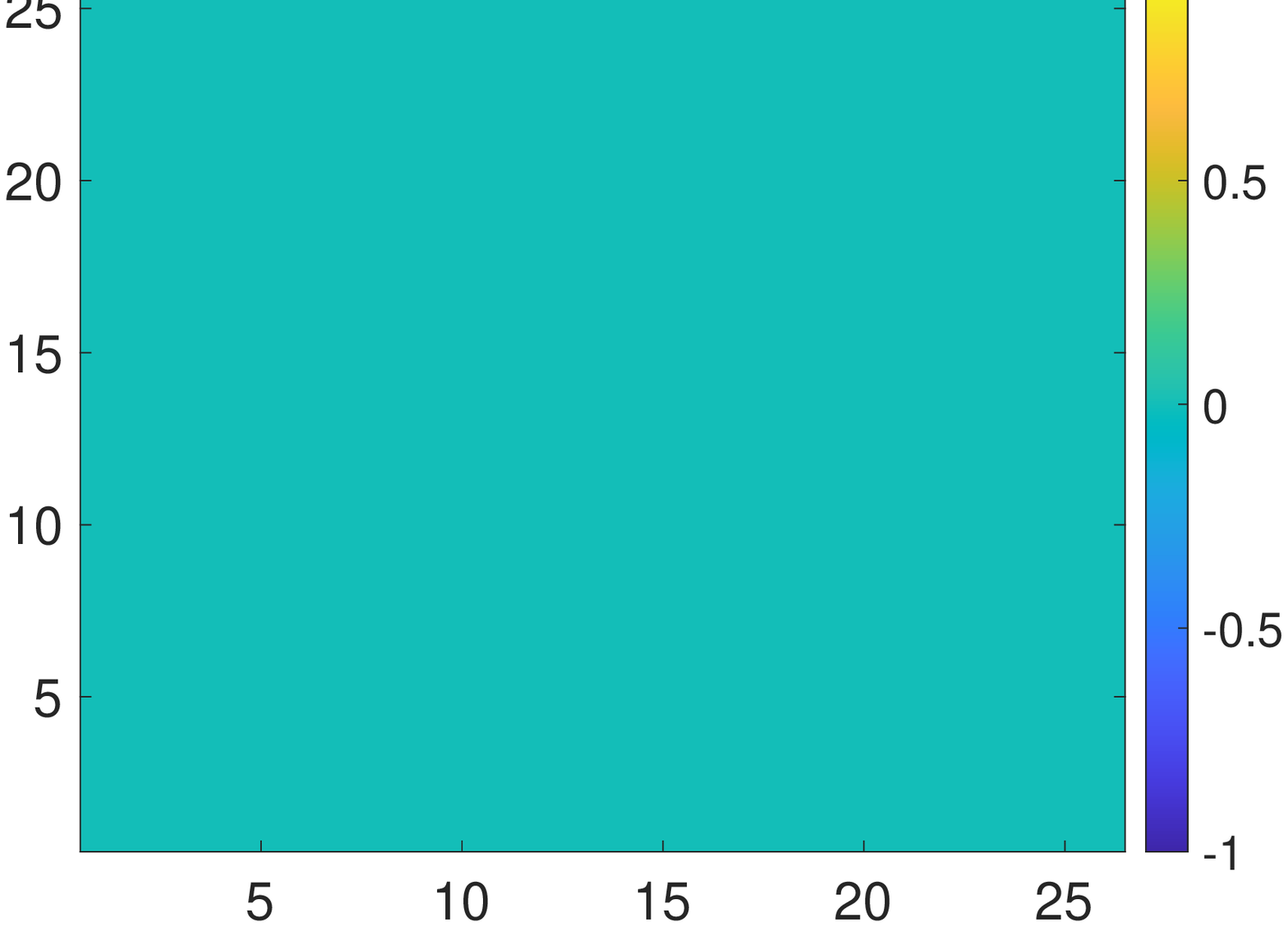} 
}
\caption{Pairing and spin density at $U=0$ on a $26\times 26$ lattice and $h=0.01$. (a) $P(k)$ at half-filling, and (b) at density 0.9. (c) $D(k)$ at half-filling and (d) at density 0.9.}
\label{U0a}
\end{figure}

   To illustrate the situation at small $U$, we start with $U=0$, Fig.\ \ref{U0a}.  The spin densities and condensates are shown at
half-filling and at density=0.9.  Here there is no AF region, and d-wave condensation begins at half-filling.  

\begin{figure}[h!]
\center
\subfigure[~half-filling]  
{
\includegraphics[scale=0.35]{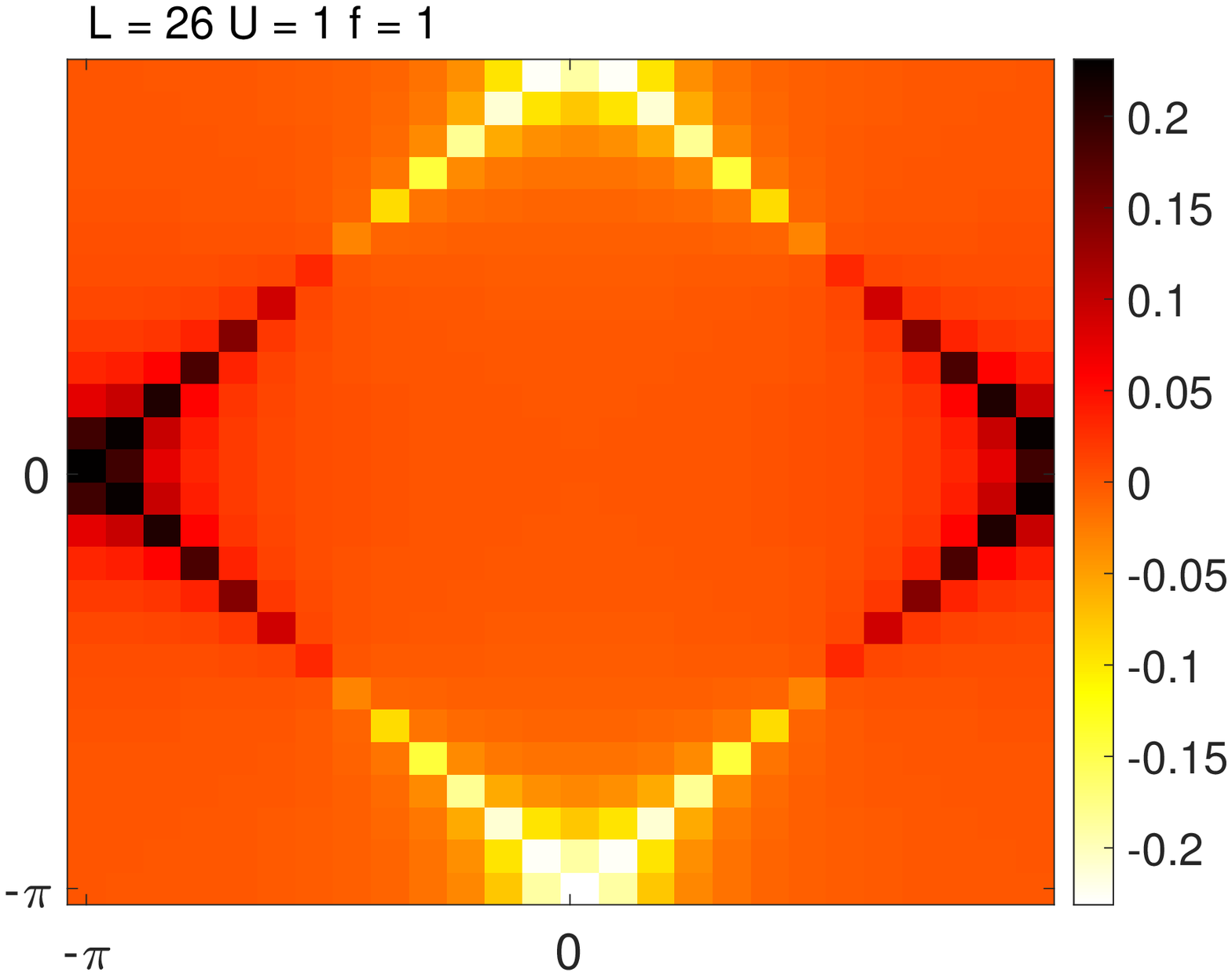} 
\label{pU1d1}
}
\subfigure[~density 0.94] 
{
\includegraphics[scale=0.35]{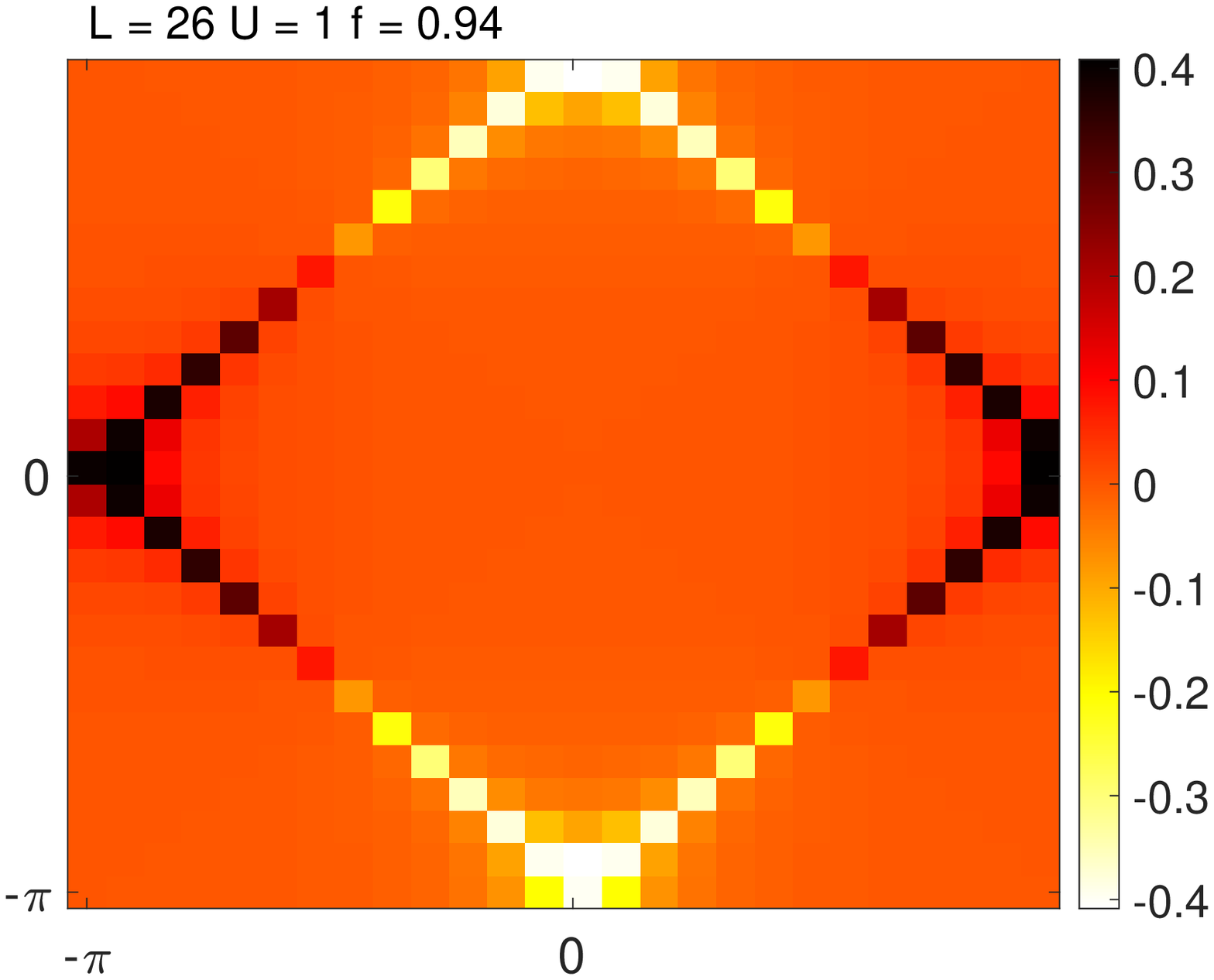} 
\label{pU1d94}
}
\subfigure[~half-filling] 
{
\includegraphics[scale=0.35]{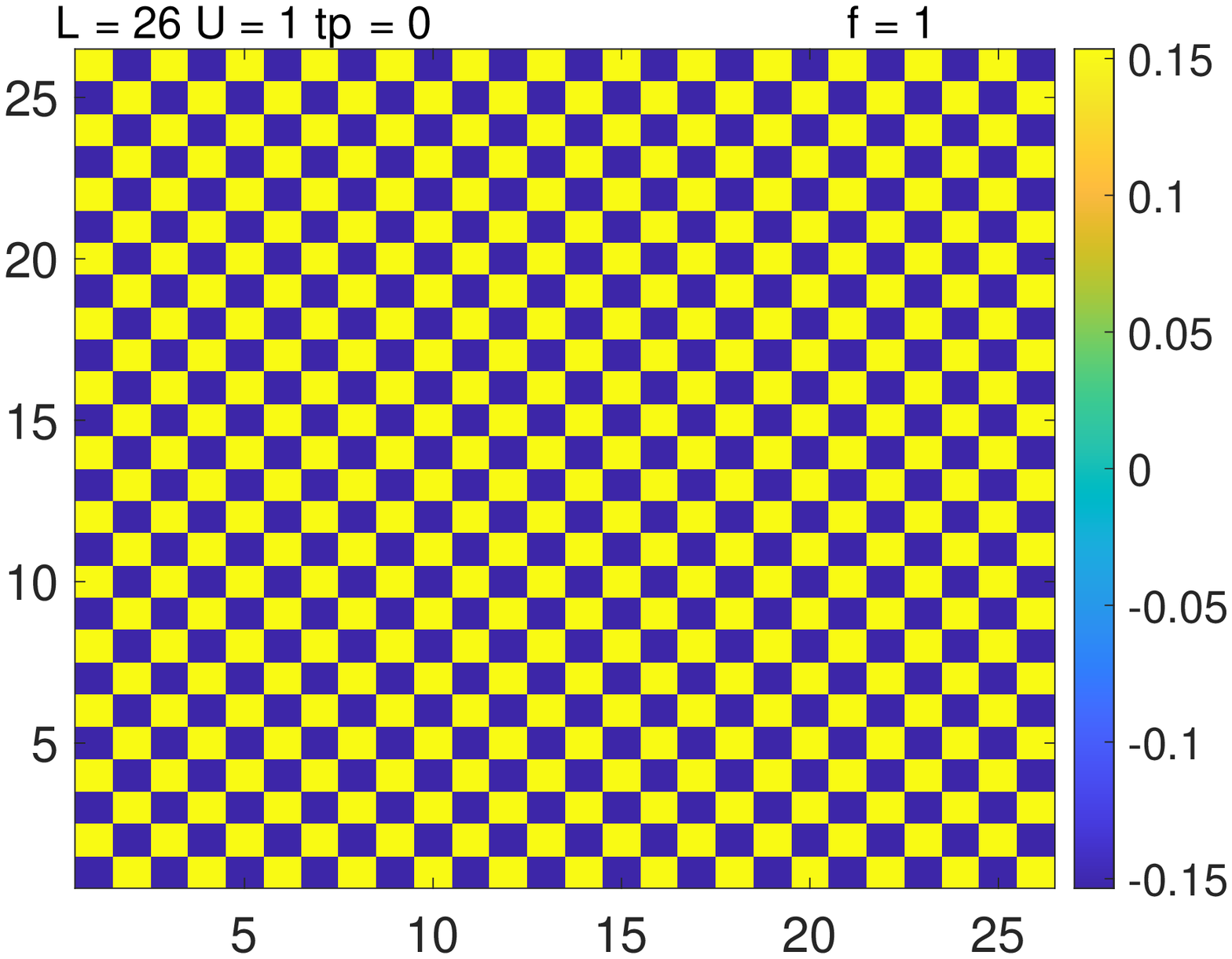} 
\label{gU1d1}
}
\subfigure[~density 0.94] 
{
\includegraphics[scale=0.35]{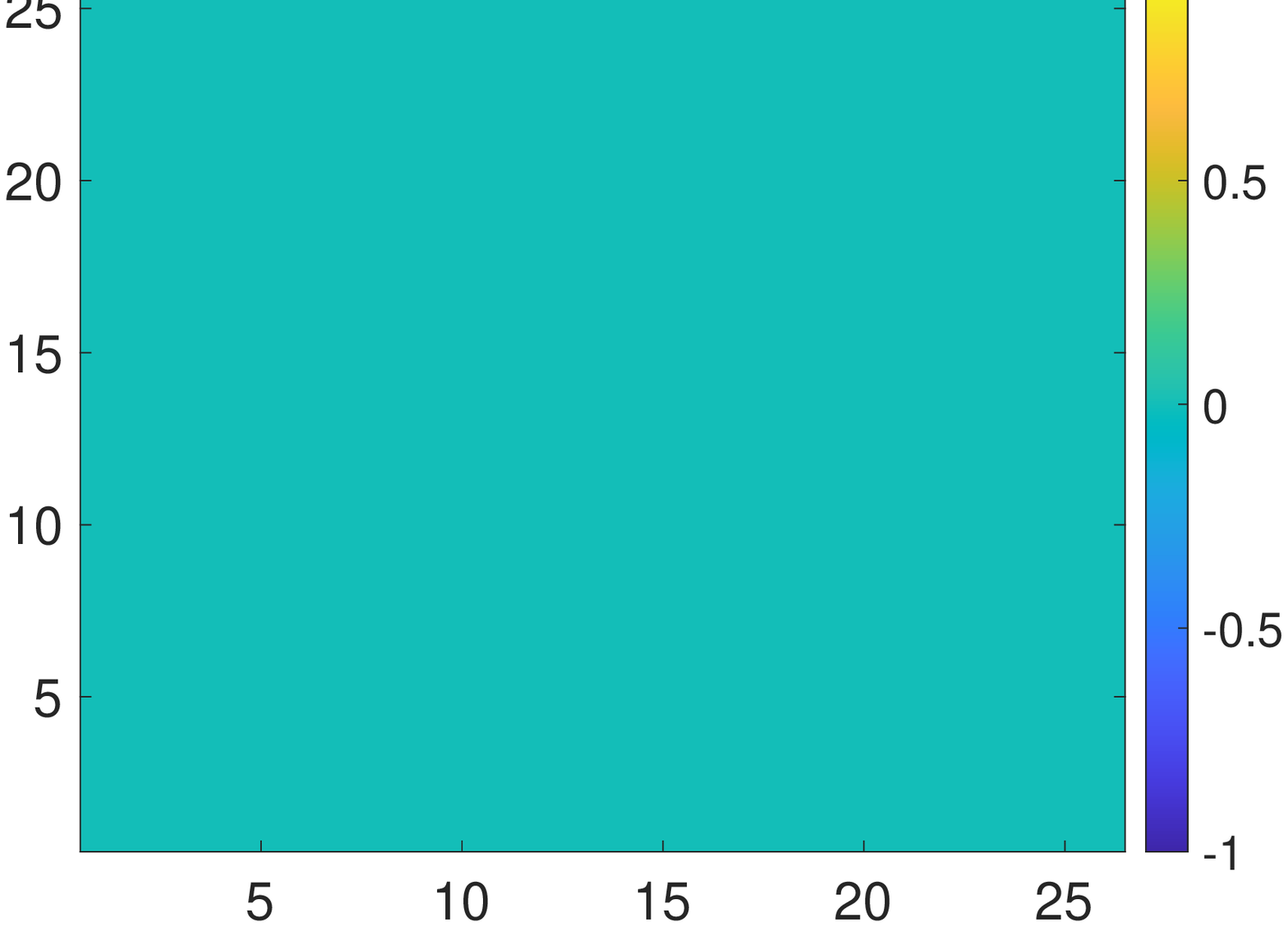} 
\label{gU1d94}
}
\caption{Pairing and spin density at $U=1$ on a $26\times 26$ lattice and $h=0.01$. (a) $P(k)$ at half-filling, and (b) at density 0.94. (c) $D(k)$ at half-filling and (d) at density 0.94.}
\label{U1a}
\end{figure}

\ni At $U=1$ (Fig.\ \ref{U1a}) a checkerboard pattern in the spin density (with non-zero AF), and a d-wave condensate coexist at half-filling, but both are of modest amplitude.  Away from half-filling, the magnitude of $m$ decreases and $P_{max}$ increases, and at filling 0.94  AF has entirely disappeared.  
 
\begin{figure*}[h!]
\center
\subfigure[~]  
{
\includegraphics[scale=0.25]{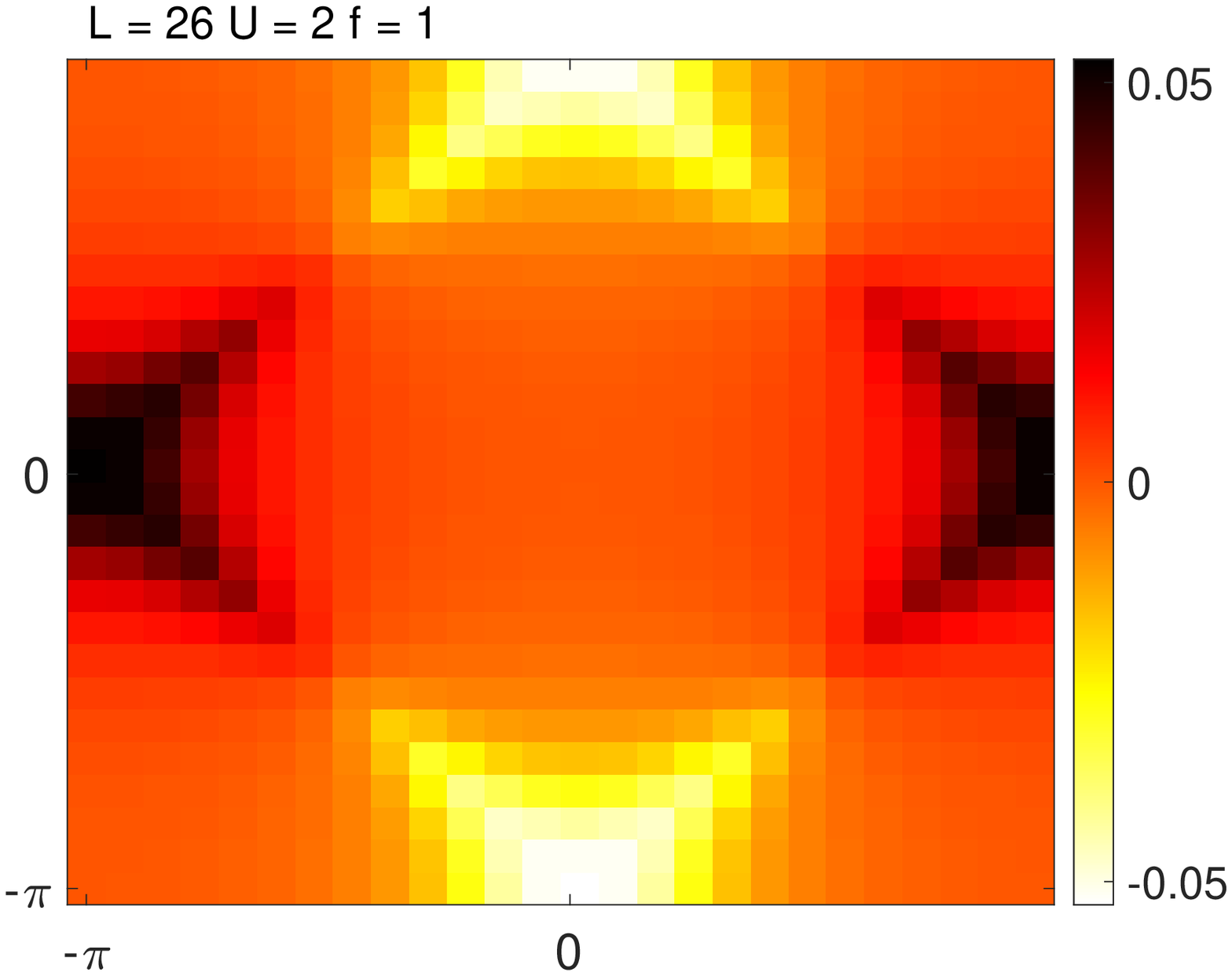} 
\label{pU2d1}
}
\subfigure[~] 
{
\includegraphics[scale=0.25]{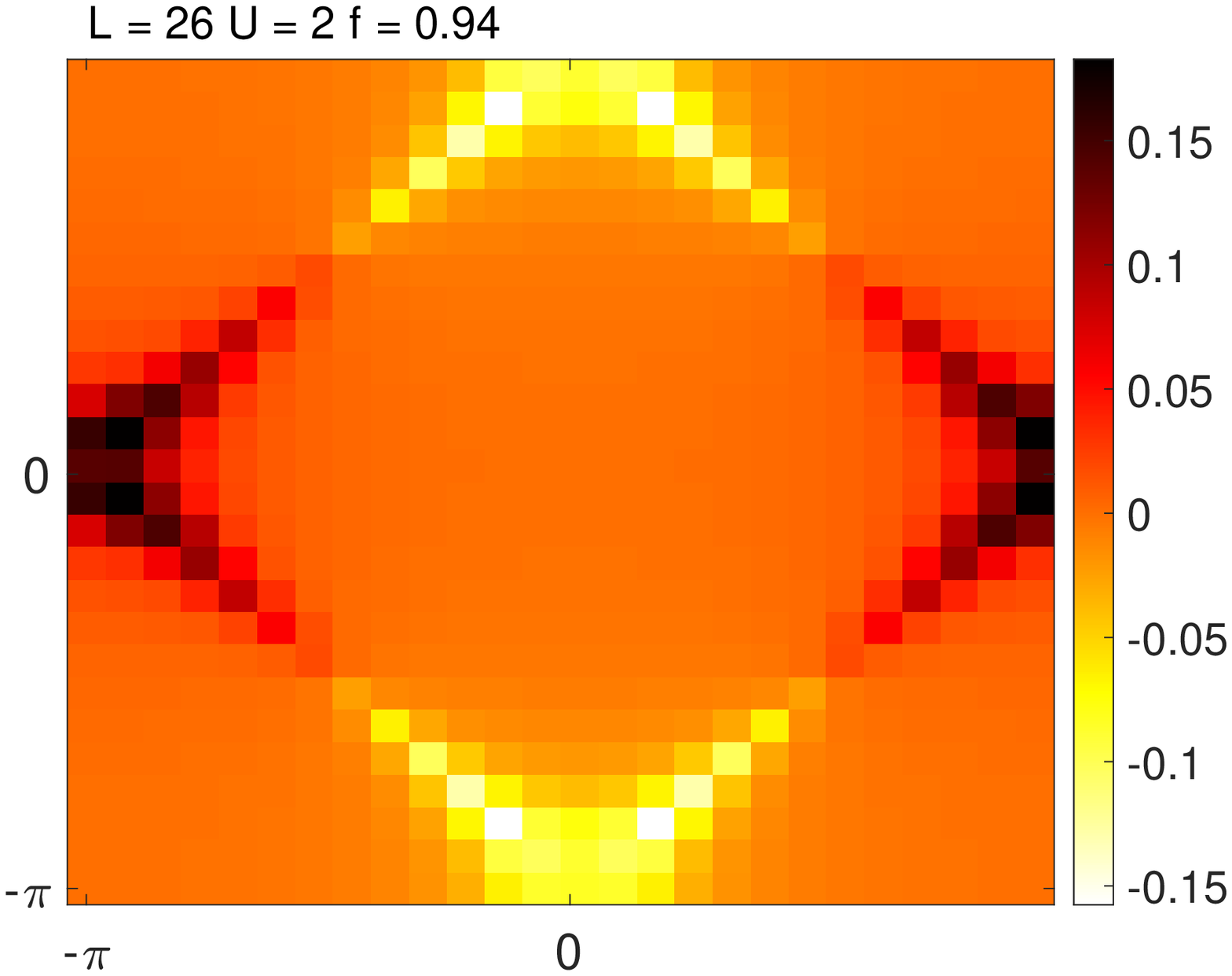} 
\label{pU2d94}
}
\subfigure[~] 
{
\includegraphics[scale=0.25]{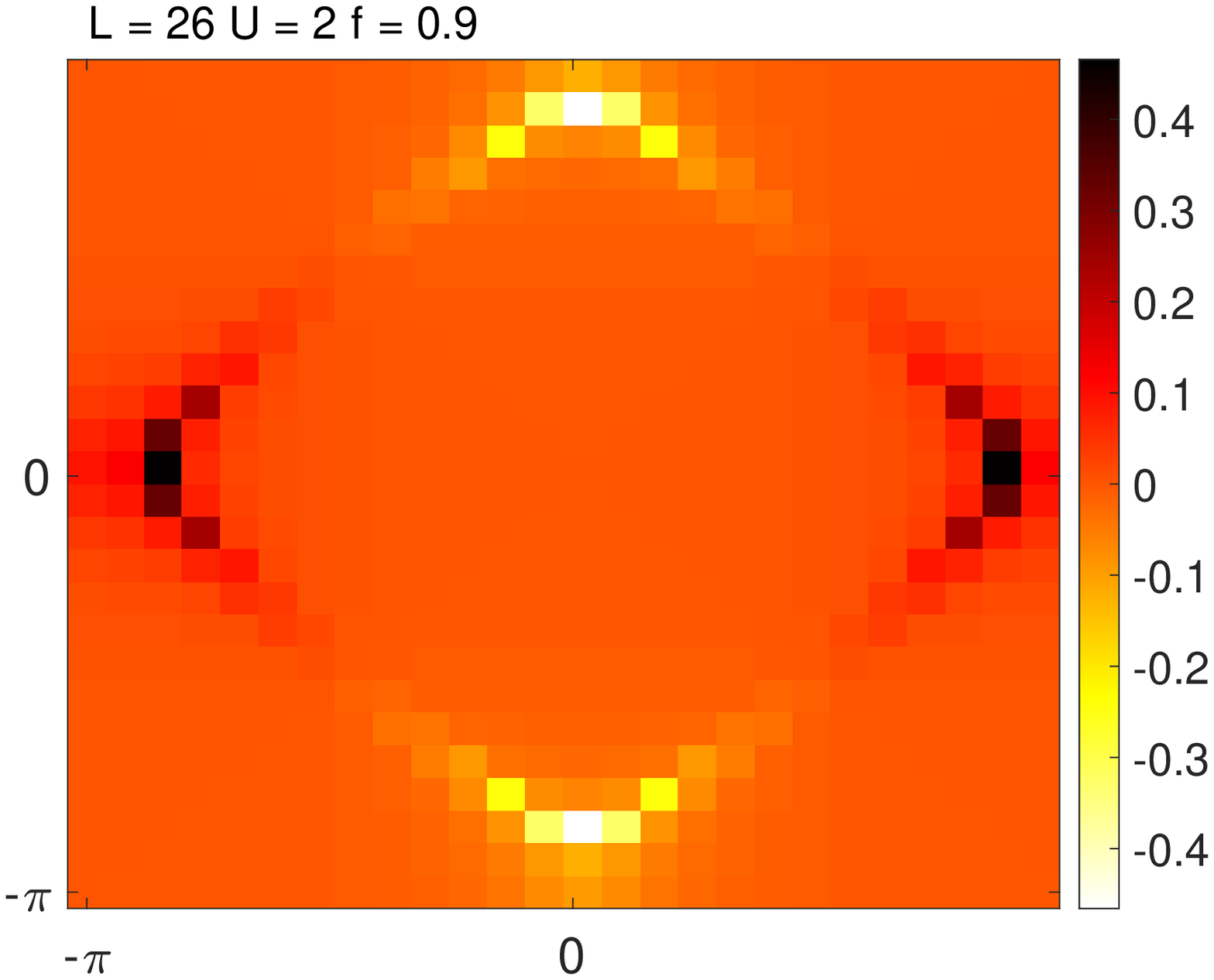} 
\label{pU2d9}
}
\subfigure[~] 
{
\includegraphics[scale=0.25]{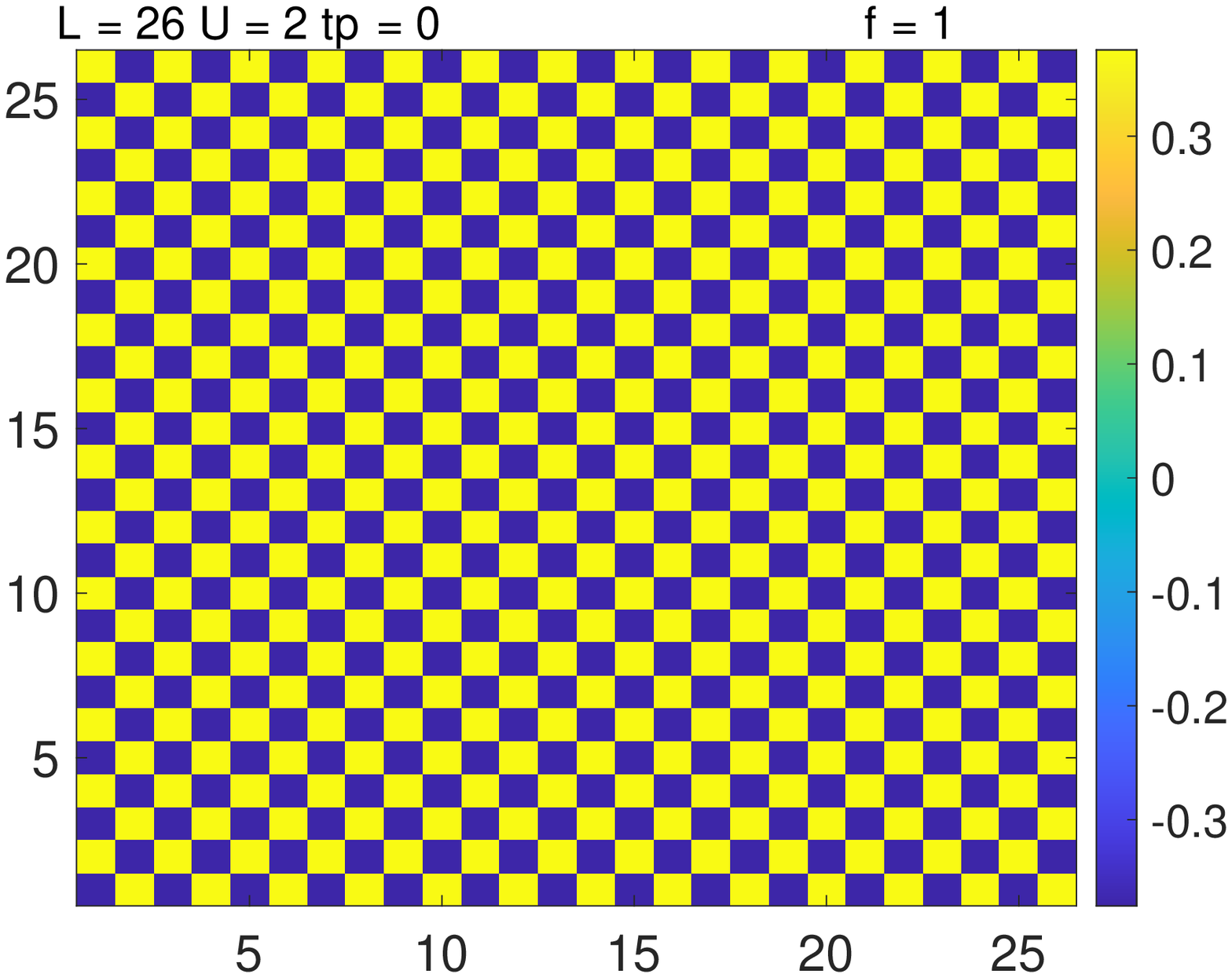} 
\label{gU2d1}
}
\subfigure[~] 
{
\includegraphics[scale=0.25]{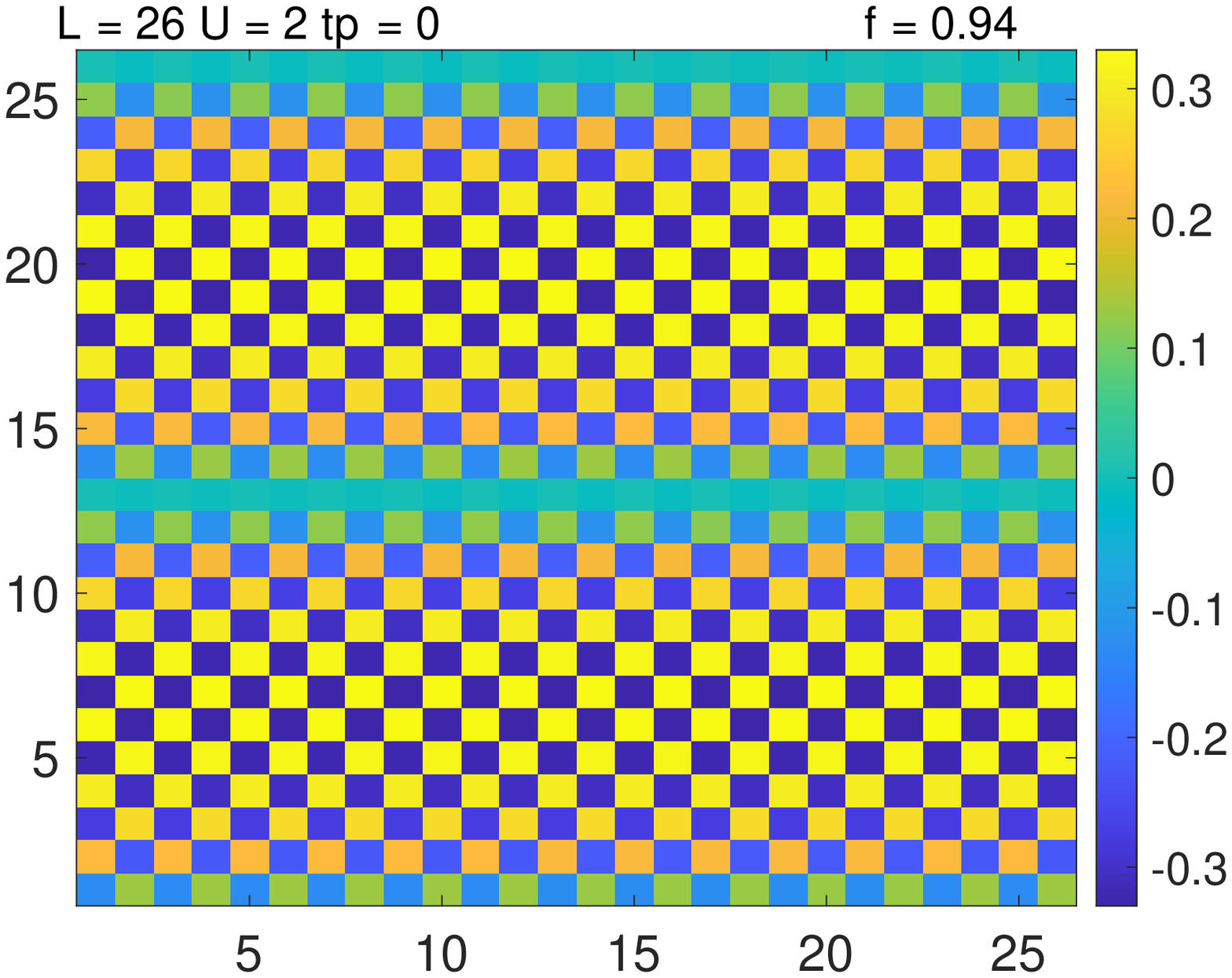} 
\label{gU2d94}
}
\subfigure[~]
{
\includegraphics[scale=0.25]{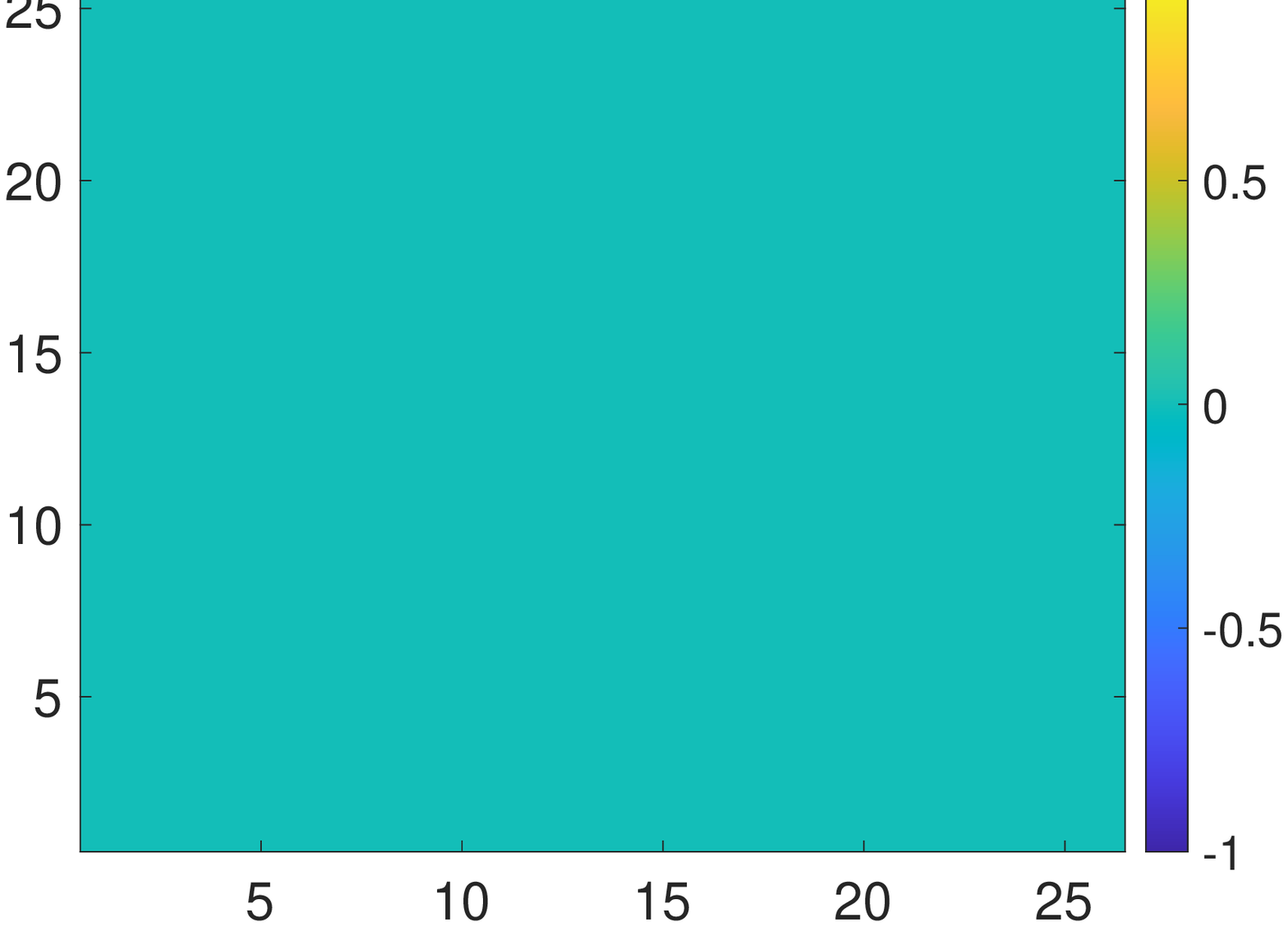} 
\label{gU2d9}
}
\caption{Pairing and spin density at $U=2$ on a $26\times 26$ lattice and $h=0.01$. (a) $P(k)$ at half-filling, and (b) at density 0.94. and (c) density 0.9. (d) $D(k)$ at half-filling and (e) at density 0.94 and (f) at density 0.9.}
\label{U2a}
\end{figure*}

Around $U=2$ (Fig.\ \ref{U2a}) one begins to see stripe patterns away from half-filling, coexisting with a d-wave pattern of small amplitude. Again, AF disappears entirely when $P_{max} \approx 0.4$.

 \begin{figure}[h]
 \center
 \includegraphics[scale=0.7]{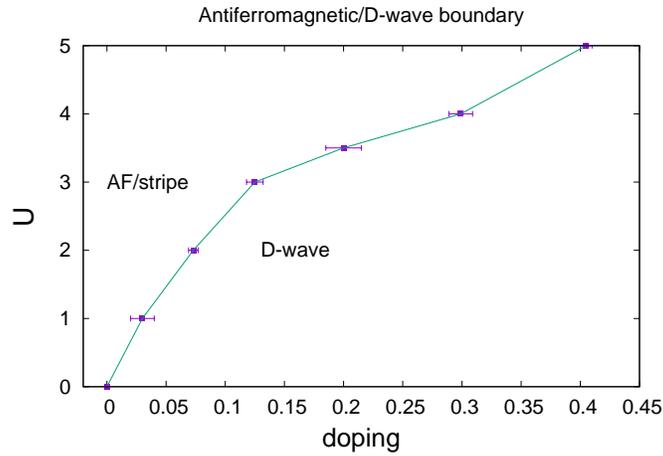}
 \caption{A rough estimate of the boundary, in the doping$-$$U/t$ plane, with d-wave condensation at $P_{max}>0.4$ to the right of the boundary, and negligible magnetization.  The region of local antiferromagnetism and, for $U \ge 2$, stripe order, exists a little to the left of the boundary (see text).  The solid points were determined numerically, at $U=1,2,3,3.5,4,5$, and the lines joining the points are to guide the eye.  Lattice volume is $26\times 26$ and $h=0.01$.} 
 \label{phase}
 \end{figure}

\ni Finally, as $U$ increases, the AF region becomes wider, and the d-wave pattern is negligible (i.e.\ $O(h)$) in most of that
region, e.g.\ Fig.\ \ref{p81}.

In Fig.\ \ref{phase} we plot a line in the density$-$$U/t$ plane where the amplitude $P_{max}$ of the d-wave condensate is 
$P_{max}> 0.4$ to the right of that line, and magnetization is negligible.  The lattice volume is $26\times 26$, and $h=0.01$. We cannot rule out the possibility that the line might shift somewhat to the left in the limit of infinite volume and $h\ra 0$.\footnote{The condensate at this volume disappears, at each $U$, at densities $\approx 0.1$, which seems unrealistically low.  It may be that this could change with increasing volume,  but it also seems likely that the mean field approximation is unreliable at small densities.}

\clearpage

\section{\label{Compare} Energy comparisons}

   We can compare the energies of the standard Hartree-Fock ground state \rf{Om1}, which we now denote
as $|\Om^S\rangle$ and the Hartree-Fock ground state with Nambu spinors \rf{Om2}, denoted $|\Om^N\rangle$.
Define
\beq
 \rho(x,\a\b)= \sum_{i=1}^{L^2} \phi_{i}^*(x,\a) \phi_{i}(x,\b)
\eeq

and 
 \bea
 \r(x,\a\b) &=& \sum_{i=1}^{L^2}  \phi_i^*(x,\a) \phi_i(x,\b) \non \\
 \r_{nn}(x,\a\b) &=&  \sum_{i=1}^{L^2}  \phi^*_i(x,\a)\bigg( \phi_i(x+\hat{e}_1,\b) + \phi_i(x-\hat{e}_1,\b) \non \\
   & & +\phi_i(x+\hat{e}_2,\b) + \phi_i(x-\hat{e}_2,\b) \bigg) \non \\
  \rho_{nnn}(x,\a \b) &=&
  \sum_{i=1}^{L^2} \phi^*_i(x,\a)\bigg( \phi_i(x+\hat{e}_1 + \hat{e}_2,\b)  \non \\
& & + \phi_i(x+\hat{e}_1- \hat{e}_2,\b)  +\phi_i(x-\hat{e}_1+\hat{e}_2,\b) \non \\
& & + \phi_i(x-\hat{e}_1- \hat{e}_2,\b) \bigg) \ ,
\label{rhon}
 \eea
 where  Greek indices represent spins $\ua,\da$ in standard Hartree-Fock, or the indices $1,2$ of
 Nambu spinors in Nambu Hartree-Fock.
\ni It is not hard to show (see the Appendix) that
\bea
\langle \Om^S |H| \Om^S \rangle  &=& -t \sum_x \{ \rho_{nn}(x,\ua \ua) + \rho_{nn}(x,\da \da) \} 
        -t' \sum_x \{ \rho_{nnn}(x,\ua \ua) + \rho_{nnn}(x,\da \da) \} \non \\
& & ~~+~U \sum_x \bigg( \r(x,\ua \ua) \r(x,\da \da) -
                  \r(x,\ua \da) \r(x,\da \ua) \bigg)
       - \m M \ , 
\label{ES}
\eea
where $M$ is the total number of electrons, while
 \bea
 \langle \Om^N |H| \Om^N \rangle &=& 
       -t \sum_x \{ \rho_{nn}(x,11) - \rho_{nn}(x,22) \} 
       -t' \sum_x \{ \rho_{nnn}(x,11) - \rho_{nnn}(x,22) \}  \non \\
 & & ~~+ U \sum_x \r(x,11)  - U\sum_x \bigg( \r(x,11) \r(x,22) -
                  \r(x,12) \r(x,21) \bigg) \non \\
  & & ~~~ - \m\sum_x \{ \rho(x,11) + 1 - \rho(x,22) \}  \ ,
 \label{EN}
 \eea
where $H$ is the full Hubbard model Hamiltonian \rf{Hub}.   

\begin{figure*}[h!]
\center
 \hspace{-10pt}
\subfigure[~$d=0.9,t'=0$] 
{
 \includegraphics[scale=0.4]{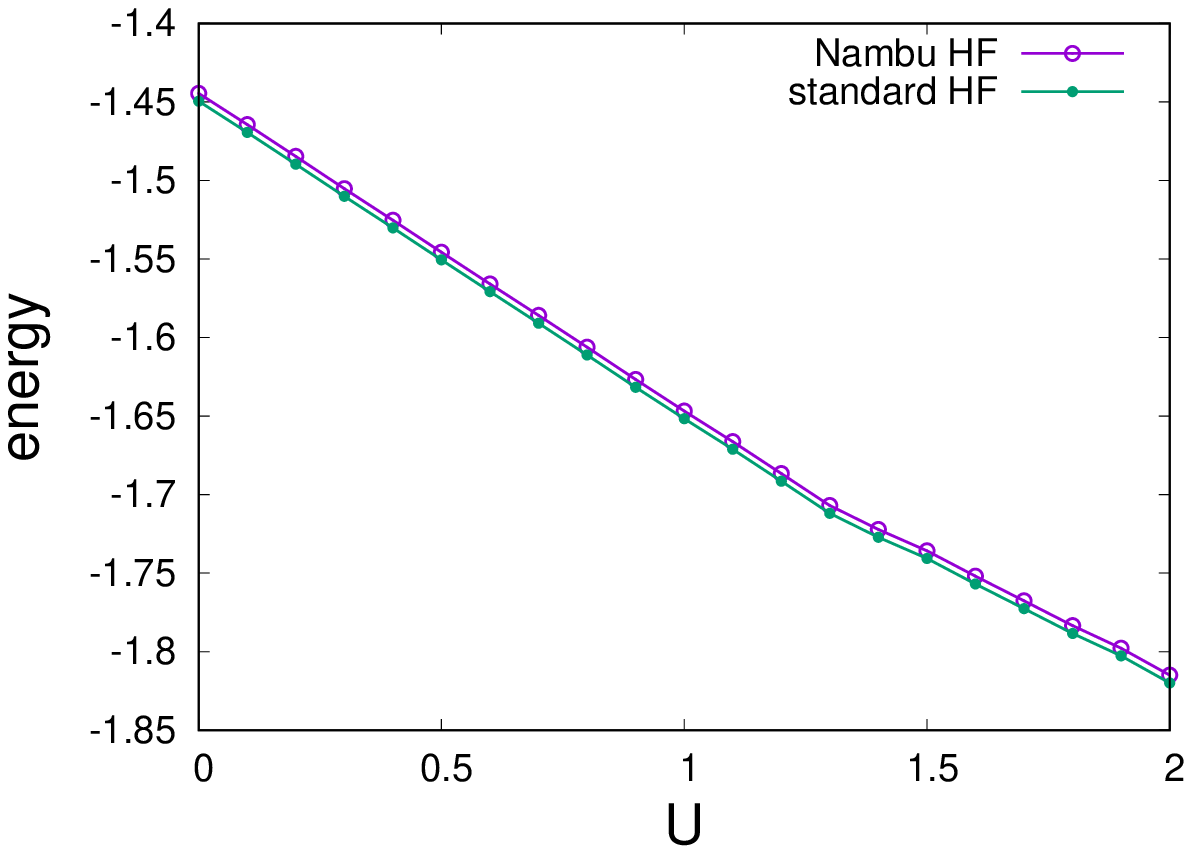} 
\label{d9t0}
}
\subfigure[~$d=0.9,t'=0.1$]  
{
 \includegraphics[scale=0.4]{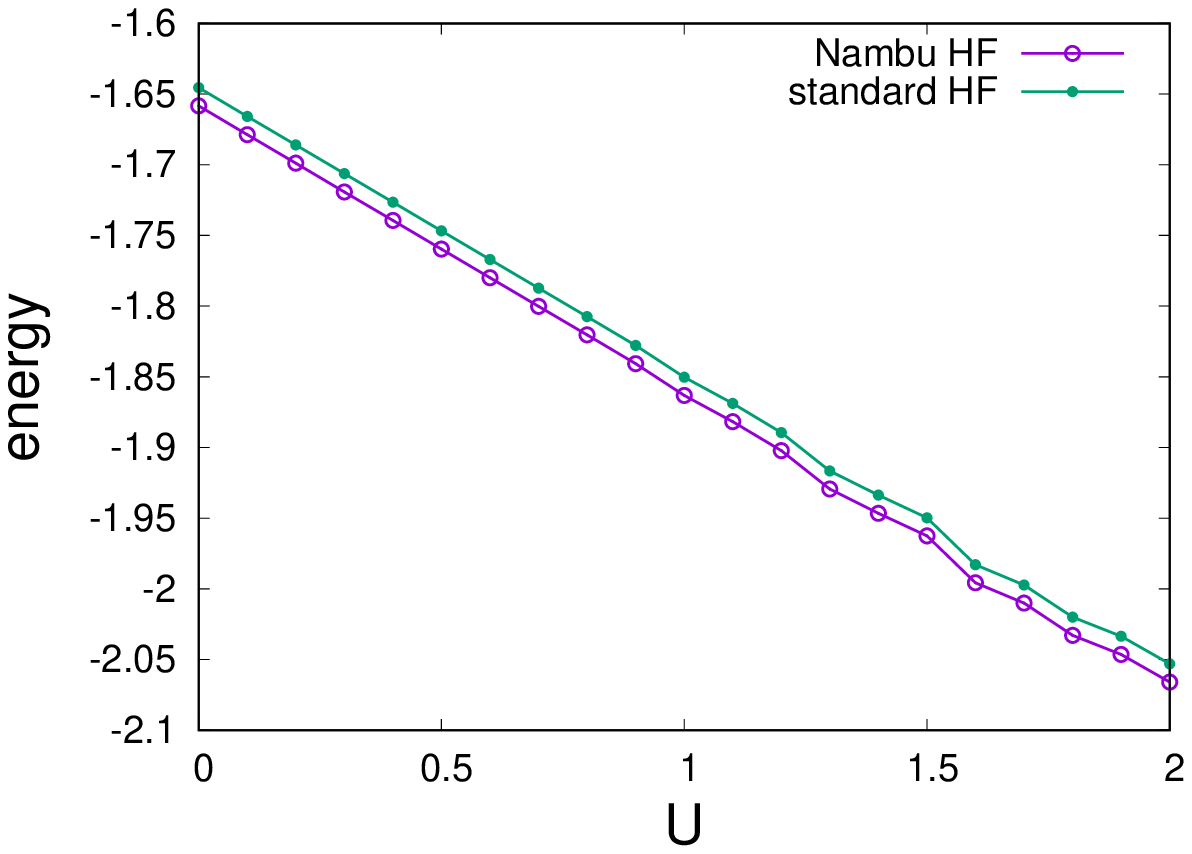} 
\label{d9t1}
}
\subfigure[~$d=0.9,t'=0.3$]
{
 \includegraphics[scale=0.4]{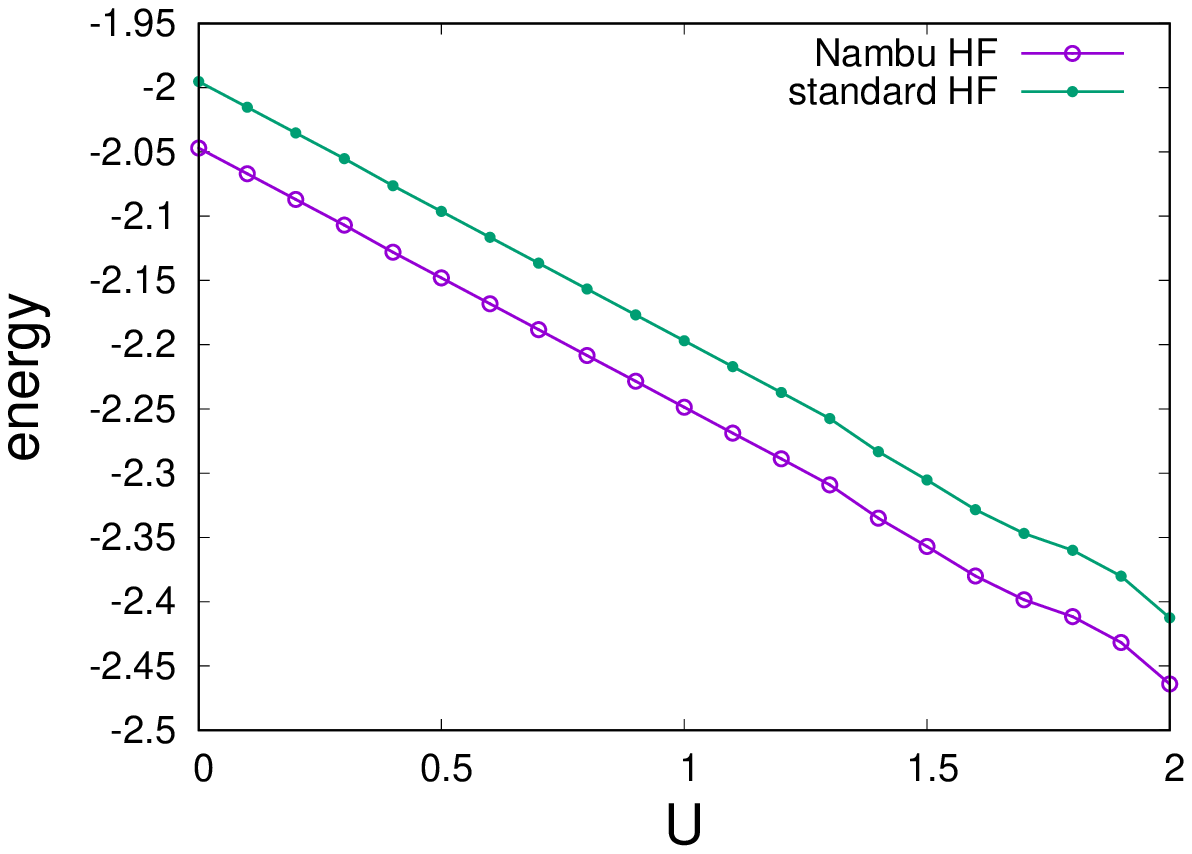}
\label{d9t3}
 }
 \hspace{5pt}
\subfigure[~$d=0.7,t'=0.1$]
{
\includegraphics[scale=0.4]{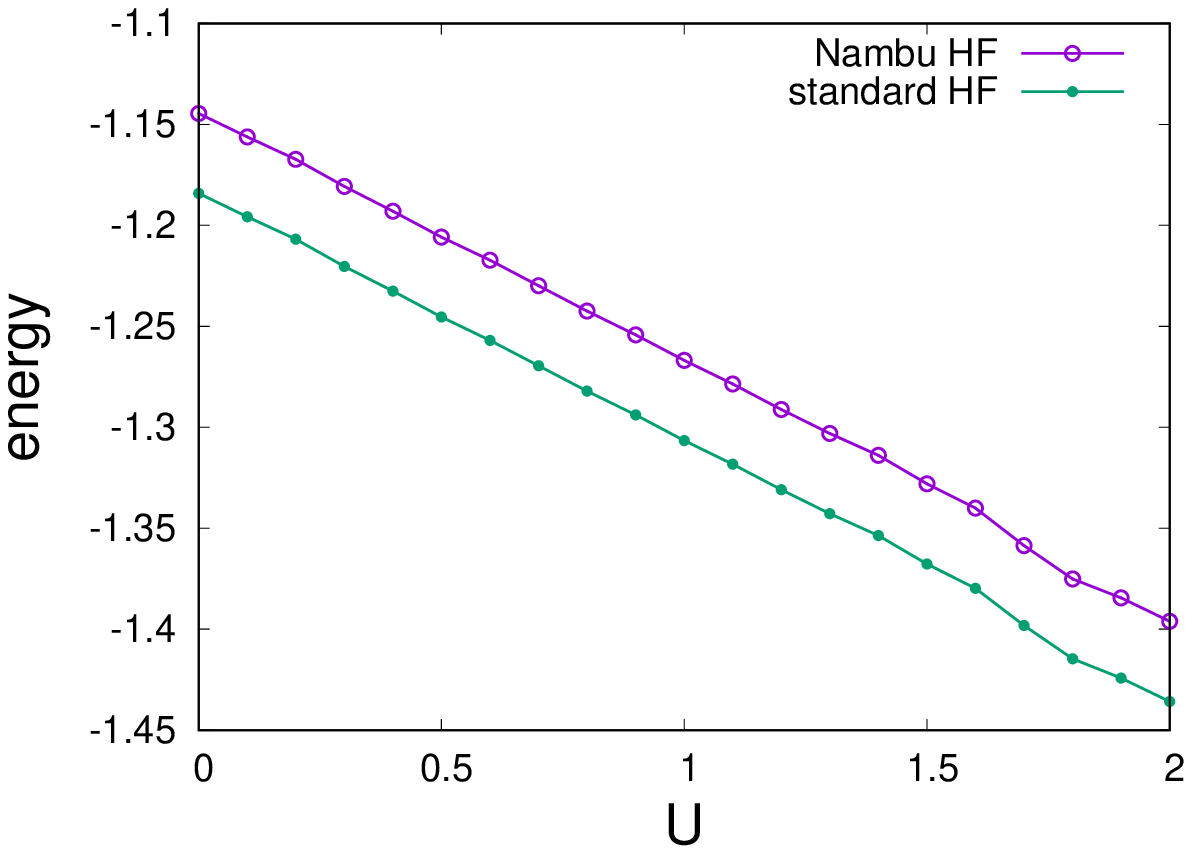}
\label{d7t1}
}
\subfigure[~$d=0.7,t'=0.3$]
{
\includegraphics[scale=0.4]{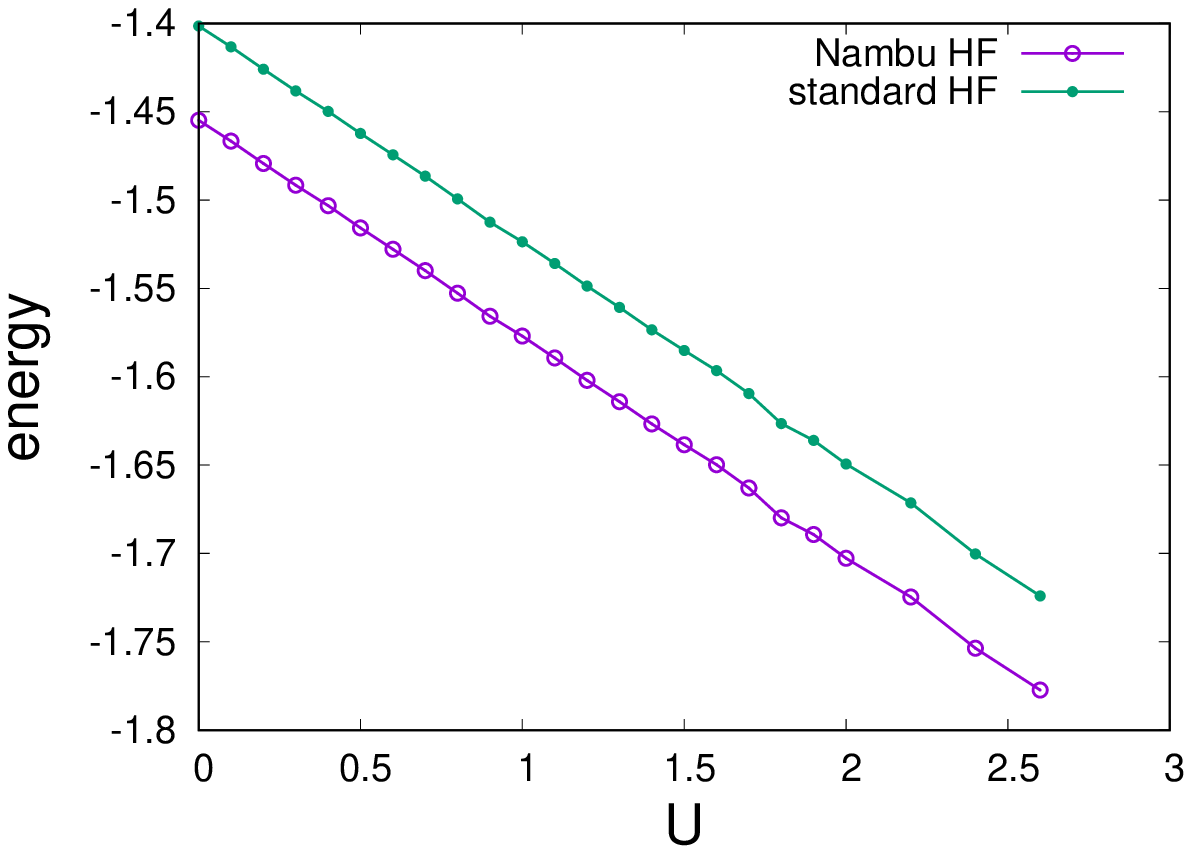}
\label{d7t3}
}
\hspace{30pt}
\subfigure[~$d=1.1,t'=0.1$]
{
\includegraphics[scale=0.4]{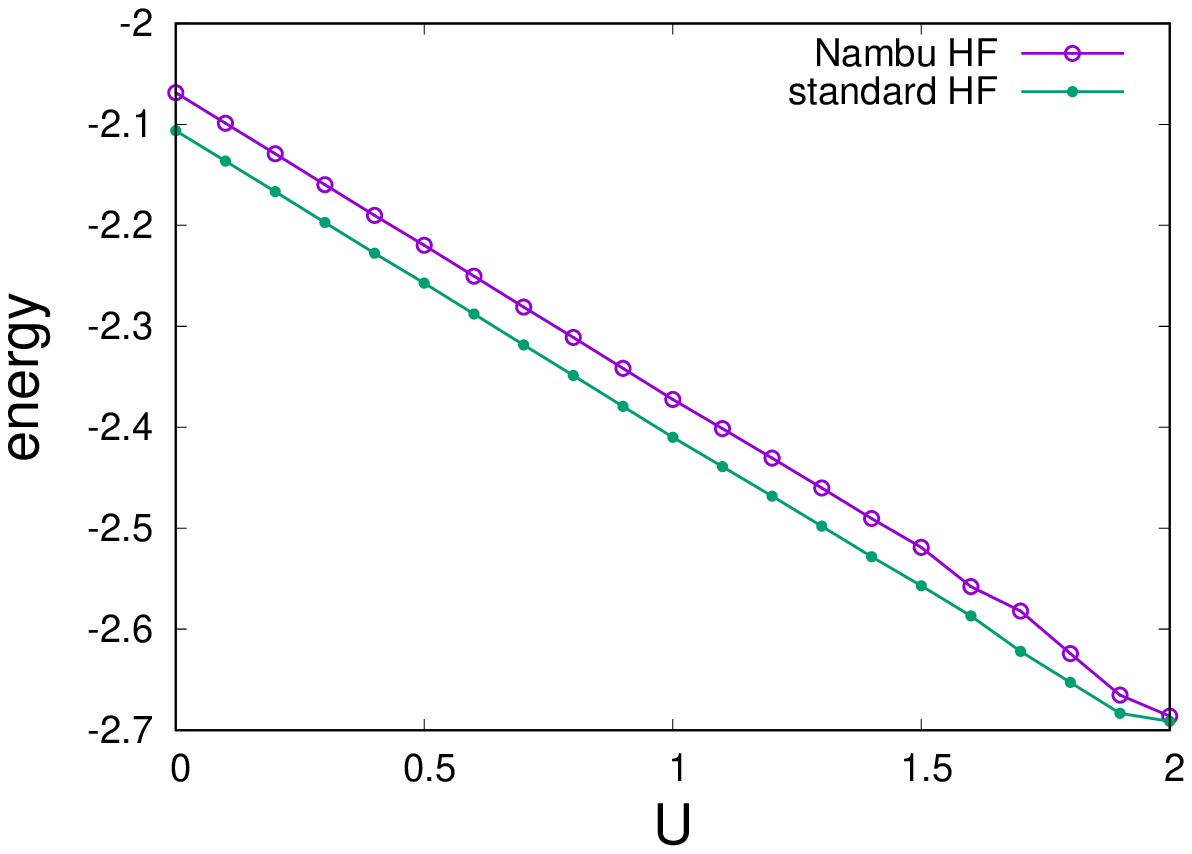}
\label{d11t1}
}
\subfigure[~$d=1.1,t'=-0.1$]
{
\includegraphics[scale=0.4]{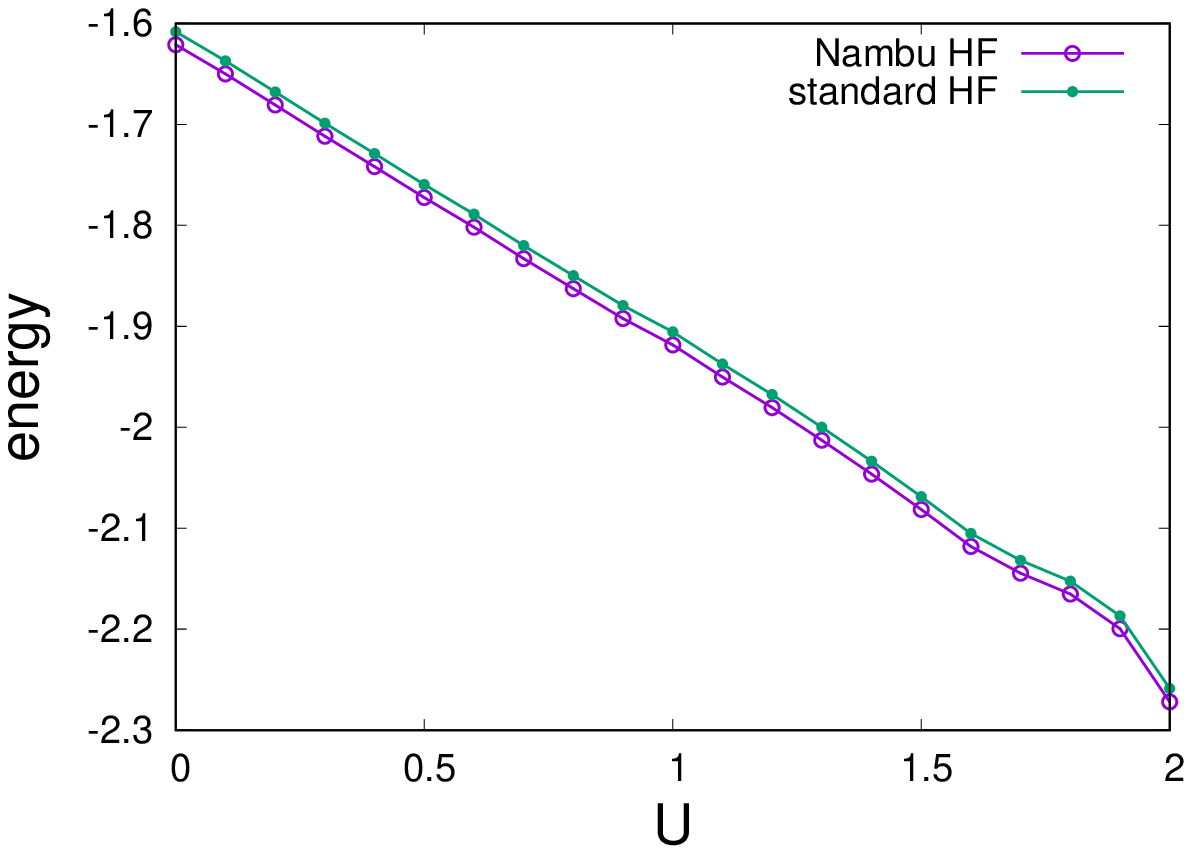}
\label{d11t1}
}
\hspace{45pt}
\subfigure[~$d=1.0,t'=0$]
{
\includegraphics[scale=0.4]{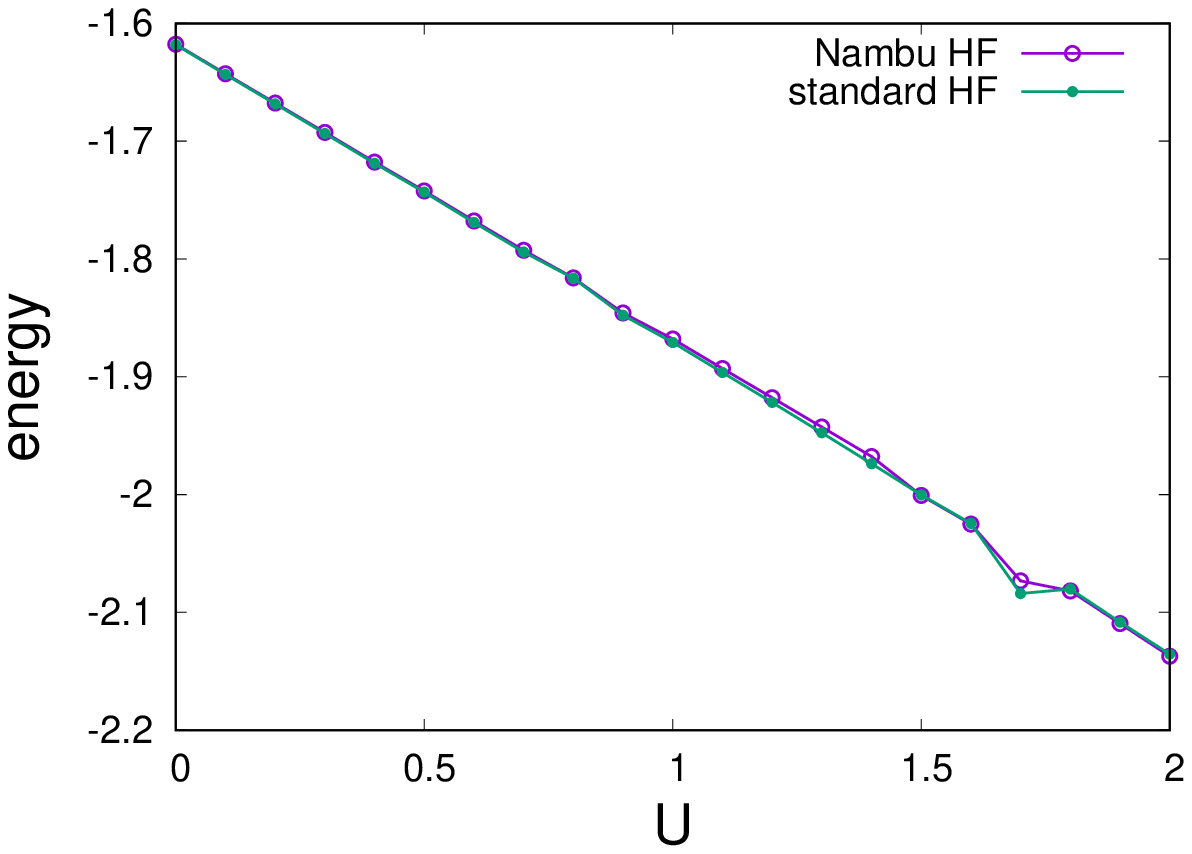}
\label{d10t0}
}
 \hspace{-15pt}
\subfigure[~$d=1.0,t'=0.1$]
{
\includegraphics[scale=0.4]{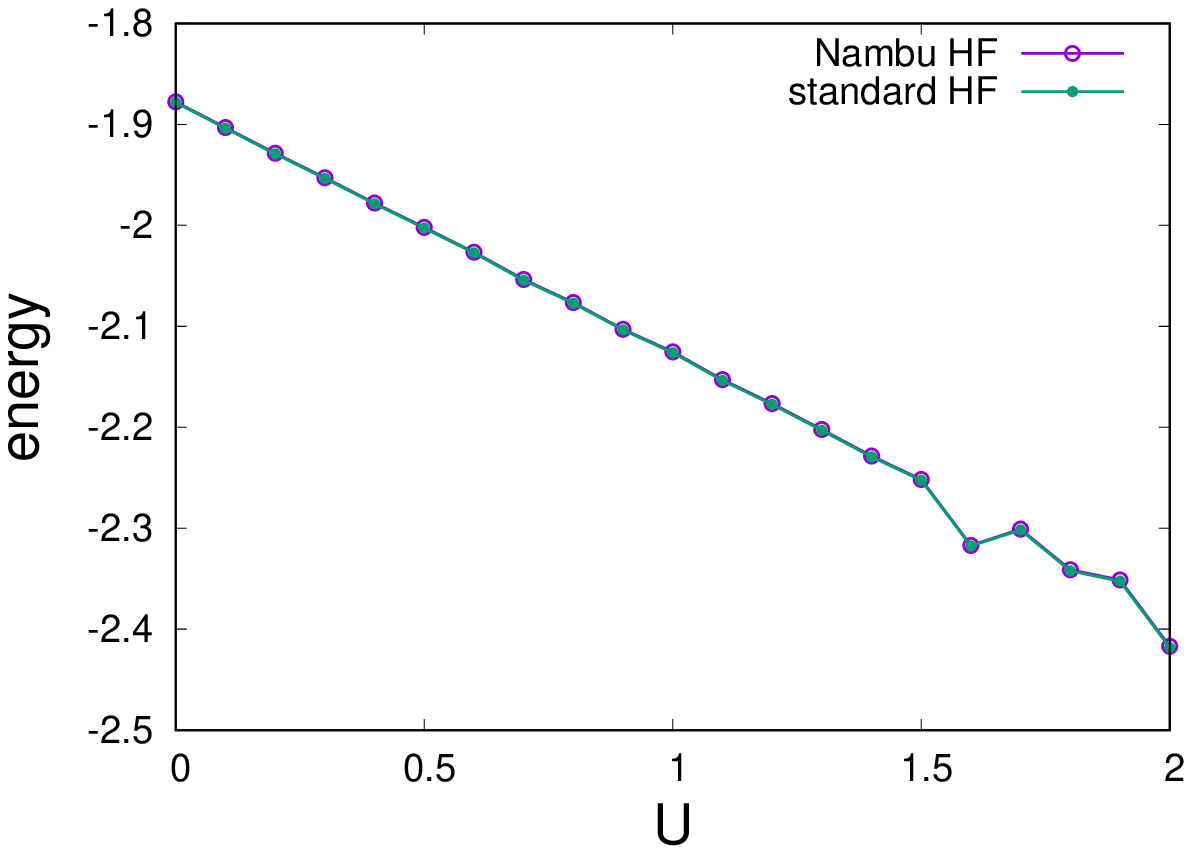}
\label{d10t1}
}
 \hspace{-15pt}
\subfigure[~$d=1.0,t'=-0.1$]
{
\includegraphics[scale=0.4]{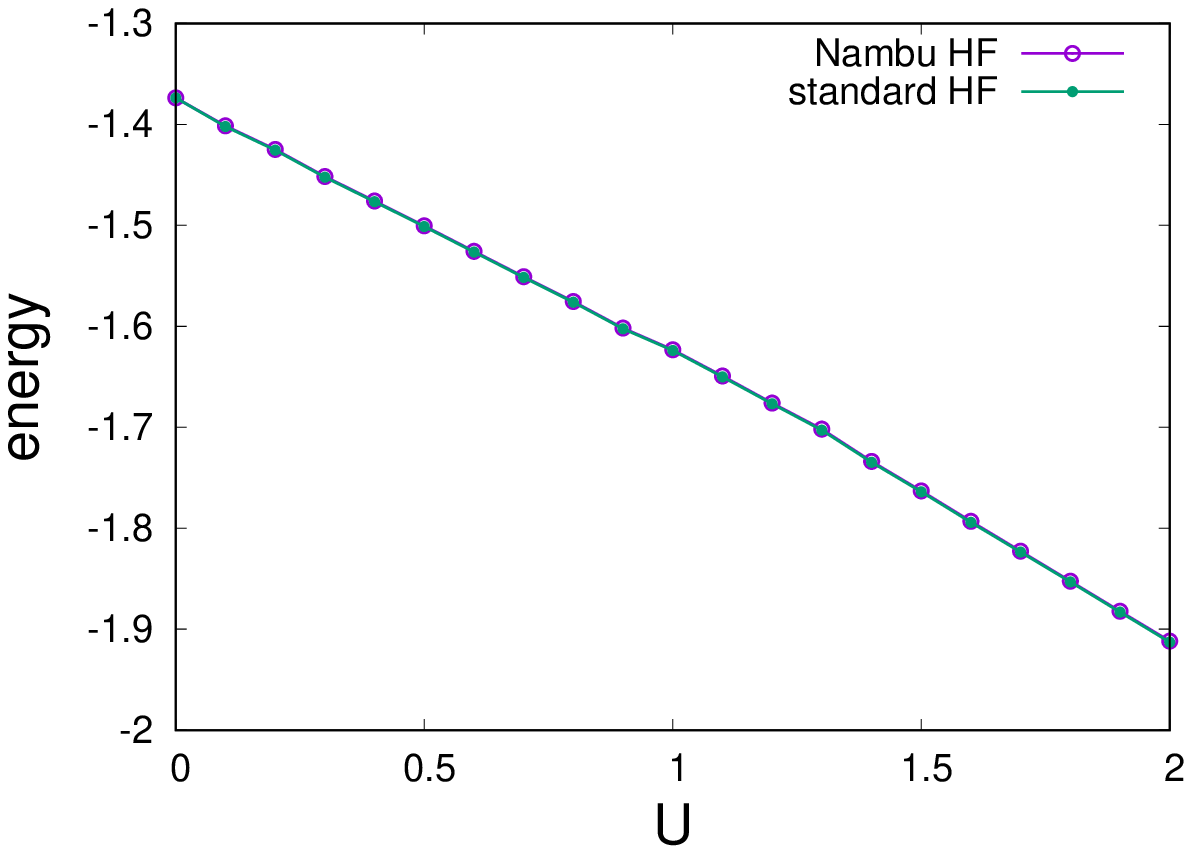}
\label{d10tn1}
}
\caption{Comparison of the energy density expectation values of the Hubbard Hamiltonian vs.\ $U$ in the
standard and Nambu Hartree-Fock ground states for a variety of parameters:  (a) density 0.9, $t'=0$; 
(b) density 0.9, $t'=0.1$. (c) density 0.9, $t'=0.3$; (d) density 0.7, $t'=0.1$; (e) density 0.7, $t'=0.3$;
(f) density 1.1, $t'=0.1$; (g) density 1.1, $t'=-0.1$. Subfigures (h-j) are all at density 1.0, showing that the energies of standard and Nambu Hartree-Fock are very nearly degenerate in this range of $U$ at each $t'=0,0.1,-0.1$.}
\label{main}
\end{figure*}

 The importance of next-nearest (and even next-next-nearest) interactions in the
context of the $t-t'-J$ model has been stressed in \cite{Kivelson,Gong,Jiang}, and
the relation between $t'$ and $T_c$ is discussed in \cite{Pavarini2001, Marki2005}.   Also there is an ARPES phenomenological study \cite{Matsuyama2013} suggesting the significance of $t'$ in fitting cuprate photoemission lineshape data in the framework of the t-J model. These studies all indicate the relevance of next-nearest neighbor interations, parametrized by $t'$, so in this
section we consider some cases with $t' \ne 0$.
 
   Figure \ref{main} compares the energy expectation values of the Hubbard Hamiltonian in the
ground state of the standard Hartree-Fock approximation, $\langle \Om^S |H| \Om^S \rangle$, with
the same quantity $\langle \Om^N |H| \Om^N \rangle$ evaluated in the ground state of
the Nambu Hartree-Fock formulation.  The energy expectation values are evaluated for a range of $U$ at $t=1$, at densities $d=0.7,0.9,1.0,1.1$ for several
values of $t'$.  For standard Hartree-Fock the ground state is an eigenstate of density, while in Nambu Hartree-Fock the density is an expectation value adjusted via the chemical potential.  The energies are compared at equal densities, and the same set of parameters $U,t',\mu$, with $t=1$.  What we find is that at $t'=0$ the Nambu and standard energy densities are very nearly degenerate,
and below half-filling there is a range of $t'>0$ where the Nambu Hartree-Fock ground state
is lower in energy than the standard Hartree-Fock ground state.  Above half-filling, our results indicate that $|\Om^N\rangle$ with d-wave condensation may be energetically favorable at $t'<0$.  At exactly half filling, the two ground state energies are essentially degenerate,
and insensitive to $t'$.
 
At this point we note that $t'>0$ below half-filling is probably unphysical.  There is evidence that,
in cuprates, $t'>0$ corresponds to electron doping (greater than half-filling), while $t'<0$ is associated
with hole doping (less than half-filling) \cite{tohyama1994role,tohyama1990physical,eskes1989effective}.  Of course the $t'>0$ condition is unnecessary for pairing in the Nambu formulation, but it does come into play if we demand that the Nambu state is energetically preferred.  
 
    It is not clear how seriously we should take this energy comparison.  As we have noted previously, both the standard and Nambu
Hartree-Fock approximations involve discarding either number-changing terms, in the standard formulation, or exchange terms,
in the Nambu formulation, which introduces a systematic error on top of the mean-field approximation.  We believe the situation could
be improved by applying a Boguliubov approach, in which all terms in the mean-field approximation are retained.  This possibility is currently under investigation.

\section{\label{Conclude} Conclusions}

    By reformulating the Hartree-Fock approach to the Hubbard model in terms of Nambu spinors, we have been able to demonstrate, within this mean field approximation, the emergence of d-wave condensation in two dimensions,  and have roughly mapped out the boundary, in the plane of hole-doping and $U/t$, between a locally antiferromagnetic region where stripes and other geometric patterns are seen, and a region where there is a d-wave condensate. Condensates of this type emerge via 
spontaneous symmetry breaking, and our approach is to introduce a small symmetry-breaking term proportional to a parameter $h$, and study the dependence of the condensate on lattice volume and on $h$.  An increase in the condensate amplitude with increasing volume, even as $h$ is reduced, is the signal that the condensate is due to symmetry breaking, rather than simply a linear response to the symmetry breaking term. 

    We have also seen that a configuration very much like a Mott insulator also emerges at half-filling, but in a somewhat unexpected way. At half-filling and large $U/t$ we find that each site is occupied with certainty by either a spin up or spin down electron, with spins arranged in a checkerboard pattern such that the nearest neighbors of an electron of a given spin are electrons of the opposite spin.  There is also an energy gap between occupied and unoccupied states.  But electron localization does not arise from highly localized one-particle eigenstates of the effective Hamiltonian.  In fact those states are unlocalized, but together conspire to produce the localization of electron number just described.  Away from half-filling, but prior to the emergence of a condensate, periodic geometric patterns such as stripes are observed in the spin density.

    There are many directions for future work.  For one thing, we have not yet systematically mapped out the boundaries between antiferromagnetic and ferromagnetic regions.  Also, since the computations in this work are all at zero temperature, the effects of finite temperature should be investigated.  Finally, as mentioned at the end of the previous section, a Boguliubov approach that avoids dropping either the exchange or the number changing terms in the Hartree-Fock approach remains to be explored.  We hope to report on these further studies at a later time.

\bigskip

\ni {\bf Acknowledgements} \\
This work is supported by the U.S.\ Department of Energy under Grant No.\ DE-SC0013682.

\appendix

\section{}

    Starting again with the Hubbard model Hamiltonian 
 \bea
 H &=& -t \sum_{<xy>} (c^\dg(x,\ua) c(y,\ua) + c^\dg(x,\da) c(y,\da)) 
        -t' \sum_{[xy]} (c^\dg(x,\ua) c(y,\ua) + c^\dg(x,\da) c(y,\da))   \non \\
     & & ~~+ U\sum_x c^\dg(x,\ua) c(x,\ua) c^\dg(x,\da) c(x,\da) 
            - \m \sum_x  (c^\dg(x,\ua) c(x,\ua) + c^\dg(x,\da) c(x,\da)) \non \\
    &=&  K + U \sum_x V_x - \m \tilde{N} \ ,
 \eea
 where $<xy>$ and $[xy]$ are nearest and next-nearest neighbors, respectively, and where
 \beq
 V_x = c^\dg(x,\ua) c(x,\ua) c^\dg(x,\da) c(x,\da) \ ,
 \eeq
we assume we have calculated the Hartree-Fock ground state
\beq
       |\Om^S\rangle = \prod_{i=1}^M \sum_{x_i,s_i} \phi_i(x_i,s_i) c^\dg(x_i,s_i)
\eeq
by the methods explained in our last paper \cite{Matsuyama:2022kam}.  This is an eigenstate of (spin up + spin down) electron number, with $M$ the total number of electrons.  It will be convenient to introduce
\beq
          c^\dg_i \equiv  \sum_{x_i,s_i} \phi_i(x_i,s_i) c^\dg(x_i,s_i) \ ,
\eeq
in which case
\beq
   |\Om^S\rangle = \prod_{i=1}^M c^\dg_i |0\rangle \equiv c^\dg_1  c^\dg_2 c^\dg_3 .... c^\dg_M|0\rangle \ ,
\eeq
and also the corresponding bra vector
\beq
  \langle \Om^S| =   \langle 0 |\prod_{i=M}^1 c_i  \equiv \langle 0 | c_M c_{M-1} c_{M-2} .... c_1 \ .
\eeq
We want to compute
\bea
 \langle \Om^S| V_x |\Om^S \rangle 
  &=& \langle \Om^S| c^\dg(x,\ua) c(x,\ua) c^\dg(x,\da) c(x,\da) |\Om^S \rangle \non \\
    &=& \langle \Om^S| c^\dg(x,\ua)  c^\dg(x,\da) c(x,\da) c(x,\ua) |\Om^S \rangle \non \\
    &=& \langle 0 |\left(\prod_{i=M}^1 c_i \right) c^\dg(x,\ua)  c^\dg(x,\da) c(x,\da) c(x,\ua) 
           \left(\prod_{i=1}^M c^\dg_i \right) |0\rangle \non \\
\label{A7}
\eea
using the anticommutation relations
\bea
         \{ c(x,s), c^\dg_i \} &=& \phi_i(x,s) \non \\
         \{ c^\dg(x,s), c_i \} &=& \phi^*_i(x,s)  \ .
\eea 
Anticommute $c^\dg(x,\ua), c^\dg(x,\da)$ to the left and $c(x,\da) c(x,\ua)$ tor the right in \rf{A7}. This leaves

{\small
\bea
 \lefteqn{ \hspace{-30pt}\langle \Om^S| V_x |\Om^S \rangle} \non \\
   &\hspace{-30pt}=&\hspace{-10pt}\sum_{i_1} \sum_{j_1} (-1)^{i_1-1} (-1)^{j_1-1} \phi^*_{j_1}(x,\ua) \phi_{i_1}(x,\ua) 
     \langle 0 |\left(\prod_{\underset{j\ne j_1}{j=M}}^1 c_i \right) c^\dg(x,\da) c(x,\da) 
        \left(\prod_{\underset{i\ne i_1}{i=1}}^M c^\dg_i \right) |0\rangle \non \\
   &\hspace{-30pt}=&\hspace{-10pt}\sum_{i_1} \sum_{j_1} \sum_{i_2 \ne i_1} \sum_{j_2 \ne j_1}
    \phi^*_{j_1}(x,\ua) \phi_{i_1}(x,\ua) \phi^*_{j_2}(x,\da) \phi_{i_2}(x,\da) 
   (-1)^{p(i_1,i_2,j_1,j_2)}  \langle 0 |\left(\prod_{\underset{j\ne j_1,j_2}{j=M}}^1 c_i \right)
        \left(\prod_{\underset{i\ne i_1,i_2}{i=1}}^M  c^\dg_i \right) |0\rangle
\eea
}
where $(-1)^{p(i_1,i_2,j_1,j_2)}$ is an overall sign which comes from anticommuting the $c^\dg(x,s)$ operators to the left and $c(x,s)$ to the right.  Now, because of the orthogonality of the $\phi_i(x,s)$
\beq
      \sum_x \sum_{s=\ua \da} \phi_j^*(x,s)  \phi_i(x,s) = \d_{ij} 
\eeq
we have
\bea
\langle 0 |\left(\prod_{\underset{j\ne j_1,j_2}{j=M}}^1 c_j \right)   
    \left(\prod_{\underset{i\ne i_1,i_2}{i=1}}^M  c^\dg_i \right) |0\rangle 
           = \d_{i_1,j_1} \d_{i_2,j_2} + \d_{i_1,j_2} \d_{i_2,j_1} \ ,
\eea
and then, for $i_1=j_1, i_2=j_2$ we have
\bea
 \hspace{-40pt}(-1)^{p(i_1,i_2,j_1,j_2)}  &=& \left\{ \begin{array}{cl} 
        (-1)^{i_1-1}  (-1)^{i_1-1}  (-1)^{i_2-1}  (-1)^{i_2-1} & i_2<i_1 \cr
         (-1)^{i_1-1}  (-1)^{i_1-1}  (-1)^{i_2-2}  (-1)^{i_2-2} & i_2>i_1 \end{array} \right.  = 1 \ ,
\eea
while for $i_1=j_2, i_2=j_1$
\bea
\hspace{-40pt} (-1)^{p(i_1,i_2,j_1,j_2)} &=& \left\{ \begin{array}{cl}
         (-1)^{i_1-1}  (-1)^{i_2-1}  (-1)^{i_2-1}  (-1)^{i_1-2} & i_2<i_1 \cr
         (-1)^{i_1-1}  (-1)^{i_2-1}  (-1)^{i_2-2}  (-1)^{i_1-1} & i_2>i_1 \end{array} \right. = -1 \ .
\eea
Therefore
\bea
 \langle \Om^S| V_x |\Om^S \rangle &=& \sum_{i_1} \sum_{i_2 \ne i_1}
 \bigg( \phi^*_{i_1}(x,\ua) \phi_{i_1}(x,\ua)  \phi^*_{i_2}(x,\da) \phi_{i_2}(x,\da)
         -  \phi^*_{i_1}(x,\da) \phi_{i_1}(x,\ua)  \phi^*_{i_2}(x,\ua) \phi_{i_2}(x,\da) \bigg) \non \\
&=& \sum_{i_1} \sum_{i_2}
 \bigg( \phi^*_{i_1}(x,\ua) \phi_{i_1}(x,\ua)  \phi^*_{i_2}(x,\da) \phi_{i_2}(x,\da)
         -  \phi^*_{i_1}(x,\da) \phi_{i_1}(x,\ua)  \phi^*_{i_2}(x,\ua) \phi_{i_2}(x,\da) \bigg) \non \\
&=&  \r(x,\ua\ua) \r(x,\da\da) - \r(x,\ua\da) \r(\da\ua)
 \eea
 where we note that in the second line the contribution from $i_2=i_1$ cancels between the
 two terms, so that the second line is the same as the first, and we have
\bea
\hspace{-10pt}\langle \Om^S|U\sum_x c^\dg(x,\ua) c(x,\ua) c^\dg(x,\da) c(x,\da) |\Om^S \rangle
 = ~U \sum_x \bigg( \r(x,\ua \ua) \r(x,\da \da) -  \r(x,\ua \da) \r(x,\da \ua) \bigg) \ .
\eea
The hopping and chemical potential terms are much simpler:
\bea 
\lefteqn{ \hspace{-30pt}\langle \Om^S |K-\m \tilde{N}| \Om^S \rangle} \non \\
   &=& -t \sum_{<xy>}   \langle 0 |\left(\prod_{j=M}^1 c_i \right) \bigg\{ c^\dg(x,\ua) c(y,\ua) +
       c^\dg(x,\da) c(y,\da) \bigg\}\left(\prod_{i=1}^M c^\dg_i \right) |0\rangle \non \\
       & & -t' \sum_{[xy]}  \langle 0 |\left(\prod_{j=M}^1 c_i \right) \bigg\{ c^\dg(x,\ua) c(y,\ua) + c^\dg(x,\da) c(y,\da)
       \bigg\}\left(\prod_{i=1}^M c^\dg_i \right) |0\rangle - \m M  \non \\
       &=& -t \sum_{<xy>}  \bigg\{ \phi^*_i(x,\ua) \phi_i(y,\ua) + \phi^*_i(x,\da) \phi_i(y,\da) \bigg\} \non \\
       & & -t' \sum_{[xy]}  \bigg\{ \phi^*_i(x,\ua) \phi_i(y,\ua) + \phi^*_i(x,\da) \phi_i(y,\da) \bigg\} - \m M\non \\
       &=& -t \sum_x \{ \rho_{nn}(x,\ua \ua) + \rho_{nn}(x,\da \da)\}
               -t'\sum_x \{\r_{nnn}(x,\ua \ua) + \r_{nnn}(x,\da\da)\} - \m M \ ,
\eea
where we have defined $\r_{nn}, \r_{nnn}$ in \rf{rhon}.
Thus our final answer, for the energy expectation value of the standard Hartree-Fock state, is
\bea 
\langle \Om^S |H| \Om^S \rangle 
 &=& -t \sum_x \{ \rho_{nn}(x,\ua \ua) + \rho_{nn}(x,\da \da) \} 
  -t' \sum_x \{ \rho_{nnn}(x,\ua \ua) + \rho_{nnn}(x,\da \da) \} \non \\
& &  + U \sum_x \{ \r(x,\ua \ua) \r(x,\da \da) - 
                  \r(x,\ua \da) \r(x,\da \ua) \bigg) \}  - \m M  \ ,
\eea
in accordance with \rf{ES}. 
The energy expectation value  $\langle \Om^N |H| \Om^N \rangle$ of the Hubbard Hamiltonian in the Nambu Hartree-Fock ground state is computed in the same way. Again start by writing down the Hubbard model Hamiltonian in terms of Nambu spinors
\bea
H &=& -t \sum_{<xy>} (c^\dg(x,\ua) c(y,\ua) + c^\dg(x,\da) c(y,\da))  
          -t'  \sum_{[xy]} (c^\dg(x,\ua) c(y,\ua) + c^\dg(x,\da) c(y,\da)) \non \\ 
     & & + U\sum_x c^\dg(x,\ua) c(x,\ua) c^\dg(x,\da) c(x,\da) 
     - \m\sum_{x} (c^\dg(x,\ua) c(x,\ua) + c^\dg(x,\da) c(x,\da))   \non \\
    &=& -t \sum_{<xy>} \{\p^\dg_1(x) \p_1(y) + \p_2(x) \p^\dg_2(y)\} 
            -t' \sum_{[xy]} \{\p^\dg_1(x) \p_1(y) + \p_2(x) \p^\dg_2(y)\} \non \\
     & &      + U\sum_x \p^\dg_1(x) \p_1(x) \p_2(x) \p^\dg_2(x) 
           - \m \sum_{x} \{\p^\dg_1(x) \p_1(x) + \p_2(x) \p^\dg_2(x)\}  \non \\
    &=& -t \sum_{<xy>} \{\p^\dg_1(x) \p_1(y) - \p^\dg_2(y) \p_2(x)\} 
     -t' \sum_{[xy]} \{\p^\dg_1(x) \p_1(y) - \p^\dg_2(y) \p_2(x)\} \non \\
    & & + U\sum_x \p^\dg_1(x) \p_1(x) (1 - \p^\dg_2(x) \p_2(x)\}  
    -  \m \sum_{x} \{\p^\dg_1(x) \p_1(x) + 1 - \p^\dg_2(x) \p_2(x)\}  \non \\
    &=& -t \sum_{<xy>} \{\p^\dg_1(x) \p_1(y) - \p^\dg_2(x) \p_2(y)\}  
    -t' \sum_{[xy]} \{\p^\dg_1(x) \p_1(y) - \p^\dg_2(x) \p_2(y)\}  \non \\
    & &+ U \sum_x \p^\dg_1(x) \p_1(x) 
       -U \sum_x \p^\dg_1(x) \p_1(x) \p^\dg_2(x) \p_2(x) \non \\
    &  &   
       - \m  \sum_{x} \{\p^\dg_1(x) \p_1(x) + 1 - \p^\dg_2(x) \p_2(x)\} \ ,
 \eea
 with the Nambu Hartree-Fock ground state
 \bea
\hspace{-40pt} |\Om^N \rangle &=& \prod_{i=1}^{L^2} \sum_{x_i,\a_i} \phi_i(x_i,\a_i) \p_{\a_i}(x_i) |0\rangle 
 = \prod_{i=1}^{L^2}  \psi_{[i]}^\dg |0\rangle \ ,\hspace{-5pt} ~~~~\mbox{where} ~~~~
 \psi_{[i]}^\dg \equiv  \sum_{x_i,\a} \phi_i(x,\a) \p_\a(x) \ ,
 \eea
 and the $\phi_i(x,\a)$ are determined numerically from the eigenvalue equation \rf{eval}, as explained previously.
 
 The algebra goes in exactly the same way as for the standard Hartree-Fock energy expectation value, and from our previous results we can essentially write down the answer from inspection:
  \bea
\lefteqn{\hspace{-30pt} \langle \Om^N |H| \Om^N \rangle} \non \\
   &\hspace{-10pt}=&-t \sum_x \{ \rho_{nn}(x,11) - \rho_{nn}(x,22) \}  
 -t' \sum_x \{ \rho_{nnn}(x,11) - \rho_{nnn}(x,22) \}  
        + U \sum_x \r(x,11) \non \\
        &\hspace{-10pt}& - U\sum_x \bigg( \r(x,11) \r(x,22) -
                  \r(x,12) \r(x,21) \bigg) 
 - \m\sum_x \{ \rho(x,11) + 1 - \rho(x,22) \} ,
 \eea
 as stated in eq.\ \rf{EN}.

\bibliography{hub}  

\begin{thebibliography}{10}
\expandafter\ifx\csname url\endcsname\relax
  \def\url#1{\texttt{#1}}\fi
\expandafter\ifx\csname urlprefix\endcsname\relax\def\urlprefix{URL }\fi
\expandafter\ifx\csname href\endcsname\relax
  \def\href#1#2{#2} \def\path#1{#1}\fi

\bibitem{Hirsch}
J.~E. Hirsch,
  \href{https://link.aps.org/doi/10.1103/PhysRevB.31.4403}{Two-dimensional
  hubbard model: Numerical simulation study}, Phys. Rev. B 31 (1985)
  4403--4419.
\newblock \href {https://doi.org/10.1103/PhysRevB.31.4403}
  {\path{doi:10.1103/PhysRevB.31.4403}}.
\newline\urlprefix\url{https://link.aps.org/doi/10.1103/PhysRevB.31.4403}

\bibitem{Penn}
D.~R. Penn, \href{https://link.aps.org/doi/10.1103/PhysRev.142.350}{Stability
  theory of the magnetic phases for a simple model of the transition metals},
  Phys. Rev. 142 (1966) 350--365.
\newblock \href {https://doi.org/10.1103/PhysRev.142.350}
  {\path{doi:10.1103/PhysRev.142.350}}.
\newline\urlprefix\url{https://link.aps.org/doi/10.1103/PhysRev.142.350}

\bibitem{Poilblanc}
D.~Poilblanc, T.~M. Rice,
  \href{https://link.aps.org/doi/10.1103/PhysRevB.39.9749}{Charged solitons in
  the hartree-fock approximation to the large-u hubbard model}, Phys. Rev. B 39
  (1989) 9749--9752.
\newblock \href {https://doi.org/10.1103/PhysRevB.39.9749}
  {\path{doi:10.1103/PhysRevB.39.9749}}.
\newline\urlprefix\url{https://link.aps.org/doi/10.1103/PhysRevB.39.9749}

\bibitem{Zaanen}
J.~Zaanen, O.~Gunnarsson,
  \href{https://link.aps.org/doi/10.1103/PhysRevB.40.7391}{Charged magnetic
  domain lines and the magnetism of high-${T}_{c}$ oxides}, Phys. Rev. B 40
  (1989) 7391--7394.
\newblock \href {https://doi.org/10.1103/PhysRevB.40.7391}
  {\path{doi:10.1103/PhysRevB.40.7391}}.
\newline\urlprefix\url{https://link.aps.org/doi/10.1103/PhysRevB.40.7391}

\bibitem{Machida}
K.~Machida, Magnetism in la2cuo4 based compounds, Physica C: Superconductivity
  158~(1-2) (1989) 192--196.

\bibitem{Schulz1}
H.~Schulz, Domain walls in a doped antiferromagnet, Journal de Physique 50~(18)
  (1989) 2833--2849.

\bibitem{Schulz2}
H.~J. Schulz,
  \href{https://link.aps.org/doi/10.1103/PhysRevLett.64.1445}{Incommensurate
  antiferromagnetism in the two-dimensional hubbard model}, Phys. Rev. Lett. 64
  (1990) 1445--1448.
\newblock \href {https://doi.org/10.1103/PhysRevLett.64.1445}
  {\path{doi:10.1103/PhysRevLett.64.1445}}.
\newline\urlprefix\url{https://link.aps.org/doi/10.1103/PhysRevLett.64.1445}

\bibitem{Ichimura}
M.~Ichimura, M.~Fujita, K.~Nakao,
  \href{https://doi.org/10.1143/JPSJ.61.2027}{Spin density wave states in two
  dimensional hubbard model}, Journal of the Physical Society of Japan 61~(6)
  (1992) 2027--2039.
\newblock \href {http://arxiv.org/abs/https://doi.org/10.1143/JPSJ.61.2027}
  {\path{arXiv:https://doi.org/10.1143/JPSJ.61.2027}}, \href
  {https://doi.org/10.1143/JPSJ.61.2027} {\path{doi:10.1143/JPSJ.61.2027}}.
\newline\urlprefix\url{https://doi.org/10.1143/JPSJ.61.2027}

\bibitem{Verges}
J.~Verg{\'e}s, E.~Louis, P.~Lomdahl, F.~Guinea, A.~Bishop, Holes and magnetic
  textures in the two-dimensional hubbard model, Physical Review B 43~(7)
  (1991) 6099.

\bibitem{Inui}
M.~Inui, P.~Littlewood, Hartree-fock study of the magnetism in the single-band
  hubbard model, Physical Review B 44~(9) (1991) 4415.

\bibitem{Dasgupta}
C.~Dasgupta, J.~Halley, Phase diagram of the two-dimensional disordered hubbard
  model in the hartree-fock approximation, Physical Review B 47~(2) (1993)
  1126.

\bibitem{Xu}
J.~Xu, C.-C. Chang, E.~J. Walter, S.~Zhang, Spin-and charge-density waves in
  the hartree--fock ground state of the two-dimensional hubbard model, Journal
  of Physics: Condensed Matter 23~(50) (2011) 505601.

\bibitem{Scalettar}
R.~Scalettar, \href{http://www.cond-mat.de/events/correl16}{An introduction to
  the hubbard model}, in: E.~Pavarini, E.~Koch, J.~van~den Brink, G.~Sawatzky
  (Eds.), Quantum Materials: Experiments and Theory, 2016.
\newline\urlprefix\url{http://www.cond-mat.de/events/correl16}

\bibitem{Powell}
B.~J. Powell, An introduction to effective low-energy hamiltonians in condensed
  matter physics and chemistry (2010).
\newblock \href {http://arxiv.org/abs/0906.1640} {\path{arXiv:0906.1640}}.

\bibitem{Lechermann}
F.~Lechermann, \href{http://www.cond-mat.de/events/correl11}{Model hamiltonians
  and basic techniques}, in: E.~Pavarini, E.~Koch, A.~Lichtenstein (Eds.), The
  LDA+DMFT approach to strongly correlated materials, 2011.
\newline\urlprefix\url{http://www.cond-mat.de/events/correl11}

\bibitem{Imada}
M.~Imada, A.~Fujimori, Y.~Tokura,
  \href{https://link.aps.org/doi/10.1103/RevModPhys.70.1039}{Metal-insulator
  transitions}, Rev. Mod. Phys. 70 (1998) 1039--1263.
\newblock \href {https://doi.org/10.1103/RevModPhys.70.1039}
  {\path{doi:10.1103/RevModPhys.70.1039}}.
\newline\urlprefix\url{https://link.aps.org/doi/10.1103/RevModPhys.70.1039}

\bibitem{Fazekas}
P.~Fazekas, \href{https://www.worldscientific.com/doi/abs/10.1142/2945}{Lecture
  Notes on Electron Correlation and Magnetism}, WORLD SCIENTIFIC, 1999.
\newblock \href
  {http://arxiv.org/abs/https://www.worldscientific.com/doi/pdf/10.1142/2945}
  {\path{arXiv:https://www.worldscientific.com/doi/pdf/10.1142/2945}}, \href
  {https://doi.org/10.1142/2945} {\path{doi:10.1142/2945}}.
\newline\urlprefix\url{https://www.worldscientific.com/doi/abs/10.1142/2945}

\bibitem{Matsuyama:2022kam}
K.~Matsuyama, J.~Greensite, {Multiplicity, localization, and domains in the
  Hartree\textendash{}Fock ground state of the two-dimensional Hubbard model},
  Annals Phys. 442 (2022) 168922.
\newblock \href {http://arxiv.org/abs/2201.05750} {\path{arXiv:2201.05750}},
  \href {https://doi.org/10.1016/j.aop.2022.168922}
  {\path{doi:10.1016/j.aop.2022.168922}}.

\bibitem{Kivelson}
H.-C. Jiang, S.~A. Kivelson,
  \href{https://link.aps.org/doi/10.1103/PhysRevLett.127.097002}{High
  temperature superconductivity in a lightly doped quantum spin liquid}, Phys.
  Rev. Lett. 127 (2021) 097002.
\newblock \href {https://doi.org/10.1103/PhysRevLett.127.097002}
  {\path{doi:10.1103/PhysRevLett.127.097002}}.
\newline\urlprefix\url{https://link.aps.org/doi/10.1103/PhysRevLett.127.097002}

\bibitem{Gong}
S.~Gong, W.~Zhu, D.~N. Sheng,
  \href{https://link.aps.org/doi/10.1103/PhysRevLett.127.097003}{Robust
  $d$-wave superconductivity in the square-lattice $t\text{\ensuremath{-}}j$
  model}, Phys. Rev. Lett. 127 (2021) 097003.
\newblock \href {https://doi.org/10.1103/PhysRevLett.127.097003}
  {\path{doi:10.1103/PhysRevLett.127.097003}}.
\newline\urlprefix\url{https://link.aps.org/doi/10.1103/PhysRevLett.127.097003}

\bibitem{Jiang}
S.~Jiang, D.~J. Scalapino, S.~R. White,
  \href{https://www.pnas.org/doi/abs/10.1073/pnas.2109978118}{Ground-state
  phase diagram of the t-t'-j model}, Proceedings of the National Academy of
  Sciences 118~(44) (2021) e2109978118.
\newblock \href
  {http://arxiv.org/abs/https://www.pnas.org/doi/pdf/10.1073/pnas.2109978118}
  {\path{arXiv:https://www.pnas.org/doi/pdf/10.1073/pnas.2109978118}}, \href
  {https://doi.org/10.1073/pnas.2109978118}
  {\path{doi:10.1073/pnas.2109978118}}.
\newline\urlprefix\url{https://www.pnas.org/doi/abs/10.1073/pnas.2109978118}

\bibitem{Pavarini2001}
E.~Pavarini, I.~Dasgupta, T.~Saha-Dasgupta, O.~Jepsen, O.~K. Andersen,
  \href{https://link.aps.org/doi/10.1103/PhysRevLett.87.047003}{Band-structure
  trend in hole-doped cuprates and correlation with
  ${\mathit{t}}_{\mathit{c}\mathrm{max}}$}, Phys. Rev. Lett. 87 (2001) 047003.
\newblock \href {https://doi.org/10.1103/PhysRevLett.87.047003}
  {\path{doi:10.1103/PhysRevLett.87.047003}}.
\newline\urlprefix\url{https://link.aps.org/doi/10.1103/PhysRevLett.87.047003}

\bibitem{Marki2005}
R.~S. Markiewicz, S.~Sahrakorpi, M.~Lindroos, H.~Lin, A.~Bansil,
  \href{https://link.aps.org/doi/10.1103/PhysRevB.72.054519}{One-band
  tight-binding model parametrization of the high-${T}_{c}$ cuprates including
  the effect of ${k}_{z}$ dispersion}, Phys. Rev. B 72 (2005) 054519.
\newblock \href {https://doi.org/10.1103/PhysRevB.72.054519}
  {\path{doi:10.1103/PhysRevB.72.054519}}.
\newline\urlprefix\url{https://link.aps.org/doi/10.1103/PhysRevB.72.054519}

\bibitem{Matsuyama2013}
K.~Matsuyama, G.-H. Gweon,
  \href{https://link.aps.org/doi/10.1103/PhysRevLett.111.246401}{Phenomenological
  model for the normal-state angle-resolved photoemission spectroscopy line
  shapes of high-temperature superconductors}, Phys. Rev. Lett. 111 (2013)
  246401.
\newblock \href {https://doi.org/10.1103/PhysRevLett.111.246401}
  {\path{doi:10.1103/PhysRevLett.111.246401}}.
\newline\urlprefix\url{https://link.aps.org/doi/10.1103/PhysRevLett.111.246401}

\bibitem{tohyama1994role}
T.~Tohyama, S.~Maekawa, Role of next-nearest-neighbor hopping in the t-t'-j
  model, Physical Review B 49~(5) (1994) 3596.

\bibitem{tohyama1990physical}
T.~Tohyama, S.~Maekawa, Physical parameters in copper oxide superconductors,
  Journal of the Physical Society of Japan 59~(5) (1990) 1760--1770.

\bibitem{eskes1989effective}
H.~Eskes, G.~Sawatzky, L.~Feiner, Effective transfer for singlets formed by
  hole doping in the high-tc superconductors, Physica C: Superconductivity
  160~(5-6) (1989) 424--430.

\end{thebibliography}

\end{document}